\documentclass{aa}

\usepackage{graphicx}
\usepackage{txfonts}
\usepackage{lscape}
\usepackage{placeins}
\usepackage{subcaption}
\usepackage{amsmath}
\usepackage{xcolor} 
\usepackage[toc,title,page]{appendix}
\nolinenumbers  

\makeatletter
\let\makeLineNumber\relax
\makeatother

\begin{document}

\title{Stardust galaxies at z>9: \\a dust-origin transition behind the excess of UV-bright galaxies}
\author{D. Burgarella\inst{1} \and 
V. Buat\inst{1} \and 
A.K. Inoue\inst{2} \and 
T.T. Takeuchi\inst{3,4} \and 
C. Aurin\inst{1} \and 
J.-C. Bouret\inst{1} \and 
P. Dayal\inst{5, 6, 7}\and 
T. Dewachter\inst{1} \and 
M. Dickinson\inst{8} \and 
C. Kobayashi\inst{9} \and 
G. P. Nikopoulos\inst{10, 11}  \and 
R. S. Somerville\inst{12}
}
\institute{$^1$Aix Marseille Universit\'e, CNRS, CNES, LAM, Marseille, France\\
$^2$Waseda Research Institute for Science and Engineering, Faculty of Science and Engineering, Waseda University, 3-4-1 Okubo,
Shinjuku, Tokyo 169-8555, Japan\\
$^3$Division of Particle and Astrophysical Science, Nagoya University, Furo-cho, Chikusa-ku, Nagoya 464-8602, Japan\\
$^4$The Research Center for Statistical Machine Learning, the Institute of Statistical Mathematics, 10--3 Midori-cho, Tachikawa, Tokyo 190--8562, Japan\\
$^5$Canadian Institute for Theoretical Astrophysics, 60 St George St, University of Toronto, Toronto, ON M5S 3H8, Canada\\
$^6$David A. Dunlap Department of Astronomy and Astrophysics, University of Toronto, 50 St George St, Toronto ON M5S 3H8, Canada\\
$^7$Department of Physics, 60 St George St, University of Toronto, Toronto, ON M5S 3H8, Canada\\
$8$SF’s National Optical-Infrared Astronomy Research Laboratory, 950 N. Cherry Ave., Tucson, AZ 85719, USA\\
$^9$Centre for Astrophysics Research, Department of Physics, Astronomy and Mathematics, University of Hertfordshire, College Lane, Hatfield AL10 9AB, UK\\
$^{10}$Cosmic Dawn Center (DAWN), Denmark\\
$^{11}$Niels Bohr Institute, University of Copenhagen, Jagtvej 155A, DK-2200, Copenhagen N, Denmark\\
$^{12}$Center for Computational Astrophysics, Flatiron Institute, 162 5th Ave, New York, NY 10010, USA\\
\email{denis.burgarella@lam.fr}}
\date{Received Jan 27, 2026; accepted ...}

\abstract{
Recent James Webb Space Telescope (JWST) observations suggest that galaxies at z $\gtrsim$ 9 may be dominated by low-opacity supernova (SNe)-produced dust before efficient interstellar-medium (ISM) grain growth becomes established. Such a transition in the dominant origin and opacity of dust could naturally explain the prevalence of galaxies with extremely low dust attenuation, as well as the excess of ultraviolet-bright (UV-bright) galaxies relative to most pre-JWST predictions. We investigate whether this transition, combined with a redshift evolution of the star-formation efficiency, can account for these observed properties.

We develop a physically motivated attenuation framework combining (i) extinction laws for SNe dust processed by reverse shocks, (ii) metallicity- and dust-to-metal-dependent opacity scalings, and (iii) porous radiative-transfer geometries allowing partial UV-photon leakage. Unlike outflow-driven scenarios requiring large-scale gas evacuation, our approach preserves gas reservoirs while reducing the effective UV opacity through dust composition and geometry. We use observed and model-motivated galaxy scaling relations to follow how dust mass, metallicity, gas content, galaxy size, and star–dust geometry regulate the far-UV attenuation as galaxies evolve from an early SNe-dominated dust regime toward a later ISM grain-growth regime.

We find that the observed relation between far-UV attenuation A$_{FUV}$, and stellar mass, M$_{\rm star}$, at z$>$9 is best reproduced by an intrinsic FUV dust opacity of $\kappa_{FUV}^{dust}\simeq10^3-10^4 {\rm cm^2.g^{-1}}$, substantially below standard ISM-like values. Within the baseline physical scenario adopted in this work, this low-opacity regime is naturally associated with SN-produced dust processed by reverse shocks and produces very low attenuation even in gas-rich galaxies. It reproduces galaxies with extremely low dust attenuation (GELDAs), which dominate the observed samples at z$\gtrsim$9.

Applied to intrinsic UV luminosity functions from semi-empirical and semi-analytic models, our SNe-dominated and hybrid attenuation prescriptions suppress mainly the brightest sources, bringing theoretical predictions into agreement with JWST measurements showing that only modest dust corrections are needed once the intrinsic UV luminosity functions are close to the JWST ones. Our results suggest that the UV-bright galaxy excess at $z\gtrsim9$ may reflect a transition in dust origin and opacity during the earliest phases of galaxy evolution. In the most metal-poor systems, the dust content of some GELDAs may trace the chemical “ashes” of the first stellar generations, potentially including Population III SNe.
}

\maketitle

\section{Introduction}
\label{sec:introduction}

The James Webb Space Telescope (JWST) has revealed a population of galaxies at redshifts z$>$9 that are unexpectedly bright in ultraviolet (UV) relative to pre-JWST predictions \citep[e.g.][]{Naidu2022, Donnan2023, Bouwens2023b, Harikane2023, Finkelstein2023}. 
Even after revised photometric and spectroscopic redshifts have removed a number of interlopers and spurious candidates \citep[e.g.][]{Zavala2023, Arrabal-Haro2023}, the abundance of bright galaxies remains significantly above most theoretical expectations \citep{Harikane2023, Yung2024}. 

Several mechanisms have been proposed to explain this excess of UV-bright galaxies at $z \gtrsim 9$ 
\citep[e.g.][]{Finkelstein2023, Mason2023, Ferrara2023, Harikane2023}. We list some of them: 
(i) stochastic bursts of star formation \citep[e.g.,][]{Tacchella2020, Shen2023}; 
(ii) unusually high star-formation efficiencies \citep[e.g.,][]{Finkelstein2023, Mason2023, Mauerhofer2025}; 
(iii) a top-heavy stellar initial mass function (IMF) \citep[e.g.,][]{Inayoshi2022, Cueto2024, Trinca2024}; 
(iv) possible contamination by faint active galactic nuclei (AGN) \citep[e.g.,][]{Kocevski2023, Maiolino2024}; and 
(v) reduced dust attenuation \citep[e.g.,][]{Cullen2023, Ferrara2023, Burgarella2025}. 
Among these, the last explanation has gained particular traction because it is quite natural while meeting the observational constraints.

Dust grains play a fundamental role in shaping the observed properties of galaxies, even at the earliest epochs. They absorb and scatter stellar radiation, attenuating and reddening the emergent light, and thus obscure the intrinsic spectral energy distribution (SED) of young galaxies. Recovering the intrinsic UV luminosities therefore requires an attenuation law that accounts for both radiative-transfer effects and the relative geometry of stars and dust \footnote{Attenuation laws describe the wavelength-dependent effect of dust mixed with stars, including scattering, clumpiness, and partial covering \citep[e.g.][]{Charlot2000, Inoue2005, Witt2000, Seon2016}. Radiative-transfer models and local analogs of high-redshift systems demonstrate that such mixed or porous geometries significantly reduce the effective UV optical depth, producing greyer attenuation curves even for intrinsically steep extinction laws \citep{Natta1984, Gordon2003, Calzetti2000, Salim2020}.}.

In the low-metallicity regime expected at $z>9$, SNe are expected to provide an important, and possibly dominant, source of dust before efficient ISM grain growth becomes established (\citealt{Todini2001, Gall2011, DwekCherchneff2011, Popping2017}). Reverse-shock processing can substantially reduce the surviving dust mass and may preferentially remove small grains, thus lowering the effective UV opacity and flattening the extinction curve \citep{Nozawa2007, Hirashita2008, Micelotta2016, Kirchschlager2019, Asano2014}. As a result, the effective attenuation in the far-UV (FUV, $\lambda \simeq 1500\,\text{\AA}$) can be extremely low ($A_{\mathrm{FUV}} \lesssim 0.3$\,mag), consistent with the ``stardust'' scenario in which SNe provide the dominant dust source while ISM grain growth remains inefficient \citep{Ferrara2023, Burgarella2025}. However, the resulting grain-size distribution is not unique: it depends on the initial ejecta grain distribution, the density structure of the ejecta and ambient medium, grain–grain collisions and shattering, and magnetic-field effects. We therefore treat low-opacity SNe-processed dust as a physically motivated interpretation of the inferred low UV opacity, rather than as a unique consequence of reverse-shock processing. Accordingly, we adopt SNe-dominated, reverse-shock-processed dust as the baseline physical scenario for the earliest low-metallicity galaxies. This choice is motivated independently by their short evolutionary timescales, low metallicities, and the expected inefficiency of ISM grain growth, rather than being inferred from the opacity value alone. This framework naturally explains the blue UV continua ($\beta_{UV}\!\lesssim\!-2.5$) of many $z>9$ galaxies observed by JWST. This is consistent with studies that find very young, metal-poor galaxies with only modest reddening, but not attenuation-free, at z$>$9: across the full range in M$_{UV}$ \citep{Cullen2023}.

Assuming that reduced dust attenuation is a major contributor to the observed excess of UV-bright galaxies, we analyse two scenarios respectively discussed by \citet{Ferrara2023} and \citet{Burgarella2025}, hereafter referred to as the \emph{attenuation-free} and \emph{stardust} scenarios. Both frameworks reproduce the JWST UV excess by effectively reducing the dust attenuation. 

In the attenuation-free scenario proposed by \citet{Ferrara2023}, the ISM could become so porous that FUV photons escape along virtually unobscured channels, even though dust grains remain present within the galaxy limits. The conceptual basis of this mechanism traces back to \citet{Ferrara1990}, who first demonstrated that radiation pressure in galactic disks can accelerate and expel dust grains from the ISM, creating dust-depleted sight lines even when metals remain bound to the system. 

In physical terms, intense stellar feedback, including radiation pressure on dust, photoionization heating, stellar winds, and early supernova explosions, injects momentum and thermal energy into dense star-forming clouds, fragmenting them and carving low-density chimneys through which UV radiation escapes. This process drastically reduces the geometric coupling between dust and the young UV-emitting population, such that the luminosity-weighted sight lines acquire very small effective optical depths. In the most extreme case suggested by \citet{Ferrara2023,Ferrara2025}, most of the dust produced by early SNe could have been efficiently expelled or diluted during the first phases of galaxy assembly: this is the \emph{attenuation-free} scenario.

In the stardust scenario, dust forms primarily in SNe and is partially destroyed by the reverse shock, producing small dust masses 2–3\,dex lower than those expected from ISM growth channels \citep{Burgarella2025}. This predicts systematically lower attenuation at lower metallicity while higher-metallicity objects would predominantly increase their dust mass via an efficient accretion of metals on dust grains. A number of models of dust formation, both semi-analytical (e.g. \citealt{Popping2017, Triani2020, Dayal2022, Mauerhofer2023, Somerville2025} and \citealt{Yung2025}) and cosmological (e.g. \citealt{Graziani2020, Lewis2023, DiCesare2023} and \citealt{Choban2024}) predict a critical metallicity in the ISM (and thus a critical stellar mass via the mass-metallicity relation, MZR) above which grain growth via gas–dust accretion becomes efficient, marking the transition from dust production dominated by stellar sources to accretion-driven dust growth in the ISM as proposed by \citet{Asano2013}. The transition between these two dust-production regimes is expected to occur around a critical metallicity of $Z_{\mathrm{crit}}\!\approx\!0.1\,Z_\odot$, corresponding to $12+\log_{10}(\mathrm{O/H})\!\approx\!7.6$ \citep[e.g.][]{Asano2013,DeVis2019,Burgarella2025}. This $Z_{\mathrm{crit}}$ value corresponds to a stellar-mass transition of roughly $8.0\!\lesssim\!\log_{10}(M_{star}/M_\odot)\!\lesssim\!9.0$ at high redshift. While this transition is supported by local-Universe observations (\citealt{RemyRuyer2014} and \citealt{DeVis2019}), it remained undetected or unidentified at redshifts $>$ 0. 

However, recent JWST and ALMA observations show that galaxies with very low measured FUV attenuation, consistent with $A_{\mathrm{FUV}}=0$ within $2\sigma_{AFUV}$ exist, while still being gas-rich yet metal-poor, a combination difficult to reconcile with dust expulsion. For instance, S04590 at z = 8.496 (\citealt{Heintz2023, Fujimoto2024}) presents a gas fraction f$_{gas}$ = M$_{gas}$/(M$_{gas}$ + M$_{star}$) $>$ 90 \% that appears to be somewhat similar to the local metal-poor galaxy I Zw 18 but also very similar to the sample of galaxies identified by \citet{Burgarella2025}. These Galaxies with an  Extremely Low Dust Attenuation (GELDAs) are naturally consistent with the stardust model, where SNe dominate dust production, and suppresses attenuation. 

In this paper, we develop a model that reproduces the observed low $A_{\mathrm{FUV}}$ values and their dependence on stellar mass and metallicity. This model has a twofold objective. First, we test whether stardust production (with various star - dust geometries) can explain the extremely low FUV attenuation, at $z\gtrsim9$. Second, we assess whether this mechanism contributes to the observed excess of UV-luminous ($z\gtrsim9$) galaxies discovered by JWST. Section~\ref{sec:attenuation-framework} introduces the attenuation framework, including the attenuation- and gas-based routes; Section~\ref{sec:comparison} compares the resulting predictions with observations, including the attenuation–mass relation and the observed UV luminosity functions; Section~\ref{sec:discussion} discusses the implications; and Section~\ref{sec:conclusions} summarizes the main results, limitations, and future prospects.
\section{The attenuation framework}
\label{sec:attenuation-framework}

We model the observed far-UV attenuation using two complementary routes. In the attenuation route, the dust optical depth is derived from the dust mass versus stellar mass relation and an adopted galaxy size, and is converted into $A_{\rm FUV}$ using one of three star–dust geometries. In the gas route, the optical depth is instead derived from the gas surface density, metallicity, and dust-to-metal ratio, while using the same attenuation kernels. The two routes therefore differ in how the intrinsic optical depth is constructed, but share the same geometrical treatment.

The SNe-dominated and ISM grain-growth prescriptions are used as limiting dust regimes. The primary fitted model is the hybrid model, which interpolates smoothly between these limits. Table~\ref{tab:model_overview} summarizes the inputs, outputs, geometries, and nomenclature of the models used throughout this paper.

\begin{table*}
\caption{Overview of the attenuation models used in this work}.
\label{tab:model_overview}
\centering
\begin{tabular}{p{2.4cm} p{4.0cm} p{3.2cm} p{3.0cm} p{3.4cm}}
\hline\hline
Route 
& Main relations entering the optical depth 
& Model output 
& Geometry / attenuation kernel 
& Name used in this paper \\
\hline

Attenuation route 
& Hybrid $M_{\rm dust}$--$M_{\rm star}$ relation and size--mass relation
& $A_{\rm FUV}(M_{\rm star})$
& Leaky screen
& Attenuation-route screen \\

Attenuation route 
& Hybrid $M_{\rm dust}$--$M_{\rm star}$ relation and size--mass relation
& $A_{\rm FUV}(M_{\rm star})$
& Leaky mixed
& Attenuation-route mixed \\

Attenuation route 
& Hybrid $M_{\rm dust}$--$M_{\rm star}$ relation and size--mass relation
& $A_{\rm FUV}(M_{\rm star})$
& Poissonian clumps
& Attenuation-route Poisson \\

Gas route 
& Gas-fraction relation, mass--metallicity relation, 
DTM--metallicity relation, and size--mass relation
& $A_{\rm FUV}(M_{\rm star})$
& Leaky screen
& Gas-route screen \\

Gas route 
& Gas-fraction relation, mass--metallicity relation, 
DTM--metallicity relation, and size--mass relation
& $A_{\rm FUV}(M_{\rm star})$
& Leaky mixed
& Gas-route mixed \\
\hline
\end{tabular}
\tablefoot{The attenuation route derives the intrinsic optical depth from the hybrid dust mass versus stellar mass relation, whereas the gas route derives it from the  gas surface density, metallicity, and dust-to-metal ratio. The SNe-dominated and ISM grain-growth prescriptions define limiting branches of the hybrid model rather than independent fitted models. The attenuation kernels are shared by the two routes where applicable. Detailed parameter values are given in Appendix~G.}
\end{table*}

\subsection{Assumptions on the stellar populations}
\label{subsec:stellar-pops}
Although our framework can be generalized to metal-poor, compact systems at later cosmic times (e.g. lower redshift starbursts undergoing their first major episodes of star formation), the primary goal here is to try and explain the origin of dust grains and bright-end excess at $z>9$, when the Universe was younger than $\sim\!500$\,Myr \citep{Planck2018}. 
Theory and simulations consistently place the onset of star formation at $t\!\sim\!100$–200\,Myr after the Big Bang (roughly $z\!\sim\!15$–30), with rapid subsequent metal enrichment by the first SNe \citep[e.g.][]{BrommYoshida2011,Bromm2013,Greif2011,Wise2012}. This is where the stellar populations of the observed GELDAs at $z>9$ should have been formed. We stress, however, that this paper does not only intend to model GELDAs but also post-GELDAs galaxies.

Motivated by these constraints and by the very blue UV continua of many $z\gtrsim9$ systems (e.g., \citealt{Cullen2023, Atek2023}), we assume that the galaxies considered here are dominated by a single, young stellar population. 

We stress, however that although galaxies in the higher stellar-mass range (M$_{star} \gtrsim 10^{8 - 9} M_\odot$) may host older stellar populations and dust produced during previous star-formation episodes, ongoing SNe activity can efficiently destroy pre-existing grains, preventing the long-term accumulation of dust. However, the net effect of this destruction would be to further reduce the dust mass and thus the FUV dust attenuation, after the reverse shock. Moreover, the observed far-UV emission is dominated by the most recent star-forming episode, so that the effective attenuation is primarily luminosity-weighted toward the youngest stellar population rather than reflecting the full star-formation history. In this sense, our assumption of a single young stellar population should be understood as a modelling approximation aimed at capturing the dominant contributors to the UV light and its attenuation, rather than as a statement about the absence of earlier star formation.

\subsection{Star–dust geometries and attenuation kernels}
\label{sec:geometry}

We adopt a physically motivated treatment that accounts for stochastic distributions of optical depth and partial covering (clumpiness). Two porosity prescriptions are implemented: leaky and Poisson. The covering fraction $f_{\rm cov}$ specifies the fraction of sight lines intersecting dust, while porosity describes the stochastic spatial distribution of dust that sets the optical-depth distribution along those sight lines. The factor $1-f_{\rm cov}$ is thus the unobscured, dust-free fraction \citep{Natta1984,Inoue2005}. Radiative transfer through clumpy media \citep[e.g.][]{Natta1984,Witt2000} significantly reduces the effective opacity produced by a given dust mass.

\begin{figure}
  \centering
\includegraphics[width=\linewidth]{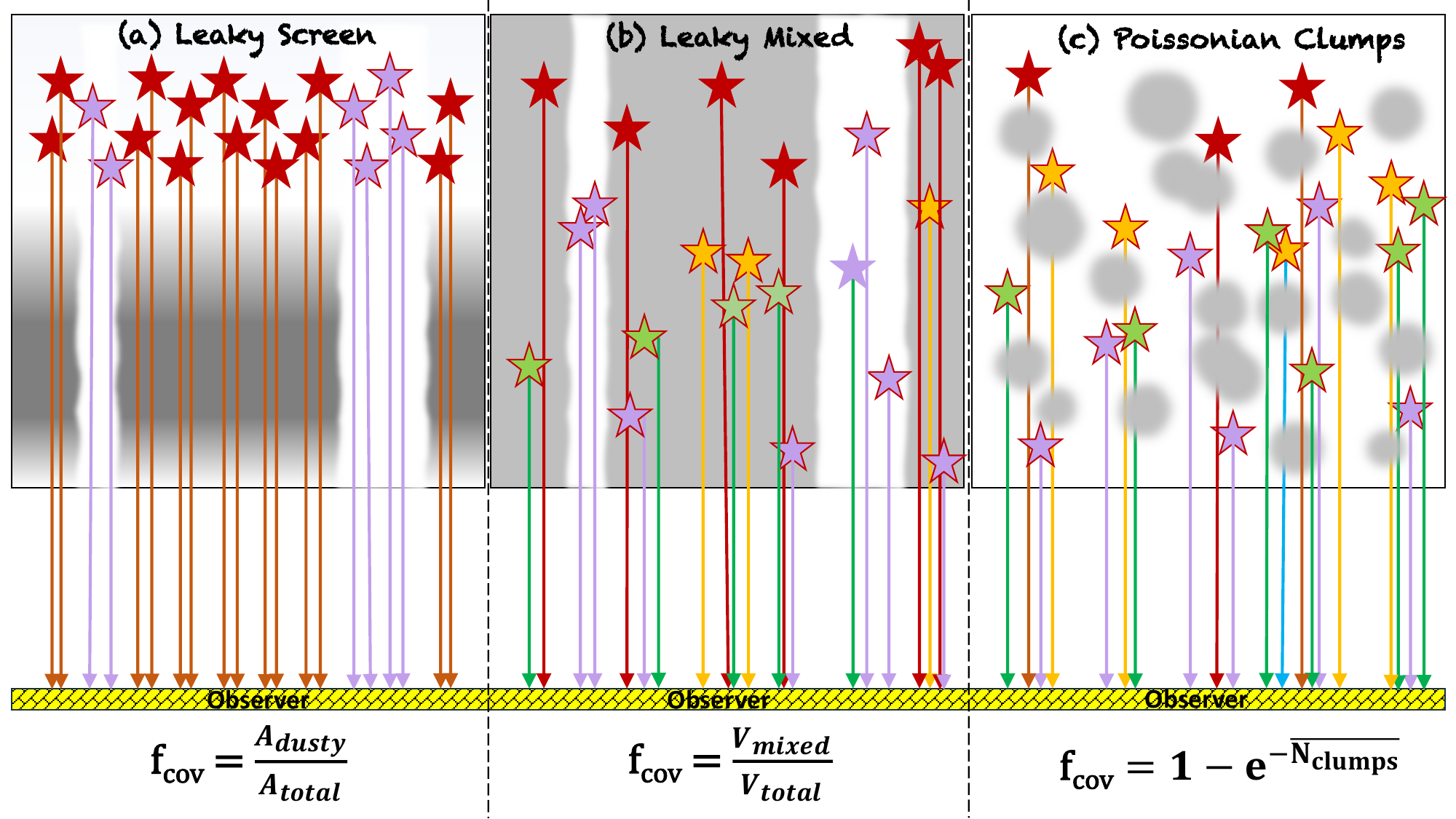}
\caption{
Schematic illustration of the three dust–star geometries adopted in this work. Arrows trace the paths of stellar radiation, with the color gradient from red to purple indicating strong to weak attenuation along each line of sight. (a) \textit{Leaky screen}: a foreground layer of dust with partial covering fraction $f_{\rm cov} = A_{\rm dusty}/A_{\rm total}$, allowing some unobscured sight lines. (b) \textit{Leaky mixed}: a partially filled medium in which dust and stars are interspersed within a mixed volume, $f_{\mathrm{cov}} = V_{\mathrm{mixed}}/V_{\mathrm{tot}}$. (c) \textit{Poissonian clumps}: an inhomogeneous distribution of discrete dusty clumps, characterized by an effective covering fraction $f_{\mathrm{cov}} = 1 - e^{-\bar{N}_{\mathrm{clumps}}}$. These geometries span the range from uniform screens to highly porous, clumpy media and are used to compute the effective attenuation laws discussed in Appendix~\ref{appendix:ext2att}.
}
  \label{fig:geometries}
\end{figure}

Figure~\ref{fig:geometries} illustrates the adopted geometries, while Table~\ref{tab:model_overview} summarizes how they are used in the different attenuation routes.

The geometrical prescriptions introduced here are independent of the route used to construct the intrinsic optical depth. The attenuation-route and gas-route calculations therefore use the same leaky-screen and leaky-mixed kernels; the Poissonian case is additionally explored for the attenuation route. This separation allows us to distinguish uncertainties associated with the dust or gas scaling relations from those associated with star–dust geometry.

\subsection{Attenuation-route model}
\label{subsec:attenuation-route}

In the attenuation route, the intrinsic dust optical depth is constructed from the dust mass and galaxy size. We adopt the SNe-dominated and ISM grain-growth relations as physically motivated low- and high-dust limiting branches, and fit a hybrid model that interpolates smoothly between them. The limiting branches are therefore not treated as independent fitted models; they define the endpoints of the hybrid attenuation model.

Our baseline assumption is that the extremely low $A_{\rm FUV}$ estimated for some GELDAs ($\sim$20 \% of galaxies at $z>4$) and most ($\sim$80 \%) at $z\gtrsim9$ arises because dust is produced by SNe (including reverse-shock grain destruction). This assumption appears across many models (\citealt{Inoue2011, Asano2013, Feldmann2015, Popping2017, Li2019, Graziani2020, Triani2020, Parente2022, Choban2024}). We define two limiting cases, illustrated in the $\log_{10} (M_{dust})$ versus $\log_{10} (M_{star})$ diagram of Fig.~4 in \citet{Witstok2023}: 
(i) Only the most massive stars ($M_{star}>8\,M_{\odot}$) contribute to the metal and dust budget via core-collapse SNe; most grains ($\gtrsim95\%$) are destroyed by the reverse shock (see however \citealt{Gall2018}). 
(ii) At the other extreme, dust growth in the ISM leads to the maximum dust mass physically allowed, assuming all metals accrete onto grains. These are representative cases; intermediate or combined origins are possible. The dust mass ratio derived assuming a stardust or an ISM-growth dust mass origin $M_{\mathrm{dust}}^{\mathrm{growth}}/M_{\mathrm{dust}}^{\mathrm{stardust}}\!\approx\!200$ is quite large but could vary by a factor of a few depending on assumptions. We thus expect that objects lying on the upper and lower branches would present large differences in their observables. 

We use the term “SNe-dominated dust” to denote an effective low-opacity dust component motivated by dust formed in SNe ejecta and subsequently processed by reverse shocks. For the wavelength dependence of this component we consider two representative prescriptions. The first is the reverse-shock-processed SNe dust model of \cite{Hirashita2008}, which is model-based and follows the modification of the grain-size distribution after shock processing. The second is the high-redshift QSO extinction curve of \cite{Maiolino2004}, which is empirically inferred and is often used as an observationally motivated proxy for non-standard, SN-like dust. 

While these curves are not assumed to provide a unique description of early-Universe dust, they are adequate for our objectives; they bracket plausible low-opacity, relatively grey UV extinction behaviors and allow us to test whether the observed $A_{\rm{FUV}} - M_{\rm{star}}$ relation requires dust properties substantially different from standard ISM-like grains.

On the other hand, “ISM grain-growth dust” denotes the high-opacity limiting case in which grain growth in the dense ISM has become efficient. This regime is represented by a larger dust mass at fixed stellar mass and by an ISM-like UV opacity. The hybrid model interpolates between these two limiting regimes as a function of stellar mass or metallicity.
 
Because we intend to model the evolution of $A_{\rm FUV}$ from a stardust-dominated regime to an ISM grain-growth dominated regime, we parameterize the dust-mass - stellar-mass relation using the two limiting branches discussed by \citet[][their Fig.~4]{Witstok2023}. We write the relation as
\begin{equation}
y \equiv \log_{10}\left(\frac{M_{\rm dust}}{M_\odot}\right)
= a + b\,x,
\qquad
x \equiv \log_{10}\left(\frac{M_\star}{M_\odot}\right).
\end{equation}
For the ISM grain-growth dominated branch, we adopt
\begin{equation}
y_{\rm grow}(x) = x - 1.296,
\end{equation}
while for the SNe-only branch we adopt
\begin{equation}
y_{\rm SNe}(x) = x - 3.422.
\end{equation}
These two prescriptions are linear approximations to the two limiting dust-production branches discussed by \citet{Witstok2023}: a low-dust SNe-only branch, in which dust production is dominated by stellar sources and a large fraction of dust grains is destroyed by reverse-shock processing, and a high-dust ISM grain-growth branch, corresponding to the maximum dust mass allowed by efficient grain growth.

\begin{figure*}[!t]
\centering
  \includegraphics[width=0.6\linewidth]{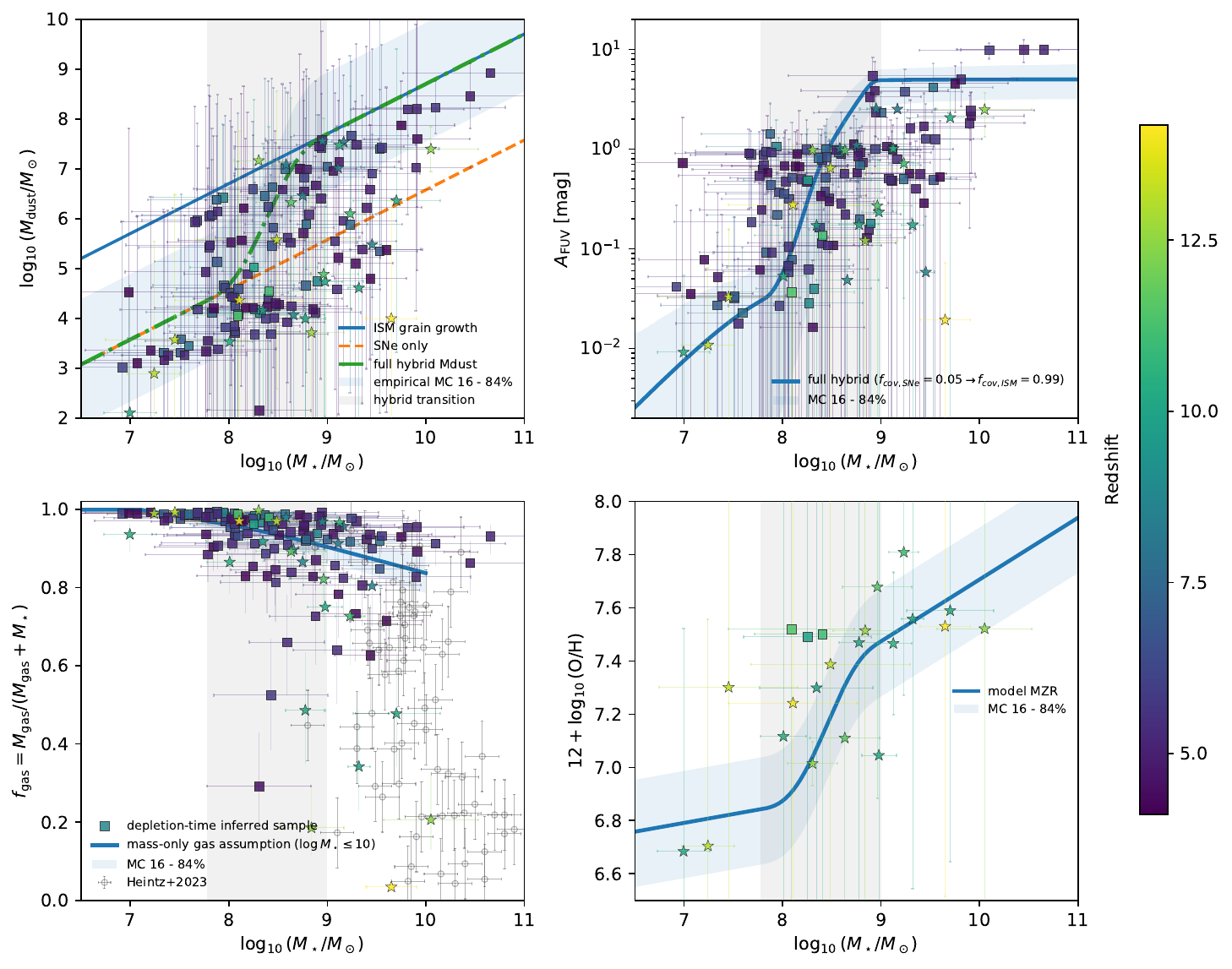}
  \caption{
Joint scaling-relation constraints and best-fitting full hybrid model. All panels are shown as a function of stellar mass. Squares denote the \citealt{Burgarella2025} sample and stars denote the this paper sample. Marker colours encode redshift according to the colour bar, while the horizontal and vertical error bars show the reported measurement uncertainties. The grey vertical region marks the fitted transition between the SNe-dominated and ISM grain-growth regimes. 
\emph{Top left:} Dust-mass - stellar-mass relation. The solid blue and dashed orange curves show the ISM grain-growth and SNe-only limiting relations, respectively. The green curve shows the full hybrid dust-mass relation obtained using the common quintic smootherstep transition. The shaded region denotes the 16th - 84th percentile Monte Carlo interval derived from the empirical dust-mass scatter.
\emph{Top right:} FUV attenuation. The solid blue curve shows the best-fitting full hybrid attenuation relation for the preferred leaky mixed geometry, and the blue shaded region denotes its
16th - 84th percentile Monte Carlo interval. The dotted orange curve shows the attenuation ceiling associated with the 95th-percentile covering-fraction relation. It is a population envelope rather than a confidence interval on the median blue relation. The measurements near $A_{\rm FUV}\simeq10$~mag lie close to the upper boundary of the SED-fitting attenuation grid and should therefore be regarded as boundary-constrained high-attenuation candidates rather than as precise measurements at exactly 10~mag. 
\emph{Bottom left:} Gas fraction, $f_{\rm gas}=M_{\rm gas}/(M_{\rm gas}+M_{\rm star})$. The solid blue curve shows the fitted mass-only gas-fraction relation and is displayed up to $\log_{10}(M_{\rm star}/M_\odot)=10$. The coloured dashed curves show depletion-time-based estimates for main-sequence galaxies at representative redshifts. They assume $M_{\rm mol}={\rm SFR}\,t_{\rm dep}$ with a constant $t_{\rm dep}=0.48$~Gyr. Coloured symbols show the depletion-time-inferred \citealt{Burgarella2025} and this paper gas fractions, including the range associated with the adopted atomic-to-molecular gas correction. Open grey circles show the Heintz et al. (2023) comparison sample. The shaded region denotes the 16th - 84th percentile interval of the fitted mass-only relation.
\emph{Bottom right:} Mass - metallicity relation. Only galaxies at $z>9$ are displayed and included in the metallicity fit. The solid blue curve is the fitted MZR and the shaded region is its
16th - 84th percentile interval. The horizontal dotted line marks $Z=0.1\,Z_\odot$. The preferred full hybrid model uses $\kappa_{\rm FUV,SNe}=10^{3}$ and $\kappa_{\rm FUV,ISM}=10^{5}\ {\rm cm^{2}\,g^{-1}}$, $f_{\rm cov,SNe}\simeq0.05$, $f_{\rm cov,ISM}\simeq0.99$, $f_{\rm dust,active}\simeq0.3$, radii of 70 - 300~pc, and a fitted transition approximately over $7.8\lesssim\log_{10}(M_{\rm star}/M_\odot)\lesssim9$.
}
\label{fig:scaling_relations}
\end{figure*}

We then define a hybrid dust-mass relation that transitions smoothly between these two limits. In the stellar-mass-based version used in Fig.~\ref{fig:scaling_relations}, the hybrid relation follows the SNe-only branch at low stellar masses and the ISM grain-growth branch at high stellar masses. The transition boundaries, $x_{\rm lo}$ and $x_{\rm hi}$, are determined by the joint fit and are shared by the dust-mass, opacity, covering-fraction, size, and metallicity relations. The gas-fraction relation uses the same characteristic transition mass but is described separately by a smooth change in slope. For the best-fitting model shown in Figure~\ref{fig:scaling_relations}, the transition extends approximately over $8 \lesssim x \lesssim 9$. The interpolation is performed with a quintic smootherstep function\footnote{The quintic smootherstep, $t(u)=6u^5-15u^4+10u^3$, is a fifth-order interpolation function commonly used to join two limiting values smoothly. When extended as t=0 below the lower boundary and t=1 above the upper boundary, the function and its first two derivatives are continuous at both boundaries. Its use here is purely a smooth parametrization of the transition and does not imply a specific physical time-evolution law.}, ensuring continuous first and second derivatives at both boundaries:
\begin{equation}
y_{\rm hyb}(x)
=
\left[1-t(x)\right]y_{\rm SNe}(x)
+
t(x)y_{\rm grow}(x),
\label{eq:hybrid_mdust}
\end{equation}
with
\begin{equation}
u(x)
=
{\rm clip}
\left[
\frac{x-x_{\rm lo}}{x_{\rm hi}-x_{\rm lo}},
0,1
\right],
\qquad
t(x)
=
u^3\left(10-15u+6u^2\right).
\label{eq:hybrid_weight}
\end{equation}
By construction, $t=0$ for $x\leq x_{\rm lo}$ and $t=1$ for $x\geq x_{\rm hi}$. The same transition function is used throughout the full hybrid model.

The observational scatter about the adopted scaling relations is propagated through 1000 Monte Carlo realizations; the detailed procedure is given in Appendix H.

In the following, we also use an analogous hybrid attenuation model in which the transition from stardust-dominated to ISM-grown dust is controlled either by stellar mass or by metallicity.

In the following, the term ``SNe-dominated branch'' refers to the limiting case in which the dust mass follows the low-dust branch expected when stellar sources dominate dust production and ISM grain growth remains inefficient. The associated low-$\kappa_{\rm FUV}$ values should therefore be interpreted as an effective opacity motivated by SNe-processed dust models. This parametrization does not assume a unique grain-size distribution; rather, it is intended to test whether the observed $A_{\rm FUV}$ - $M_\star$ relation requires dust with substantially lower UV opacity than standard ISM-like dust.

The SNe-dominated and ISM grain-growth prescriptions introduced above define two limiting regimes. In the SNe-dominated regime, the dust mass follows the low-dust branch of the $M_{\rm dust}$ - $M_{\rm star}$ relation, the intrinsic FUV opacity is fixed to the low-opacity endpoint $\kappa_{\rm FUV,SNe}=10^3\,{\rm cm^2\,g^{-1}}$, and the covering fraction and effective radius approach their low-mass values. In the ISM grain-growth regime, the dust mass follows the high-dust branch, the intrinsic FUV opacity approaches $\kappa_{\rm FUV,ISM}=10^5\,{\rm cm^2\,g^{-1}}$, and the covering fraction and effective radius approach their high-mass values.

The hybrid model provides a continuous interpolation between these two limits. The transition is controlled by the quintic smootherstep function defined Eq.~\eqref{eq:hybrid_weight}, with common boundaries $x_{\rm lo}$ and $x_{\rm hi}$. The same transition function is used for the dust mass, intrinsic FUV opacity, covering fraction, effective radius, and metallicity relations. The gas-fraction relation is fitted independently but its change in slope is anchored to the midpoint of the same transition interval.

The transition boundaries and the remaining effective model parameters are constrained jointly by the dust-mass, $A_{\rm FUV}$, gas-fraction, and $z>9$ mass - metallicity constraints shown in Figure~\ref{fig:scaling_relations}. The two dust-mass endpoint relations and the two opacity endpoints are kept fixed as physically motivated limiting prescriptions, whereas the transition boundaries, covering-fraction endpoints, active-dust fraction, and the parameters of the gas-fraction and metallicity relations are determined by the joint fit.


\subsubsection{Leaky-screen and leaky-mixed attenuation}
\label{subsec:scenarioA}

\paragraph{Star - dust assumed geometries}
\hfill \break
The leaky geometry is mathematically equivalent to the ``picket-fence'' prescription, in which opaque clumps leave open escape channels \citep[e.g.][]{Witt2000,Seon2016}.

We define  \(C_{\rm abs}\) and \(C_{\rm sca}\) as the absorption and scattering cross sections, and the extinction cross section \(C_{\rm ext}\equiv C_{\rm abs}+C_{\rm sca}\). The radiation - pressure cross section is
\begin{equation}
  C_{\rm pr} = C_{\rm abs} + (1-g)\,C_{\rm sca}
  = C_{\rm ext} - g\,C_{\rm sca},
  \label{eq:CprLi}
\end{equation}
with \(g\) the scattering asymmetry parameter. Dividing by the dust-grain mass \(M_{\rm dust}\) defines the corresponding opacities
\(\kappa_{\rm abs}=C_{\rm abs}/M_{\rm dust}\),
\(\kappa_{\rm sca}=C_{\rm sca}/M_{\rm dust}\), and
\(\kappa_{\rm ext}=\kappa_{\rm abs}+\kappa_{\rm sca}\).
We then define an effective opacity based on the radiation-pressure cross section:
\begin{equation}
     \kappa_{\rm eff} \equiv \frac{C_{\rm pr}}{M_{\rm dust}}
  = \kappa_{\rm abs} + (1-g)\,\kappa_{\rm sca}
  = \kappa_{\rm ext} - g\,\kappa_{\rm sca}.
  \label{eq:kappaeff_basic}
\end{equation}
Introducing the single-scattering albedo \(\omega\equiv\kappa_{\rm sca}/\kappa_{\rm ext}\), this can be written compactly as

\begin{equation}
 \kappa_{\rm eff} = (1-\omega g)\,\kappa_{\rm ext}
  \label{eq:kappatr_final}
\end{equation}
where all quantities are wavelength-dependent, but the wavelength index is omitted here for clarity. 

We use $\kappa_{\rm eff}$ as an effective approximation for the attenuation of the directed UV radiation field, accounting for the reduced impact of forward-peaked scattering in a clumpy, porous medium. Forward-peaked scattering ($g>0$) therefore reduces the effective attenuation of the directed radiation field relative to that implied by the extinction opacity alone by a factor $(1-\omega g)$. Typical FUV values are \(\omega\sim0.3\) - 0.5 and \(g\sim0.6\) - 0.8 \citep{Witt2000,Weingartner2001,Draine2003,Inoue2005,Seon2016}.

We stress that the use of the effective opacity $\kappa_{\rm eff} = (1 - \omega g)\kappa_{\rm ext}$ provides only an approximate transport correction for anisotropic forward scattering and does not constitute a full radiative-transfer treatment. It accounts for the fact that strongly forward-scattered photons are not removed from the directed radiation field as efficiently as in the case of isotropic scattering. However, this approach does not include photons that are scattered into the line of sight from other directions, nor does it account for multiple-scattering effects. A full angle-dependent radiative-transfer calculation would require dedicated methods (e.g. Witt \& Gordon 2000; Seon \& Draine 2016), which are beyond the scope of the present work.

In this framework, the attenuation-free limit advocated by \citet{Ferrara2023,Ferrara2025} corresponds to the most extreme leaky configuration, in which the covering fraction is negligible so that the emergent flux is fully open:
\begin{equation}
\begin{aligned}
  T_\lambda &\xrightarrow[\;f_{\rm cov}\to 0\;]{} 1,
  \quad
  &\Rightarrow\quad
  A_\lambda \approx 0 .
\end{aligned}
\label{eq:leaky_extreme}
\end{equation}

Here, we adopt mass-conserving porosity prescriptions that encode partial covering and clump statistics \citep[e.g.][]{Natta1984,Witt2000,Varosi1999,Nenkova2008,Seon2016}. We define the transmission as
\[
T_\lambda \equiv \frac{F_{\lambda,\mathrm{obs}}}{F_{\lambda,\mathrm{int}}},
\]
that is, the ratio of observed to intrinsic flux at wavelength $\lambda$.

The corresponding attenuation laws for leaky (picket-fence) geometries are given by:
\begin{subequations}\label{eq:porosity_TLeak}
\begin{align}
T_{\mathrm{L,screen}}(\tau_\lambda,f_{\rm cov})
&=(1-f_{\rm cov})
  + f_{\rm cov}\,
   \exp\!\left[-(1-\omega_\lambda g_\lambda)\,
               \frac{\tau_\lambda}{f_{\rm cov}}\right], \\[2pt]
T_{\mathrm{L,mixed}}(\tau_\lambda,f_{\rm cov})
&=(1-f_{\rm cov})
  + f_{\rm cov}\,
   \frac{1-\exp(-\tau_\lambda/f_{\rm cov})}{\tau_\lambda/f_{\rm cov}},
\end{align}
\end{subequations}
whose derivation is provided in Appendix~\ref{appendix:Eqs5a5b}.

We define $\tau_\lambda=\kappa_{\rm ext,\lambda}\,\Sigma_{\rm dust}$ as the extinction optical depth, $\omega_\lambda$ and $g_\lambda$ are the single-scattering albedo and asymmetry factor, and $f_{\rm cov} > 0$ is the covering fraction\footnote{A uniformly mixed spherical medium of stars and dust could produce optical depths up to a factor of about 3 larger than those of a shell geometry, because dust grains located closer to the radiation sources subtend a larger solid angle as seen by the emitters \citep{Inoue2020}.}. All channels conserve total dust mass and reduce to the uniform limits as $f_{\rm cov}\!\to\!1$.

For the initial SN-based dust tracks, the total optical depth is computed using empirical dust scaling relations, which link dust opacity to global galaxy properties such as stellar mass and metallicity. More specifically, we start with the $M_{\mathrm{dust}}$ versus $M_{star}$ relations from \citet[][]{Witstok2023} (see Sect.~\ref{subsec:attenuation-route}) which is converted into a surface density using an empirical stellar mass - size relation appropriate for high-redshift galaxies. The effective radius is prescribed as a function of stellar mass: $R_{\rm e}(M_\star)$ and the corresponding dust surface density is then
\begin{equation}
\Sigma_{\rm dust}(M_{\rm star}) \equiv \frac{M_{\rm dust}}{\pi R_{\rm e}^2}.
\end{equation}

The compact galaxy sizes entering the column-density calculation are represented by a fixed bounded size - mass relation,
\begin{equation}
r_{\rm pc}(M_{\rm star})
=
r_{\rm lo}
+
t(M_\star)\left(r_{\rm hi}-r_{\rm lo}\right),
\end{equation}
where $r_{\rm lo}=70$~pc and $r_{\rm hi}=300$~pc, corresponding to the size of galaxies observed in the early Universe (e.g. \citealt{Burgarella2025}, and references therein). The interpolation uses the same quintic transition function $t(M_{\rm star})$ and the same fitted transition boundaries as the dust-mass, opacity, covering-fraction, gas-fraction, and metallicity relations. The radius remains equal to $r_{\rm lo}$ below the transition and to $r_{\rm hi}$ above it.

Finally, assuming a wavelength-dependent dust opacity, the optical depth at FUV is computed as
\begin{equation}
\tau_\lambda(M_{\rm star}) = \kappa_{\rm ext,\lambda} \, \Sigma_{\rm dust}(M_\star),
\end{equation}
where $\kappa_{\rm ext,\lambda}$ is the dust extinction coefficient at the considered wavelength. These optical depths are then used as inputs to the leaky screen and leaky mixed attenuation kernels (Eq.~\ref{eq:porosity_TLeak}:):
\begin{equation}
  A_{\rm FUV}^{\rm screen} =
  -2.5 \log_{10} T_{\mathrm{L,screen}}
  \bigl( \tau_{\rm FUV}, f_{\rm cov}, \text{porosity} \bigr),
\end{equation}
\begin{equation}
  A_{\rm FUV}^{\rm mixed} =
  -2.5 \log_{10} T_{\mathrm{L,mixed}}
  \bigl( \tau_{\rm FUV}, f_{\rm cov}, \text{porosity} \bigr),
\end{equation}
with the same covering fraction $f_{\rm cov}$ and stars/dust geometry model (porosity). 

\subsubsection{Poissonian-clump attenuation}
\label{subsec:scenarioB}

The Poisson-clump model treats the ISM as an inhomogeneous ensemble of discrete clouds. Each sight line intercepts a Poisson-distributed number of clumps with mean $\bar{N}=-\ln(1-f_{\rm cov})$, so that the clear-sight-line probability is $P(0)=e^{-\bar{N}}$ \citep{Natta1984,Witt2000}. If each clump has optical depth $\tau_\lambda/\bar{N}$, the ensemble-averaged transmission is shown in Eq.~\ref{eq:porosity_TPois}:

\begin{subequations}\label{eq:porosity_TPois}
\begin{align}
T_{\mathrm{P}}(\tau_\lambda,f_{\rm cov})
&=\exp\!\left[-\bar N\!\left(1-e^{-\tau_\lambda/\bar N}\right)\right], \quad \bar N \equiv -\ln(1-f_{\rm cov}).
\end{align}
\end{subequations}

which conserves dust mass and accounts for stochastic shadowing \citep{Varosi1999,Nenkova2008} and the corresponding FUV attenuation is
\begin{equation}
  A_{\rm FUV}^{\rm Poisson} \;=\;
  -2.5\log_{10} T_{\mathrm{P}}(\tau_{\rm FUV}, f_{\rm cov}) \, .
\end{equation}

The Poisson geometry generalises these configurations by averaging over a stochastic distribution of clump encounters along each sight line: in the limit $\bar N\!\to\!1$ it approaches a leaky-screen geometry, while for $\bar N\!\gg\!1$, corresponding to many optically thin clumps per sight line, it converges toward the leaky-mixed limit.

All attenuation prescriptions operate directly on $\tau_\lambda$ and conserve the total dust mass, thereby isolating the geometric greying effect from changes in the intrinsic extinction law. Geometry, rather than the extinction law alone, therefore governs the emergent attenuation slope at fixed dust mass. 

The prescriptions introduced in this section generalize the treatment of dust attenuation by explicitly accounting for porous star–dust geometries and stochastic line-of-sight effects, while conserving the total dust mass and adopting a fixed intrinsic extinction law. In this framework, differences in the effective FUV attenuation arise solely from geometric effects and from the redistribution of optical depths, rather than from changes in dust composition or dust mass. This separation allows us to isolate the role of geometry in shaping the emergent attenuation, and to explore its impact independently of uncertainties in dust production and extinction properties.

\subsection{Gas-route attenuation model}
\label{subsec:gas-route}

As a complementary test, we construct the optical depth directly from gas surface density, metallicity, and dust-to-metal ratio. This gas route uses nebular metallicity as the physical quantity controlling the transition from SNe-dominated dust to efficient ISM grain growth, rather than constructing the optical depth directly from the $M_{\rm dust} - M_\star$ relation. The same leaky-screen and leaky-mixed attenuation kernels introduced in Sect. 2.2 are then applied to the resulting optical depth. Because the Poissonian case gives conclusions similar to the leaky geometries in the attenuation route, we do not repeat it for the gas route.

We note that throughout this work, $\kappa^{(\mathrm{dust})}$ denotes opacities per unit dust mass, while $\kappa^{(\mathrm{gas})}=\kappa^{(\mathrm{dust})}\,(\mathrm{DTM}\times Z)$ denotes the corresponding opacity per unit gas mass entering the optical depth, where $Z$ is the metallicity expressed in units of the solar metal mass fraction, and DTM is the dust-to-metal ratio. The optical depth at wavelength $\lambda$ is defined as
\begin{equation}
  \tau_\lambda \;=\; \kappa_\lambda^{(\mathrm{gas})}\,\Sigma_{\mathrm{gas}}
  \;=\; \kappa_\lambda^{(\mathrm{dust})}\,\Sigma_{\mathrm{dust}},
  \label{eq:tau_defs}
\end{equation}
where

\begin{equation}
\kappa_\lambda^{(\mathrm{gas})}
= \kappa_\lambda^{(\mathrm{dust})}\,(\mathrm{DTM}\times Z),
\qquad
\Sigma_{\mathrm{dust}}
= \Sigma_{\mathrm{gas}}\,(\mathrm{DTM}\times Z) .
\end{equation}

Here, $\kappa_\lambda^{(\mathrm{dust})}$ is the extinction coefficient per unit dust mass (cm$^2$\,g$^{-1}$), $Z$ is the metallicity expressed in units of the solar metal mass fraction, and DTM is the dust-to-metal ratio.

The FUV per - gas opacity is therefore
\begin{equation}
  \kappa_{UV}^{(\mathrm{gas})}
  \;=\;
  \kappa_{UV}^{(\mathrm{dust})}\,
  (\mathrm{DTM}\times Z)
  \label{eq:kappaFUV_gas_master}
\end{equation}
which provides a direct and physically motivated link between the intrinsic opacity per unit dust mass and the effective opacity per unit gas mass through the metallicity and dust-to-metal ratio. It is used below to evaluate how the combined evolution of these quantities affects the predicted FUV attenuation.

We first construct a full gas-based attenuation optical depth,
\begin{equation}
  \tau_{\rm gas}(M_{star}) \;=\;
  \kappa_{\rm gas}\,\Sigma_{\rm gas},
\end{equation}
where the gas surface density $\Sigma_{\rm gas}$ follows from the gas mass and a size-mass relation defined above and $\kappa_{\rm gas}$ incorporates both the metallicity and dust-to-metal scaling. The gas mass is computed as: 
\begin{equation}
M_{\rm gas} = \frac{f_g}{1 - f_g}\, M_\star .
\end{equation}

The gas fraction is constrained jointly with the other scaling relations. We adopt a continuously decreasing mass-only relation with a smooth change in slope around the midpoint of the common transition interval. Defining
\begin{equation}
x_{\rm b}=\frac{x_{\rm lo}+x_{\rm hi}}{2},
\qquad
S(x)=w\ln\left[
1+\exp\left(\frac{x-x_{\rm b}}{w}\right)
\right],
\end{equation}
the fitted relation is
\begin{equation}
f_{\rm gas}(x)
=
f_{\rm b}
+s_{\rm low}(x-x_{\rm b})
+\left(s_{\rm high}-s_{\rm low}\right)S(x),
\end{equation}
with the result restricted to the physical interval $0\leq f_{\rm gas}<1$. The normalization $f_{\rm b}$, the low- and high-mass slopes, and the smoothing width $w$ are determined by the joint fit. For comparison, gas fractions for the samples from \citet{Burgarella2025} and from this paper are also inferred from $M_{\rm mol}={\rm SFR}\,t_{\rm dep}$, adopting $t_{\rm dep}=0.48$~Gyr and a fiducial $M_{\rm atom}/M_{\rm mol}=2$.

To calibrate the metallicity scaling in the gas-route attenuation model, we adopt a broken high-redshift MZR designed to reproduce the trends and normalizations inferred from recent JWST measurements. Empirical studies indicate that the MZR slope depends on both stellar mass and redshift, with reported values for the MZR slopes ($s_{MZR}$) spanning $s_{MZR}\simeq0.14$ - 0.39 depending on the adopted abundance calibration. \citet{Li2023} find $s_{MZR}=0.14\pm0.04$ for $M_{star}\simeq10^{6.5}$ - $10^{9.5}\,M_\odot$ at $z\simeq2$ - 3, \citet{Sanders2021} report $s_{MZR}\simeq0.30$ at higher masses, and \citet{Chemerynska2024} measure $s_{MZR}\simeq0.39\pm0.02$ at $z\simeq6$ - 8 using their fiducial R3 calibration (decreasing to $s_{MZR}\simeq0.32$ under alternative strong-line calibrations).

We implement a smoothed broken high-redshift MZR,
\begin{equation}
\begin{aligned}
\log_{10} Z(M_\star) &=
(1 - t)\,\bigl[\log_{10} z_{9,\mathrm{low}} 
+ s_{\rm low}(\log M_\star - 9)\bigr] \\
&\quad + t\,\bigl[\log_{10} z_{9,\mathrm{high}} 
+ s_{\rm high}(\log M_\star - 9)\bigr] ,
\end{aligned}
\end{equation}

where $\log M_\star \equiv \log_{10}(M_\star)$. The normalizations and slopes of the low- and high-metallicity branches are determined jointly with the other model parameters.
The interpolation parameter $t$ is the common quintic smootherstep function defined in Eq.~(\ref{eq:hybrid_weight}), using the transition boundaries determined by the joint fit. Only galaxies at $z>9$ are included in the MZR likelihood shown in Fig.~\ref{fig:scaling_relations}.

This prescription adopts a steeper low-mass branch and a mildly steeper high-mass branch than some lower-redshift determinations, while lowering the low-mass normalization. It therefore brackets the empirical range, allows for lower metallicities in low-mass galaxies, and transitions toward the higher metallicities expected for massive galaxies at $z\gtrsim6$ \citep[e.g.][]{Maiolino2019,Curti2023,Sanders2023,Chemerynska2024}. The fitted relation reproduces the average normalization and slope of JWST-based MZR measurements at $z \simeq 9$ 
\citep{Nakajima2023,Curti2023}, and remains consistent with cosmological simulation predictions \citep{Torrey2019}.

The dust-to-metal ratio is obtained from an empirical broken interpolation in $12+\log(\mathrm{O/H})$:
\begin{align}
\mathrm{DTM} = (1-t)\times\mathrm{DTM}_\mathrm{low}(Z) + t\times\mathrm{DTM}_\mathrm{high}(Z),
\end{align}

with metallicity anchored at $(12+\log\mathrm{O/H})_\mathrm{low}=7.4$ and $(12+\log\mathrm{O/H})_\mathrm{high}=8.0$, and local power-law slopes $\alpha_\mathrm{low}=1.2$ and $\alpha_\mathrm{high}=-0.5$. The normalization follows the empirical median relation from \citet{Burgarella2025}, with first- and third-quartile variants bracketing the observed scatter. 

The conversion between oxygen abundance and total metallicity follows the calibration adopted in Burgarella et al. (2025). The total metal mass fraction is first defined as
\begin{equation}
\log_{10} Z = [12 + \log_{10}({\rm O/H})] - 10.410,
\end{equation}
which assumes that oxygen traces the total metal content with approximately fixed abundance ratios. The corresponding solar-normalized metallicity is then obtained by dividing by the solar metal mass fraction, $\frac{Z}{Z_\odot}$, where we adopt $Z_\odot = 0.0142$ (Asplund et al. 2009)\footnote{We note that this conversion assumes near-solar abundance ratios and may be affected by $\alpha$-enhancement at high redshift.}.

Because f$_{\rm dust,active}$ multiplies the dust column entering the optical depth, it is degenerate with the intrinsic dust opacity. The attenuation data therefore constrain primarily the product f$_{\rm dust,active}.\kappa_{\rm FUV}^{\rm dust}$, together with the adopted dust mass and galaxy size, rather than independently determining f$_{\rm dust,active}$ and $\kappa_{\rm FUV}^{\rm dust}$. The intrinsic opacity values discussed below should consequently be interpreted as conditional on the fitted active-dust fraction.

The hybrid model is therefore an empirical interpolation between two limiting dust regimes, rather than a time-dependent dust-evolution calculation. It is designed to test whether a
transition from a low-dust-mass, low-opacity regime to a high-dust-mass, ISM-like regime can reproduce the observed scaling relations and FUV attenuation.

The SNe-dominated and ISM grain-growth endpoints adopted here are motivated by physically plausible limiting dust populations. However, not every combination of opacity, covering fraction, geometry, extinction-curve shape, albedo, and asymmetry parameter explored in the sensitivity tests should be interpreted as a unique, internally self-consistent grain model. Some quantities are varied separately in order to isolate their respective effects on the predicted attenuation and to identify the associated degeneracies. Our results therefore constrain effective FUV opacity and star - dust porosity, rather than uniquely determining the underlying grain-size distribution or composition. A fully self-consistent treatment would require coupling the attenuation calculation to a dust-evolution model that predicts the grain-size distribution, opacity, extinction curve, albedo, and scattering phase function jointly.

A related limitation is that the present analysis treats the normalization of the FUV opacity as the primary variable, while the wavelength dependence of the intrinsic extinction law is explored through a limited set of representative curves. In a self-consistent grain model, changes in grain size and composition would generally modify both the absolute FUV opacity and the shape of the extinction curve.

However, a greyer intrinsic extinction curve does not necessarily imply that a greyer attenuation curve will be observed. The emergent attenuation law also depends on the spatial distribution of stars and dust, covering fraction, clumpiness, and scattering, so different intrinsic extinction curves can produce similar effective attenuation curves. Conversely, geometrical effects can modify or grey the attenuation curve without requiring a corresponding change in the intrinsic grain extinction law.

There is an additional observational difficulty in the regime considered here. The galaxies of primary interest have very low FUV attenuation, so dust only weakly modifies the intrinsic stellar spectrum. As $A_{\rm FUV}$ approaches zero, the wavelength-dependent imprint of the dust becomes correspondingly small and is increasingly difficult to distinguish from uncertainties in the intrinsic stellar continuum, star-formation history, metallicity, nebular emission, and photometric or spectroscopic calibration. The proposed low-opacity dust regime may therefore be intrinsically difficult to test through the UV attenuation-curve shape alone.

More informative constraints may require combining high-precision rest-frame UV spectroscopy with independent measurements of dust emission, metallicity, gas content, and star - dust geometry. Such joint constraints could help determine whether the low attenuation is primarily caused by intrinsically low-opacity grains, porous geometry, or a combination of both.

The fitted opacity, covering fraction, and geometry should be interpreted as effective descriptors. They are partly degenerate and do not uniquely determine the underlying grain-size distribution or composition; we discuss these limitations in Appendix G and Sect. 4.

\section{Comparison with observations and physical interpretation}
\label{sec:comparison}

\subsection{The galaxy sample}
\label{subsec:sample}

Figure~\ref{fig:scaling_relations} summarizes the observational constraints and the best-fitting full hybrid model. The four relations are fitted jointly, using common transition boundaries for the change from the SNe-dominated regime to the ISM grain-growth regime. The intrinsic FUV dust-opacity endpoints are fixed to 
$\kappa_{\rm FUV,SNe}=10^{3}\ {\rm cm^{2}\,g^{-1}}$
and
$\kappa_{\rm FUV,ISM}=10^{5}\ {\rm cm^{2}\,g^{-1}}$.
The preferred model shown in Fig.~\ref{fig:scaling_relations} uses a leaky mixed geometry, with $f_{\rm cov,SNe}\simeq0.05$, $f_{\rm cov,ISM}\simeq0.99$, $f_{\rm dust,active}\simeq0.3$, and an adopted radius varying from 70 to 300~pc. The fitted transition extends approximately from $\log_{10}(M_\star/M_\odot)\simeq8$ to 9.

We first use the galaxy sample defined in \citet{Burgarella2025}. This sample is composed of 173 objects with JWST NIRSpec prism data in the redshift range 4 $<$ z $<$ 11.5 extracted from the CEERS survey (\citealt{Finkelstein2023}). From this parent sample, we define a sub-sample of 49 GELDAs. 

GELDAs are defined as galaxies whose FUV attenuation, $A_{\rm FUV}$, inferred with CIGALE, is consistent with zero within twice its estimated uncertainty. Most GELDAs have $\log_{10}(M_{\rm star}/M_\odot)\lesssim 9.0$, although higher-mass examples may exist. Note that the selection at $\log_{10}\,(M_{\rm star})<9.0$ applied in \citet{Burgarella2025} is no longer used here.

In order to focus on the specific question of the origin of the z$>$9 UV excess, we built a sample of 26 objects with JWST NIRSpec prism data in the redshift range 8 $<$ z $<$ 15, including all of the objects with available prism data available at z $>$ 10, when this work was initiated. This second sample is defined in Tab.~\ref{tab:uhz} and contains 18 GELDAs, that is 69\,\% of the sample in the redshift range 8 $<$ z $<$ 15. However, GELDAs form 84\,\% of the objects at z$\geq$9.0 and only 28\,\% of the objects at z$<$9.0, in excellent agreement with the proportion found by \citet{Burgarella2025}. This type of objects clearly becomes dominant in the ultra-high redshift universe, as suggested by \citet{Burgarella2025}.

\begin{table*}
\centering
\setlength{\tabcolsep}{3pt}
\caption{The sample of ultra-high redshift galaxies (UHZ) that complements the 173 galaxies from \citet{Burgarella2025}.} 
\label{tab:uhz}

\begin{tabular}{|l|l|l|c|r|c|c|c|c|c|l|}
\hline
\multicolumn{1}{|c|}{id} &
\multicolumn{1}{c|}{RA (h:m:s)} &
\multicolumn{1}{c|}{DEC (d:m:s)} &
\multicolumn{1}{c|}{$z$} &
\multicolumn{1}{c|}{age} &
\multicolumn{1}{c|}{$\mu$} &
\multicolumn{1}{c|}{$A_{\rm FUV}$} &
\multicolumn{1}{c|}{$\log_{10}\,(Z_{\rm gas}$)} &
\multicolumn{1}{c|}{$\log_{10}\,(M_{star}$)} &
\multicolumn{1}{c|}{GELDA} &
\multicolumn{1}{c|}{ref} \\
\hline
CEERS5\_3       & 14:20:01.25 & +52:59:47.69 & 8.01 & 654.05 & --- & $2.518\pm0.090$ & $-2.566\pm0.447$ & $9.165\pm0.329$ & N & a \\
CEERS8\_1149    & 14:20:21.53 & +52:57:58.26 & 8.19 & 634.37 & --- & $1.008\pm0.038$ & $-2.470\pm0.192$ & $9.062\pm0.245$ & N & a \\
CEERS7\_1029    & 14:20:52.50 & +53:04:11.50 & 8.62 & 592.26 & --- & $\leq 0.283$ & $-2.450\pm0.212$ & $9.414\pm0.090$ & Y & a \\
CEERS11\_80083  & 14:19:50.71 & +52:50:32.51 & 8.64 & 590.60 & --- & $\leq 1.205$ & $-2.522\pm0.411$ & $8.743\pm0.243$ & Y & a \\
DDT28           & 14:19:45.27 & +52:54:42.30 & 8.76 & 579.27 & --- & $1.012\pm0.004$ & $-2.573\pm0.214$ & $8.564\pm0.144$ & N & b \\
CEERS5\_2       & 14:19:58.66 & +52:59:21.76 & 8.81 & 575.37 & --- & $\leq 0.094$ & $-2.509\pm0.533$ & $8.657\pm0.259$ & Y & a \\
CEERS5\_7       & 14:20:02.81 & +52:59:17.89 & 8.88 & 569.34 & --- & $2.525\pm0.005$ & $-3.027\pm0.316$ & $8.941\pm0.049$ & N & a \\
DDT69           & 14:19:26.78 & +52:54:16.57 & 9.74 & 501.91 & --- & $\leq 0.179$ & $-2.926\pm1.033$ & $8.313\pm0.463$ & Y & b \\
CEERS12\_80026  & 14:19:14.84 & +52:44:13.60 & 10.01 & 483.52 & --- & $\leq 0.365$ & $-2.717\pm0.889$ & $9.295\pm0.110$ & Y & a \\
DDT64           & 14:19:41.47 & +52:54:41.50 & 10.10 & 477.64 & --- & $\leq 0.366$ & $-2.743\pm0.974$ & $8.758\pm0.191$ & Y & b \\
CEERS12\_80041  & 14:18:55.81 & +52:45:29.12 & 10.15 & 474.42 & --- & $\leq 0.383$ & $-3.036\pm1.062$ & $8.959\pm0.173$ & Y & a \\
CEERS8\_18136   & 14:20:20.79 & +52:59:14.14 & 10.20 & 471.24 & --- & $\leq 2.723$ & $-2.556\pm0.895$ & $9.716\pm0.440$ & Y & a \\
UNCOVER37126    & 03:35:24.40 & -30:21:35.07 & 10.25 & 468.09 & 1.270 & $\leq 0.010$ & $-3.398\pm0.799$ & $6.999\pm0.216$ & Y & h,k \\
GZ-z10-0        & 03:32:38.12 & -27:46:24.57 & 10.38 & 460.07 & --- & $\leq 0.103$ & $-2.863\pm1.014$ & $7.969\pm0.219$ & Y & f \\
GN-z11          & 12:36:25.46 & +62:14:31.40 & 10.60 & 447.01 & --- & $1.009\pm0.002$ & $-3.052\pm0.262$ & $9.120\pm0.181$ & N & d \\
DDT10           & 14:19:37.59 & +52:56:43.81 & 11.04 & 422.67 & --- & $\leq 0.676$ & $-2.391\pm0.427$ & $9.172\pm0.144$ & Y & b \\
DDT1            & 14:19:46.36 & +52:56:32.79 & 11.40 & 404.35 & --- & $\leq 0.368$ & $-2.584\pm0.648$ & $8.912\pm0.280$ & Y & b \\
GZ-z11-0        & 03:32:39.54 & -27:46:28.65 & 11.58 & 395.67 & --- & $\leq 0.951$ & $-2.876\pm0.939$ & $8.610\pm0.319$ & Y & f \\
GHZ2            & 00:13:58.30 & -30:20:14.10 & 12.34 & 362.25 & 1.3 & $0.977\pm0.008$ & $-3.463\pm0.081$ & $8.360\pm0.153$ & N & c,j \\
UNCOVER38766    & 03:30:48.83 & -30:21:24.48 & 12.39 & 360.21 & 1.520 & $2.959\pm0.715$ & $-3.048\pm0.949$ & $9.947\pm0.181$ & N & h,i \\
GZ-z12-0        & 03:32:39.92 & -27:49:17.60 & 12.63 & 350.71 & --- & $\leq 0.330$ & $-2.705\pm0.959$ & $8.667\pm0.203$ & Y & f \\
GS-z13-1-LA     & 03:32:15.54 & -27:53:24.86 & 13.01 & 336.49 & --- & $\leq 0.014$ & $-3.369\pm0.793$ & $7.309\pm0.242$ & Y & e \\
UNCOVER13077    & 03:24:15.13 & -30:24:05.71 & 13.08 & 333.98 & 2.278 & $\leq 0.668$ & $-2.956\pm1.048$ & $8.460\pm0.769$ & Y & h,i \\
GZ-z13-0        & 03:32:35.97 & -27:46:35.40 & 13.20 & 329.74 & --- & $\leq 0.033$ & $-3.101\pm1.138$ & $7.502\pm0.666$ & Y & f \\
GZ-z14-1        & 03:19:17.83 & -27:53:09.34 & 13.90 & 306.69 & --- & $\leq 0.313$ & $-3.080\pm1.067$ & $8.083\pm0.598$ & Y & g \\
GZ-z14-0        & 03:19:19.91 & -27:51:20.27 & 14.18 & 298.22 & --- & $0.261\pm0.051$ & $-2.846\pm0.942$ & $9.477\pm0.126$ & N & g \\
\hline
\end{tabular}
\tablefoot{Note that a few objects are in both samples: DDT1, DDT10, DDT64 and CEERS8\_18136. The sample contains all galaxies identified by the authors at $z>10$ when this work was initiated. Coordinates are given as (RA\_h, RA\_m, RA\_s) and (Dec\_d, Dec\_m, Dec\_s). Redshifts $z$ and the age of the Universe are rounded to 2 decimals; all other derived quantities are rounded to 3 decimals. $\mu$ is the amplification factor for the UNCOVER galaxies. References: (a) \citet{Finkelstein2025}; (b) Arrabal Haro et al.\ (2023Natur.622..707A); (c) \citet{Castellano2024}; (j) Zavala et al.\ (2025NatAs...9..155Z); (d) Bunker et al.\ (2023A\&A...677A..88B); (e) Witstok et al.\ (2025Natur.639..897W); (f) Curtis-Lake et al.\ (2023NatAs...7..622C); (g) Carniani et al.\ (2024Natur.633..318C); (h) Bezanson et al.\ (2024ApJ...974...92B); (i) Wang et al.\ (2023ApJ...957L..34W); (k) Suess et al.\ (2024ApJ...976..101S).}
\end{table*}

\subsection{Attenuation - stellar mass relation}
\label{subsec:AfuvMstar}

\begin{figure*}[!t]
\centering
\includegraphics[width=\textwidth]{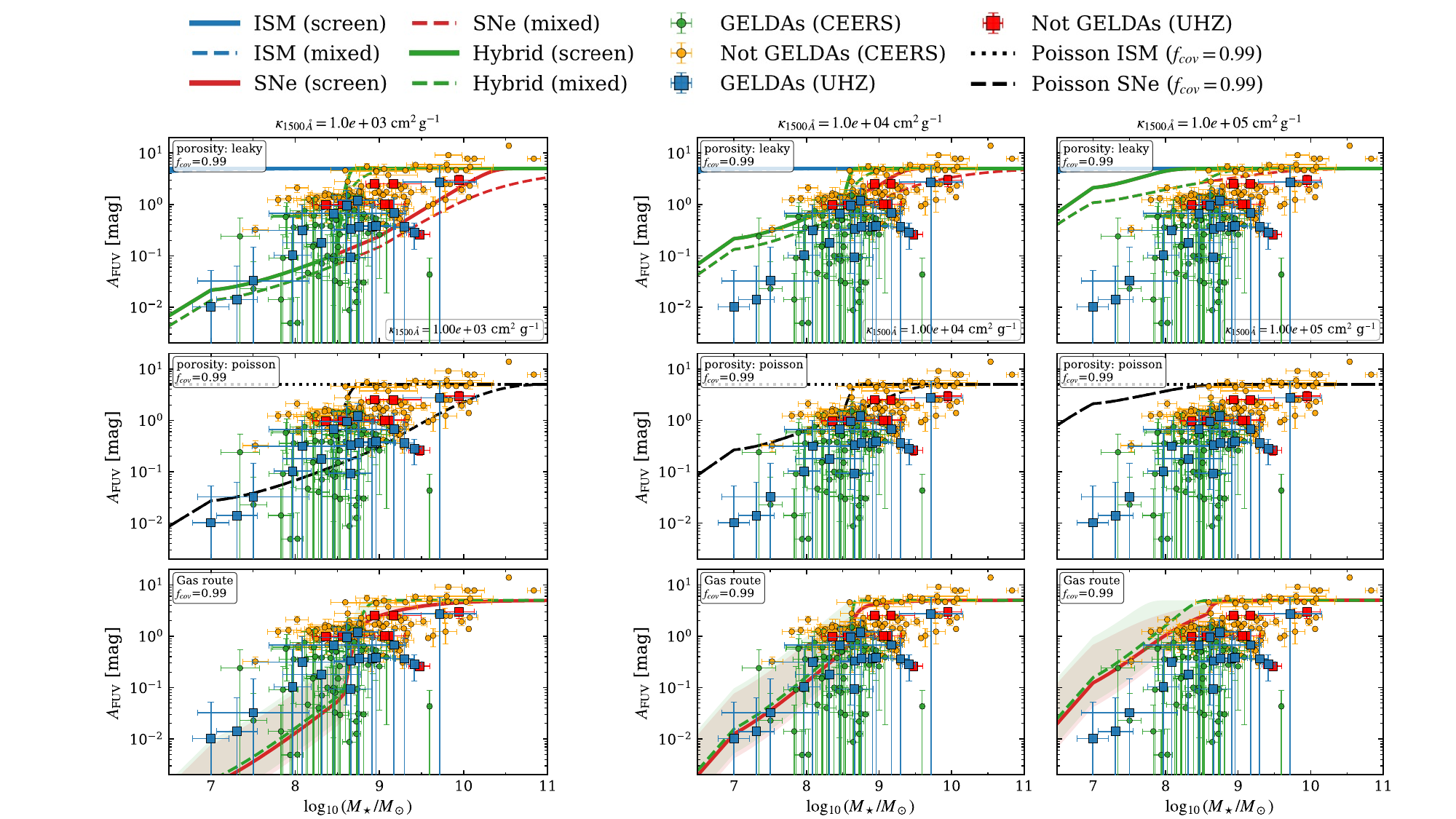}
\caption{
Predicted far-UV attenuation, $A_{\rm FUV}$, as a function of stellar mass for three intrinsic far-UV dust opacities, $\kappa_{\rm FUV}^{(\rm dust)}=10^3$, $10^4$, and $10^5\,{\rm cm^2\,g^{-1}}$. The left part of the figure shows the fiducial $\kappa_{\rm FUV}^{(\rm dust)}=10^3\,{\rm cm^2\,g^{-1}}$ models, while the additional panels show the corresponding predictions for $10^4$ and $10^5\,{\rm cm^2\,g^{-1}}$. Rows correspond to different attenuation geometries: leaky-screen and leaky-mixed configurations, Poisson-clump geometry, and the gas-based attenuation route. For the leaky and Poisson cases, the SNe-dominated and ISM grain-growth prescriptions are shown as limiting branches, together with the hybrid model interpolating between them. Observed galaxies are overplotted for comparison: GELDAs are shown in green for the CEERS sample and in blue for the ultra-high-redshift sample, while the remaining CEERS and ultra-high-redshift galaxies are shown in orange and red, respectively.
For the fiducial $\kappa_{\rm FUV}^{(\rm dust)}=10^3\,{\rm cm^2\,g^{-1}}$ comparison, the models are shown for $f_{\rm cov}=0.99$. In the opacity-comparison panels, the covering fractions are held fixed at $f_{\rm cov}=0.99$ for the ISM-dominated prescription and $f_{\rm cov}=0.50$ for the SNe-dominated prescription, with the hybrid model interpolating between these values. These fixed choices are used only to isolate the effect of intrinsic opacity and geometry and should not be interpreted as independently fitted or uniquely preferred covering fractions.
The gas-based route with $\kappa_{\rm FUV}^{(\rm dust)}=10^4\,{\rm cm^2\,g^{-1}}$ remains compatible with a substantial fraction of the observed $A_{\rm FUV}$--$M_\star$ distribution because the effective opacity per unit gas mass is reduced by metallicity and dust-to-metal ratio, $\kappa_{\rm FUV}^{(\rm gas)}=\kappa_{\rm FUV}^{(\rm dust)}({\rm DTM}\times Z)$. By contrast, models with $\kappa_{\rm FUV}^{(\rm dust)}=10^5\,{\rm cm^2\,g^{-1}}$ generally predict excessive attenuation. The comparison therefore favours a low intrinsic FUV-opacity range of approximately $10^3$--$10^4\,{\rm cm^2\,g^{-1}}$, rather than uniquely selecting $10^3\,{\rm cm^2\,g^{-1}}$.
}
\label{fig:afuv_mstar}
\end{figure*}

Figure~\ref{fig:afuv_mstar} presents the model predictions for the far-UV attenuation, $A_{\rm FUV}$, computed in Sect.~\ref{sec:attenuation-framework}, as a function of stellar mass. The different curves illustrate the impact of the adopted dust prescriptions and porous geometries, as introduced in Fig.~\ref{fig:geometries} and Table~\ref{tab:model_overview}, and described in Sect.~\ref{sec:attenuation-framework}. The dust-mass, attenuation, gas-fraction, and mass–metallicity constraints entering the joint model are shown together in Fig.~\ref{fig:scaling_relations}.

We account for the intrinsic scatter in the adopted scaling relations using a Monte Carlo approach, and show the resulting 16$^{th}$ - 84$^{th}$ percentile range as shaded regions around the fiducial model predictions.

The Monte Carlo envelopes shown in Fig.~\ref{fig:scaling_relations} quantify the propagation of the empirical scatter around the adopted SNe-only and ISM grain-growth limiting branches, together with the scatter in the other scaling relations. They do not encompass every possible alternative slope, curvature, or functional form of the $M_{\rm dust}$ - $M_{\rm star}$ relation. Our conclusion should therefore be understood as applying within the observationally motivated family of scaling relations explored in this work. Within this range, the preference for an effective FUV opacity below the standard ISM-like value remains, although the precise normalization and transition shape of the attenuation relation remain model dependent.

To further disentangle the respective roles of geometry, covering fraction, and intrinsic dust properties, we present in Appendix~\ref{appendix:geometry} a series of model grids exploring these parameters over a wide range of values (Figs.~\ref{fig:AppendixC1}-\ref{fig:AppendixC3}). These figures allow us to directly assess the degeneracies between star - dust geometry, covering fraction, and dust opacity, and to evaluate their relative impact on the predicted $A_{\rm FUV}$-$M_{\rm star}$ relation. These degeneracies are not complete within the parameter space explored here. In the dust-mass-based route, models with $\kappa_{\rm FUV}^{\rm dust}=10^{3}\ {\rm cm^{2}.g^{-1}}$ generally provide the closest match to the lowest-attenuation systems, whereas increasing the opacity to $10^{4}\ {\rm cm^{2}.g^{-1}}$ often worsens the agreement. The gas-based route is less restrictive and can accommodate $\kappa_{\rm FUV}^{\rm dust}=10^{4}\ {\rm cm^{2}.g^{-1}}$ for sufficiently low metallicities and dust-to-metal ratios. Standard ISM-like values near $10^{5}\ {\rm cm^{2}.g^{-1}}$ generally predict excessive attenuation. The inferred opacity range remains conditional on the adopted dust- or gas-scaling relations, active-dust fraction, galaxy size, covering fraction, and geometry.

We find that the covering fraction $f_{\rm cov}$ primarily controls the overall normalization of the $A_{\rm FUV}$- $M_{\rm star}$ relation, with decreasing $f_{\rm cov}$ lowering the attenuation by increasing the fraction of unobscured sightlines. The choice of geometry (leaky screen, leaky mixed, or Poisson clumps) introduces only second-order variations, mainly affecting the curvature and dispersion of the relation at fixed optical depth.

In contrast, the intrinsic dust opacity $\kappa_{\rm UV}$ has a much stronger impact, shifting the predicted attenuation by more than an order of magnitude between $\kappa_{\rm UV}=10^5$ and $10^3\,{\rm cm^2\,g^{-1}}$. This shows that variations in the intrinsic opacity produce larger systematic shifts in the predicted attenuation than the differences
among the geometries explored at fixed optical depth.

Although partial degeneracies exist between geometry, covering fraction, and opacity, they are not complete. In particular, no combination of geometry and covering fraction explored here reproduces the observed distribution when standard ISM-like opacities are adopted. The direct result of this comparison is therefore a requirement for intrinsically low effective FUV dust opacities.

Within the baseline physical scenario adopted in this work, these low opacities are naturally interpreted as arising from dust produced predominantly by SNe before efficient ISM grain growth becomes established and subsequently processed by reverse shocks. This interpretation is independently motivated by the young ages and low metallicities of galaxies at $z\gtrsim9$, and is consistent with the broad class of early dust-evolution models discussed in Sect.~\ref{subsec:attenuation-route}.

While the exact microphysical pathway leading to these low opacities remains uncertain, the observational requirement for reduced effective UV opacity is robust.

We tested the four star-formation histories listed in
Appendix~\ref{appendix:cigale_params}, using the CIGALE parameter grids adopted for the galaxies in Table~\ref{tab:uhz}. We find no substantial systematic variation in the inferred parameters among these prescriptions. We therefore adopt, for each galaxy, the median $A_{\rm FUV}$ obtained from the delayed, periodic, delayed-plus-stochastic, and exponential-plus-stochastic models.

To assess the sensitivity of the CIGALE-derived $A_{\rm FUV}$ to the assumed attenuation-curve shape, we repeated the SED fitting while varying the slope $\delta$ of the power-law modification to the attenuation curve over $\delta =$-0.6,-0.3,0.0,0.3,and 0.6. We find that $\delta$ is poorly constrained, with the Bayesian uncertainties generally spanning most or all of the explored range. We therefore adopt $\delta=0$, corresponding to an unmodified Calzetti attenuation law, for the fiducial CIGALE fits. Allowing $\delta$ to vary does not produce a significant systematic shift in the inferred $A_{\rm FUV}$ values relative to their uncertainties. The comparison with the attenuation-route and gas-route models is therefore not driven by the assumed CIGALE attenuation-curve slope. This choice is also consistent a posteriori with the attenuation-curve analysis shown in Fig.~\ref{fig:Ext2Att}, where the effective curves produced by the adopted dust geometries can reproduce a Calzetti-like attenuation law over the relevant UV wavelength range.

Over almost the entire stellar mass range of the horizontal top branch in Fig.~\ref{fig:afuv_mstar}, we find that A$_{FUV}$ reaches some kind of asymptotic value at A$_{FUV}$ $\approx$ 2 - 4. This is similar to the IR-selected sample in the more local (1.0 $\lesssim$ z $\lesssim$2.2) values found by \citet{Buat2012} with a mean dust attenuation equal to 1.6, 2.2, and 2.6 mag., for the galaxies without IR data (that is UV-selected galaxies), with MIPS data only, and with MIPS and PACS data (IR-selected), respectively.

For all panels, the saturation at A$_{FUV}$ $\gtrsim$ 2 - 4 means that the galaxies become optically thick. This constraint comes from the top galaxies that lie on a horizontal line. This puts a strong constraint on the covering factor for these objects that must be very close to 1.0. With any value of f$_{cov}\lesssim0.98$, the models would not reach these objects. The constraint is weaker for GELDAs .

The comparison with the higher opacity models shown in Figure~\ref{fig:afuv_mstar} indicates that the data do not select a single value of $\kappa_{\rm FUV}^{(\rm dust)}$ independently of the adopted attenuation route. In the attenuation-based route, the SNe-dominated and hybrid prescriptions with $\kappa_{\rm FUV}^{(\rm dust)}=10^3\, {\rm cm^2\,g^{-1}}$ provide the closest overall match to the very low attenuation of GELDAs, whereas increasing the opacity to $>10^4\,{\rm cm^2\,g^{-1}}$ generally raises the predicted attenuation and worsens the agreement for several of the dust-mass-based tracks.

The gas-based route is less restrictive. In this case, the effective optical depth also depends on the metallicity and dust-to-metal ratio through $\kappa_{\rm FUV}^{(\rm gas)} =\kappa_{\rm FUV}^{(\rm dust)}({\rm DTM}\times Z)$. Consequently, the gas-route model with $\kappa_{\rm FUV}^{(\rm dust)}=10^4\, {\rm cm^2\,g^{-1}}$ can still reproduce a substantial part of the observed $A_{\rm FUV}$ - $M_{\rm star}$ distribution, particularly for galaxies with low metallicities and dust-to-metal ratios. By contrast, standard ISM-like opacities of order $10^5\,{\rm cm^2\,g^{-1}}$ systematically predict excessive attenuation unless additional extreme leakage is introduced.

We therefore infer a preferred low-opacity range of $\kappa_{\rm FUV}^{(\rm dust)}\sim10^3$ - $10^4\, {\rm cm^2\,g^{-1}}$, rather than a unique value of $10^3\,{\rm cm^2\,g^{-1}}$. The precise value remains degenerate with the adopted dust-mass or gas scaling relations, covering fraction, galaxy size, and star - dust geometry.

Figure~A2 in \citet{Inoue2020} presents the mass absorption coefficients for single-size dust grains with various radii and grain compositions (graphite, astronomical silicates, amorphous carbon, and SiC). For small grains ($a\lesssim0.1~\mu$m), the UV mass absorption coefficient rises steeply, reaching $\kappa_{0.15\,\mu{\rm m}}\sim10^5~{\rm cm^2\,g^{-1}}$. However, when the grain size becomes comparable to or larger than the radiation wavelength, the opacity becomes strongly grain-size dependent. In particular, for large grains ($a=1~\mu$m), \citet{Inoue2020} show that the UV mass absorption coefficient decreases by nearly an order of magnitude, reaching $\kappa_{0.15\,\mu{\rm m}}$ of 10$^3$ to 10$^4 {\rm cm^2 g^{-1}}$ (Fig.~\ref{fig:kappa_grains}).

The curve in Fig.~A2 of \citet{Inoue2020} corresponding to single-size grains with radius $a=1\,\mu{\rm m}$ reaches $\kappa_{0.15\,\mu{\rm m}}\sim10^3$ - $10^4\,
{\rm cm^2\,g^{-1}}$. This illustrates that the low FUV opacities favoured by our attenuation models are physically plausible for grain populations dominated by sufficiently large grains. Such large-grain populations may also produce flatter UV extinction curves and lower effective FUV attenuation. However, the comparison is illustrative and does not imply that the SNe-dominated model adopted here consists uniquely of $1\,\mu{\rm m}$ grains.

As shown in Appendix~\ref{appendix:geometry} (Figs.~\ref{fig:AppendixC1}-\ref{fig:AppendixC3}), A$_{\rm FUV}$ depends on several  parameters that affect the $A_{\rm FUV}$-$M_{\rm star}$ relation in different ways: dust mass and size regulate the optical depth, $f_{\rm cov}$ sets the normalization via unobscured sightlines, and geometry introduces secondary variations in shape and dispersion. 

However, these degeneracies are not complete. Variations in geometry and covering fraction alone cannot reproduce the observed low-attenuation systems if standard ISM-like opacities ($\kappa_{\rm UV} \gtrsim 10^{4}\ \mathrm{cm^2\,g^{-1}}$) are adopted. In contrast, low-opacity models ($\kappa_{\rm UV} \sim 10^{3}\ \mathrm{cm^2\,g^{-1}}$) reproduce the observed trends across all geometries, demonstrating that intrinsic dust opacity is the dominant parameter.

As mentioned before, the conversion of dust and gas masses into surface densities relies on the adopted stellar mass - size relation. In our framework, the optical depth scales as $\tau \propto \Sigma \propto M / (\pi R_{\rm e}^2)$, so variations in the effective radius directly affect the normalization of the $A_{\rm FUV}$-$M_{\rm star}$ relation. 

Assuming a single characteristic size for stars, gas, and dust is a simplification, as observations suggest that gas can be more extended while dust may be comparable to or slightly more extended than the stellar component. This choice should therefore be interpreted as an effective, luminosity-weighted attenuation scale. 

Varying the adopted size primarily shifts the normalization of the relation and is partially degenerate with dust opacity, covering fraction, dust mass, and geometry. Because ($\tau_{\rm FUV}\propto \kappa_{\rm FUV}M_{\rm dust}/R_e^2$, adopting larger effective radii reduces the predicted attenuation and therefore permits correspondingly larger opacity values, whereas adopting more compact dust distributions strengthens the preference for low opacity. The quoted opacity range should therefore be regarded as conditional on the adopted 70 - 300 pc effective-radius prescription.

In summary, the observed $A_{\rm FUV}$ - $M_{\rm star}$ relation at $z>9$ favours intrinsically low effective FUV dust opacities, $\kappa_{\rm FUV}^{(\rm dust)}\sim10^3$ - $10^4\, {\rm cm^2\,g^{-1}}$, together with porous star - dust geometries. Standard ISM-like prescriptions with $\kappa_{\rm FUV}^{(\rm dust)}\sim10^5\, {\rm cm^2\,g^{-1}}$ systematically overpredict the  attenuation. Such low opacities are consistent with dust produced in SNe and subsequently processed by reverse shocks, although this interpretation is not unique and depends on the assumed grain properties and dust-evolution history.

For the low dust-to-metal ratios and metallicities measured in GELDAs, this opacity range corresponds to very low effective gas opacities, $\kappa_{\rm UV}^{(\rm gas)}\sim10^{-3}$ - $10^{-1}\,{\rm cm^2\,g^{-1}}$. Non-GELDAs reach higher effective gas opacities because of their larger dust-to-metal ratios and metallicities. Within the adopted scaling relations, this
comparison indicates that the increase in effective opacity reflects the combined growth of the dust content and metal enrichment, rather than metallicity alone.

In summary, the observed $A_{\rm FUV} - M_{\rm star}$ relation directly favours a low effective intrinsic FUV opacity, $\kappa_{\rm FUV}^{(\rm {dust})}\sim10^3-10^4\ {\rm cm^2,g^{-1}}$, together with porous star–dust geometries. Standard ISM-like prescriptions with $\kappa_{\rm FUV}^{(\rm dust)}\sim10^5 {\rm cm^2,g^{-1}}$ generally overpredict the attenuation.

The connection of this low-opacity regime with large-grain-dominated dust and, more specifically, with reverse-shock-processed SN dust is the physical interpretation adopted and tested in this work. The attenuation analysis alone does not constitute a unique determination of the dust-production channel; nevertheless, the SNe-dominated scenario is our preferred baseline because it is independently supported by the short available timescales, low metallicities, low dust-to-metal ratios, and expected inefficiency of ISM grain growth in these systems.

\subsection{High-$z$ UV luminosity function}
\label{subsec:UVLFs}

\begin{figure}[t]
    \centering
    \includegraphics[width=\linewidth]{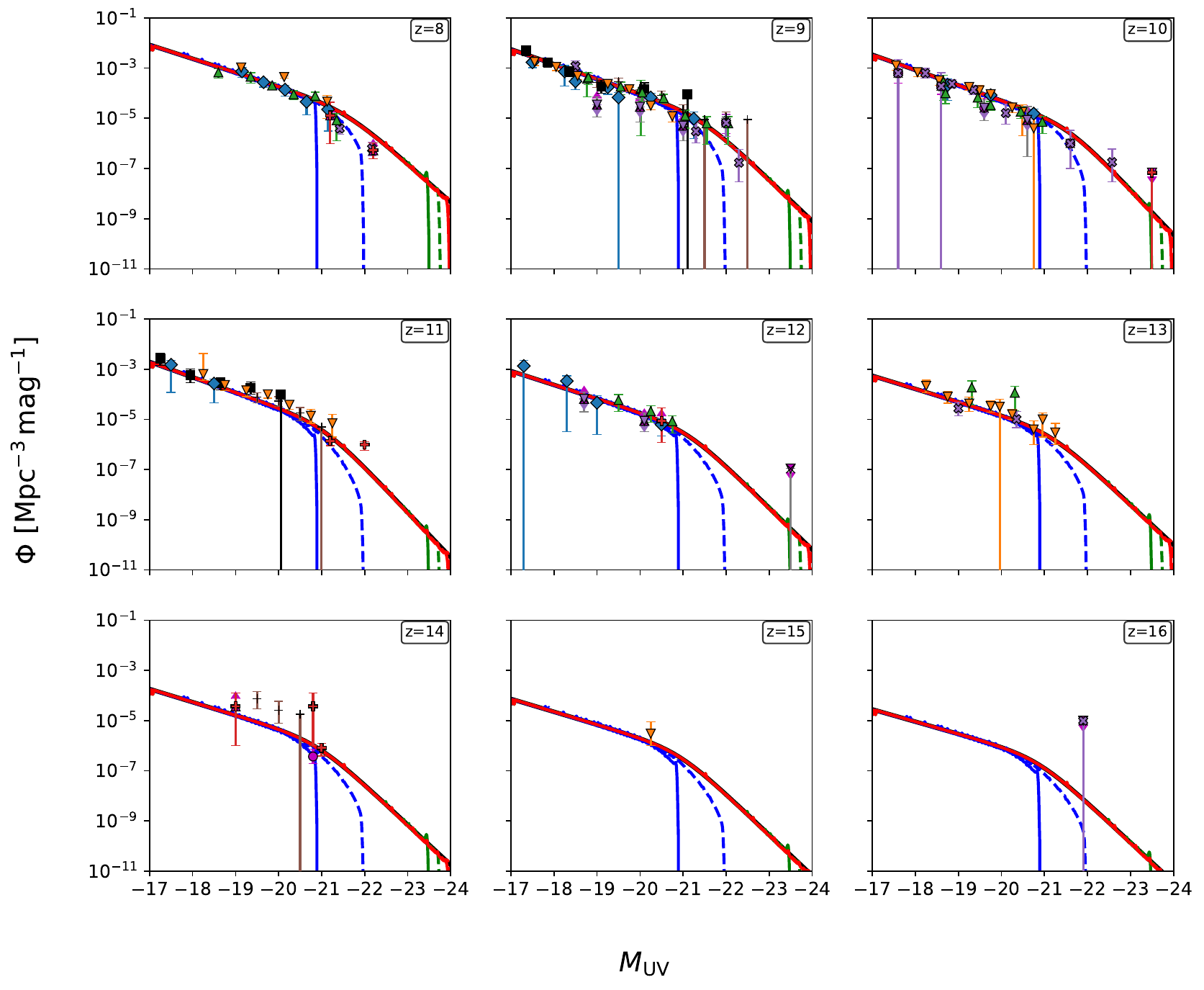}
    \caption{
        Observed rest-frame UV luminosity functions from the literature are overplotted using distinct symbols and colors. The intrinsic UVLFs assume some redshift evolution of the SFE, consistent with the constraints derived from our samples (see text). ND24 intrinsic UVLF and six attenuation scenarios compared with the observed UVLF observations measurements. Each panel shows ND24 intrinsic (black) together with the hybrid (green), pure–ISM (blue), and pure–SNe (red) attenuation curves, each in both screen and mixed geometries. Because the pure-SNe UVLFs are almost identical to the intrinsic ones, the black intrinsic UVLFs are hidden under those corrected with pure–SNe (red) attenuations. Filled symbols indicate direct detections of the UV luminosity function, while downward-pointing triangles denote upper limits and upward-pointing triangles denote lower limits, as reported in the original works. Different colors and marker shapes are used to distinguish independent observational datasets: \citet{Harikane2023}: magenta circle, \citet{Harikane2022}: black square, \citet{Adams2023}: blue diamond, \citet{Bouwens2023a}: green up-triangle and orange down-triangle, \citet{Castellano2023}: red plus, \citet{Donnan2023}: purple X, \citet{Finkelstein2023}: brown plus, \citet{Harikane2023}: gray star, \citet{Leung2023}: black square, \citet{Perez-Gonzalez2023}: blue diamond, \citet{Adams2024}: green up-triangle, \citet{Donnan2024}: orange down-triangle, \citet{Harikane2024}: red plus, and \citet{Harikane2024}: purple. 
    }
    \label{fig:uvlf_nd24}
\end{figure}

\begin{figure}[t]
    \centering
    \includegraphics[width=1.1\linewidth,height=0.3\textheight]{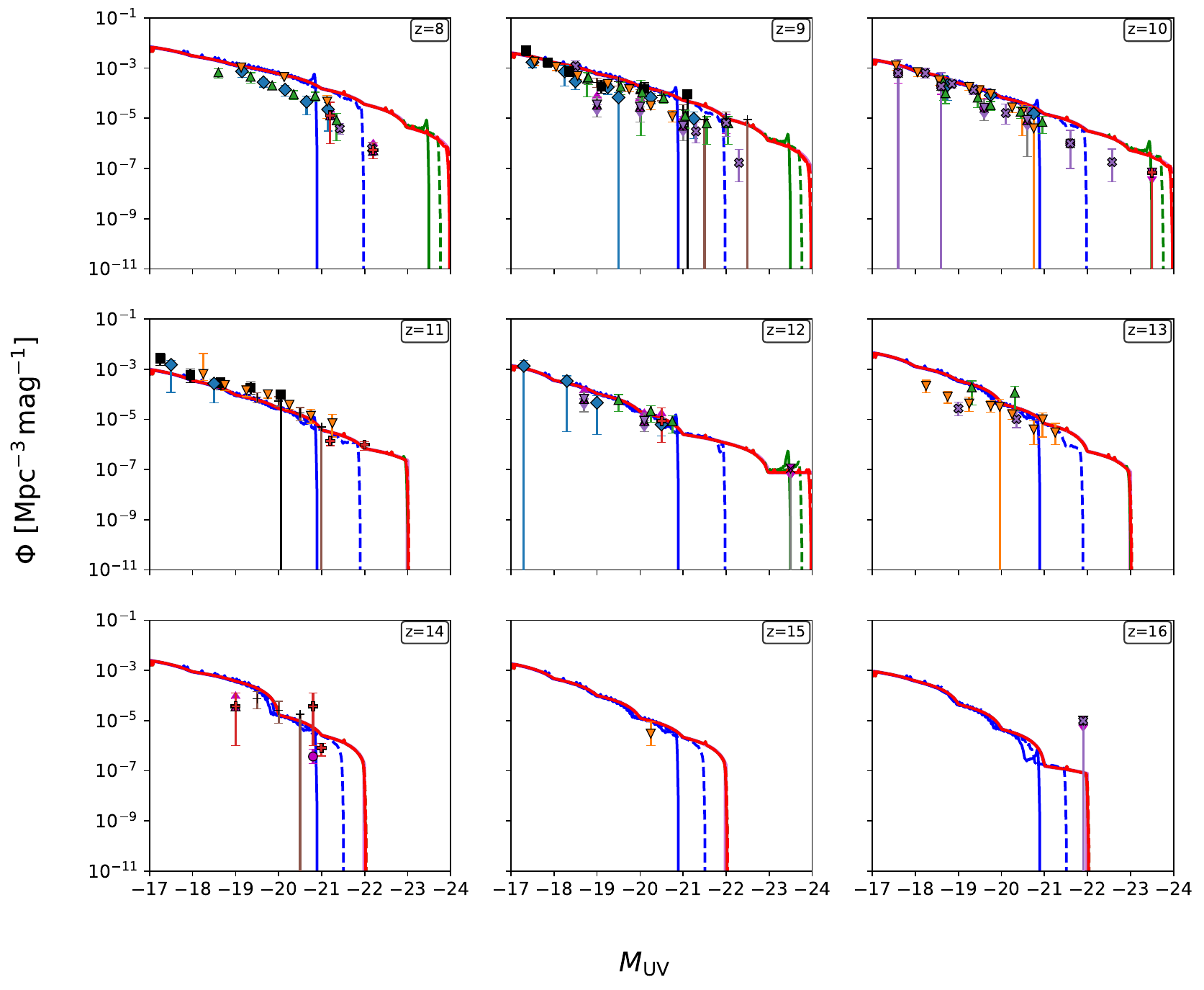}
    \caption{
        Same as Fig.\ref{fig:uvlf_nd24} but for dust-free UVLF predictions from the Santa Cruz SAM (pink curves). For the Santa Cruz models, \citet{Somerville2025} built new density-modulated SFE models with dense gas fraction f$_{\rm dense}$ = 0.1, 0.5, and 1.0 for the models labelled {\it cloudfp1}, {\it cloudfp5} and {\it cloudfd1}, respectively, with a so-called {\it basekenn} version that relies on a standard Kennicutt–Schmidt-like star formation law, in which the SFR surface density is proportional to the total cold ISM surface density. Here, the labels indicate which model is used for which redshift.
    }
    \label{fig:uvlf_scsam}
\end{figure}

Figs.~\ref{fig:uvlf_nd24} and \ref{fig:uvlf_scsam} illustrate the impact of dust attenuation on the rest-frame UV luminosity function (UVLF) over the redshift range $8 \le z \le 16$, using two independent intrinsic galaxy population models: the semi-empirical model of \citet{Nikopoulos2024} (ND24) and the Santa Cruz semi-analytic model of \citet{Somerville2025} (SCSAM). In both cases, dust attenuation is implemented through a deterministic, number-conserving remapping of the intrinsic UV magnitude distribution, allowing us to isolate how different dust prescriptions reshape the bright end of the UVLF while preserving the underlying galaxy abundance.

In the ND24 model, the intrinsic UVLF displays a smooth power-law slope at the faint end and a gradual turnover at the bright end across all redshifts shown (Fig.~\ref{fig:uvlf_nd24}). This behaviour reflects the explicit redshift evolution of the star formation efficiency (SFE) adopted in the model, $\mathrm{SFE}(z)=10^{0.13z-3.5}$, which increases by approximately a factor of eight between $z=8$ and $z=15$. This scaling is required to reproduce the observed normalization of the UVLF at high redshift and is consistent with recent observational constraints. In particular, \citet{Burgarella2025} find that galaxies in the CEERS sample ($4<z\lesssim11.5$) broadly follow a Kennicutt - Schmidt-like star formation law, while at $8<z\lesssim15$ there is evidence for an increase in the star formation rate surface density $\Sigma_{\rm SFR}$ at fixed gas surface density. At $\log(\Sigma_{\rm gas}/M_\odot\,{\rm pc}^{-2})\simeq4$, $\Sigma_{\rm SFR}$ increases by a factor of a few from the CEERS sample to the ultra-high-redshift regime, in qualitative agreement with the ND24 prescription.

\citet{Somerville2025} introduce density-modulated star formation efficiency (DMSFE) models characterized by different dense gas fractions, $f_{\rm dense}=0.1$, 0.5, and 1.0, corresponding to the {\it cloudfp1}, {\it cloudfp5}, and {\it cloudfd1} variants, respectively, in addition to a baseline Kennicutt - Schmidt model ({\it basekenn}). We find that the {\it cloudfp1} model provides the best agreement with observed UVLF constraints at $6\lesssim z\lesssim12$, whereas the higher dense-gas-fraction models (cloudfp5 and cloudfd1) systematically overpredict the number of bright UV sources when combined with the same simplified dust treatment. For our analysis, we therefore adopt the best-fit intrinsic SCSAM model at each redshift and examine how different dust prescriptions modify its UVLF (Fig.~\ref{fig:uvlf_scsam}).

In each redshift panel of Figs.~\ref{fig:uvlf_nd24} and \ref{fig:uvlf_scsam}, we show the intrinsic UVLF together with six attenuation prescriptions based on hybrid attenuation curves computed for leaky-screen and leaky-mixed geometries, as well as the limiting ``pure ISM'' and ``pure SNe'' dust branches, each evaluated for both screen and mixed configurations. As discussed previously, Poissonian clump geometries yield results qualitatively similar to the leaky cases; therefore, the UVLF analysis focuses exclusively on the leaky-screen and leaky-mixed geometries.

Across both intrinsic models, dust attenuation primarily affects the bright end of the UVLF, with the degree of suppression depending sensitively on the assumed dust origin and geometry. ISM-dominated dust produces the strongest attenuation at fixed redshift, resulting in a sharp cutoff at bright magnitudes that is increasingly inconsistent with the observed abundance of luminous sources. Hybrid prescriptions yield intermediate suppression, while the pure SNe dust branch produces minimal attenuation. This weak suppression arises because SNe-processed dust is characterized by low dust masses and intrinsically low far-UV opacities, owing to reverse-shock destruction and the dominance of large grains.

As seen in both Figs.~\ref{fig:uvlf_nd24} and \ref{fig:uvlf_scsam}, the observed bright-end constraints are equally well reproduced by hybrid and pure SNe dust prescriptions when combined with porous (leaky) geometries, as both leave the UVLF close to its intrinsic form. In this sense, any attenuation model characterized by very small dust corrections which is fully consistent with SNe-dominated dust production, together with an increasing SFE toward high redshift, provides a viable explanation for the excess of UV-bright galaxies revealed by JWST. Discriminating between hybrid and pure SNe dust scenarios will require larger and more statistically robust samples of extremely UV-luminous galaxies at $z\gtrsim10$, where the differences between attenuation prescriptions become more pronounced.

We stress that the good agreement with the observed UV luminosity functions does not uniquely validate the adopted intrinsic galaxy models or their underlying star-formation prescriptions. Because our attenuation framework predicts very low dust attenuation, it leaves the intrinsic UVLF largely unchanged and is therefore compatible with any intrinsic galaxy population already close to the observations. Instead, our results highlight that low effective UV attenuation is a necessary ingredient to preserve the bright end of the UVLF. While alternative physical mechanisms (e.g., bursty star formation or variations in star-formation efficiency) may also shape the intrinsic UVLF, they must still be consistent with this requirement.

\section{Discussion}
\label{sec:discussion}

\subsection{The origin of dust in the early Universe: supernovae as primary producers}

The presence of dust in galaxies at $z \gtrsim 8$, when the Universe was younger than $\sim 650$ Myr, poses a well-known challenge. At these early epochs there is insufficient time for asymptotic giant branch (AGB) stars to contribute significantly, as they require several hundred Myr to evolve off the main sequence (e.g.\ \citealt{Valiante2009, Schneider2024}). Grain growth in the ISM is likewise disfavoured: it becomes efficient only above a critical metallicity $Z_{\rm crit} \gtrsim 0.1\,Z_\odot$ (e.g.\ \citealt{Asano2013, Popping2017, Triani2020, Dayal2022, Mauerhofer2023, Graziani2020, Lewis2023, DiCesare2023, Choban2024}), which most galaxies at these redshifts have not yet reached. Neither AGB stars nor ISM accretion can plausibly account for the observed dust. Instead, both theoretical work and emerging observational evidence increasingly point to core-collapse SNe as the dominant dust factories in the early Universe (e.g.\ \citealt{Gallerani2010, McKinney2025, Todini2001, Nozawa2003}).

Numerous studies have shown that the ejecta of SNe, including pair-instability events (PISNe), can efficiently condense dust grains at very high redshift \citep[e.g.][]{Todini2001, Schneider2024, Hirashita2008}. This scenario is supported by both theoretical dust-formation models and observations of nearby SN remnants such as Cassiopeia~A and SN~1987A, whose far-infrared and submillimetre emission reveal cold dust masses of $0.1$ - $0.5\,M_\odot$ \citep[e.g.][]{DeLooze2017}. The short evolutionary timescales of massive stars naturally accommodate the rapid emergence of dust timescales shorter than $\sim$ 30 Myr favoring dust production by supernovae (\citealt{Langeroodi2024}).

In summary, as shown in Appendix~\ref{appendix:otherkappas} (Fig.~\ref{fig:kappa_grains}), the low UV opacity required by our model corresponds to dust populations dominated by large grains. The balance of evidence suggests that dust in the early Universe is primarily forged in the ejecta of core-collapse SNe and subsequently amplified through grain growth in the dense ISM once galaxies reach sufficient metallicities and gas densities \citep[e.g.][]{Mancini2015, Popping2017, Dayal2022, Mauerhofer2023}. This two-phase framework provides a natural explanation for the pronounced diversity in dust attenuation among high-$z$ systems: some galaxies exhibit substantial obscuration already at 4 $<$ z $<$ 9, while others remain almost dust-free despite comparable stellar masses and star-formation rates. 

Within this context, the first phase, when galaxies contain only ``stardust'' freshly produced in SNe ejecta and have not yet experienced significant ISM grain growth, naturally corresponds to the GELDA population identified with JWST by \citet{Burgarella2025}. These same objects have also been described as ``blue monsters'' because their UV luminosities exceed pre-JWST expectations \citep[e.g.][]{Finkelstein2022, BoylanKolchin2023}. 

\subsection{Do dust-rich low-metallicity galaxies at $z < 9$ and in the local Universe imply surviving pockets of ``high-$z$'' conditions?}

Our analysis suggest that GELDAs still exist (representing a lower $\sim$ 20\% of the total sample) even at z $\lesssim$ 9. Whether the existence of low-metallicity systems with an extremely low dust attenuation, that is GELDAs, exist at intermediate and low redshift is an interesting question. This would imply that some low-redshift regions of the Universe continue to sustain physical conditions similar to those at $z>9$. These systems lie well below the dust-to-gas and dust-to-metal scaling relations followed by the bulk of the galaxy population \citep[e.g.,][]{RemyRuyer2014,DeVis2019}, and instead occupy a region of the dust-mass-metallicity plane that overlaps with the high-$z$ SNe-dominated sequence.

Such galaxies may represent environments in which the ISM remains too metal-poor or too diffuse for efficient grain growth, thereby preserving conditions where dust production is driven mainly by core-collapse SNe rather than ISM accretion \citep{Gall2011,Graziani2020,Mauerhofer2023}. Similar conclusions have been reported for extremely metal-poor dwarfs in the local Universe \citep[e.g.,][]{Galliano2022}. The existence of these low-$z$ analogues suggests that pockets of the Universe may persistently undergo “high-redshift–like’’ dust evolution, with inefficient grain growth and dominant SNe dust sources, even several billion years after the epoch when such conditions were thought to be common. This continuity in dust-formation pathways provides a natural bridge between the properties of early galaxies and the diverse dust contents observed at later cosmic times.

Even more telling is the existence of local analogs. The samples compiled by \citet{RemyRuyer2014,DeVis2019} include nearby dwarf galaxies with low metallicities, strong bursty star-formation, and dust-to-gas ratios well below those of typical spirals. Many of these dwarfs show properties (metallicity, stellar mass, DTM) that resemble what is expected for early-Universe objects. Do such galaxies genuinely reproduce the physical conditions of the high-$z$ Universe?  In many respects, the answer appears to be yes.  Low-redshift dwarf starbursts reside in shallow potential wells, have inefficient metal retention. These factors mirror the conditions expected for primordial galaxies. Consequently, they may represent ``fossil environments'' that have retained a mode of ISM regulation and dust evolution reminiscent of the first few hundred Myr after the Big Bang.

However, the analogy might not be exact. Local analogs and $z\lesssim6$ starbursts exist in a Universe that is already enriched, ionised, and permeated by evolved stellar populations. Their gas accretion rates could differ from those at very high redshift. Thus, while these objects provide crucial laboratories for understanding dust production and feedback cycles, they probably do not fully recreate the cosmological context of $z > 9$ galaxies.

This opens up the possibility to directly observe some of these GELDAs in the intermediate redshift range 1 $\lesssim$ z $\lesssim$ 6 in the far-infrared and (sub-)millimeter range with IRAM/NOEMA and ALMA, providing a direct measure of their dust mass.

To assess the recoverability of the parameters derived with CIGALE, we performed a mock-catalogue analysis following Boquien et al. (2019). For each galaxy, the best-fitting model was adopted as a reference model with known physical parameters. Its model photometry was perturbed using Gaussian noise consistent with the observational uncertainties, and the resulting mock SED was refitted using the same parameter grid and analysis procedure as for the observed data. We then compared the input parameters of the reference models with the Bayesian estimates recovered from the mock catalogue. 
This mock analysis evaluates parameter recoverability for the available photometry and the adopted model grid.

For stellar mass, the comparison gives $r^2>0.95$, indicating that more than 95\% of the variance in the mock input stellar masses is reproduced by the recovered estimates. Stellar mass is therefore well constrained internally within the adopted CIGALE configurations. For the FUV attenuation, the comparison gives $r^2>0.77$, indicating that more than 77\% of the variance in the mock input A$_{FUV}$ values is reproduced. The lower value and the substantial individual uncertainties show that A$_{FUV}$ is less tightly constrained than stellar mass. Many of the low-attenuation measurements should consequently be interpreted as upper limits rather than precise determinations.

However, given the large uncertainties on $A_{\rm FUV}$, many of the measurements should be interpreted as upper limits rather than precise determinations. We therefore do not attempt to infer a statistically robust object-by-object correlation between these quantities.

From a physical perspective, such a correlation is not necessarily expected. The transition from a SNe-dominated dust regime to an ISM grain-growth regime does not follow a single evolutionary track, but instead depends on when each galaxy reaches the critical metallicity required for efficient grain growth. This naturally leads to a diversity of evolutionary pathways and contributes to the large apparent scatter observed in this plane.

Instead, the figure highlights a population-level behaviour: several low-mass, metal-poor systems are consistent with very low attenuation, supporting the existence of SNe-dominated dust regimes and identifying candidates for possible Population~III descendants (see Section~\ref{subsec:popIII_descendants}).

\subsection{Could some GELDAs be direct descendants of population~III stellar populations?}
\label{subsec:popIII_descendants}

Direct detection of truly metal-free Population~III stars remains elusive. A promising alternative is to search for their ``ashes'' in the earliest galaxies, that is, for signatures of primordial metal and dust production imprinted on galaxy emission \citep[e.g.][]{Schneider2004,Nozawa2007,Cherchneff2009}.

GELDAs at $z\gtrsim9$ are natural candidates, as they likely formed close to the onset of cosmic star formation. Simulations place the first stars at $\Delta t\sim100$ - $200~\mathrm{Myr}$ after the Big Bang \citep{Klessen2023}, followed by rapid enrichment by the first supernovae \citep{Wise2012}. In this early stardust phase, dust production is dominated by massive stars, before efficient ISM grain growth sets in. GELDAs may therefore trace systems whose dust reservoirs were produced by the first stellar generations, potentially enriched primarily by Population~III SNe. Although the stars themselves still remain unobservable, the dust content and attenuation properties of what remains after their explosion could probe the earliest stages of chemical enrichment \citep{Schneider2004,Nozawa2007,Hirashita2008}.

However, only a subset of GELDAs are probably direct descendants of Population~III stars  since continued star formation and enrichment should rapidly drive most systems toward Population~II regimes with efficient ISM dust growth.

Predictions of the fraction of JWST-bright galaxies\footnote{Here, ``JWST-bright'' refers to galaxies sufficiently luminous to yield a usable NIRSpec/PRISM spectrum at CEERS-like depth ($t_{\rm exp}\simeq3.1$ ks), corresponding to a $5\sigma$ continuum limit of $m_{\rm AB}\simeq26.5$ at $\lambda\simeq1.5\,\mu$m and/or to unresolved emission-line fluxes $F_{\rm line}\gtrsim(2$ - $7)\times10^{-18}\,\mathrm{erg\,s^{-1}\,cm^{-2}}$. Over the redshift range $8<z<15$, this corresponds approximately to $M_{\rm UV}\lesssim-20.5$ to $-21.5$, depending on redshift and spectral slope.} that could host Population~III vary widely because of uncertainties in cooling, feedback, metal mixing, and the IMF \citep[e.g.][]{Bromm2004,Wise2012,Hirano2015,Tanaka2021}. Semi-analytic models coupled to JWST limits suggest that the probability of detecting galaxies with a non-negligible Population~III contribution peaks at $z\simeq9$ - 10, reaching a few to $\sim10\%$, and declines rapidly thereafter \citep{Sarmento2018}. Detailed studies of spectroscopic diagnostics and search strategies for Population~III systems with JWST likewise indicate that only a small fraction of accessible galaxies should display identifiable Population~III signatures \citep{Trussler2023}. Simulations resolving chemical inhomogeneities find that Population~III star formation may persist in partially enriched galaxies, with hosting fractions as high as $f_{\rm PopIII|gal}\sim0.3$ - 0.7 at $z\gtrsim9$ for $M_{star}\sim10^5\,-\,10^8\,M_\odot$ systems \citep{Yajima2023}. More conservative semi-empirical/semi-analytic models calibrated to UV luminosity functions predict Population~III mass fractions $\lesssim1\%$ by $z\sim10$ in luminous galaxies, implying rare observable signatures \citep[e.g.][]{Trinca2024,Jaacks2019,Xu2016,Valiante2016,Katz2025}, consistent with evolutionary scenarios in which enrichment rapidly enables ISM dust
processing and grain growth \citep{Schneider2024}.

Taken together, these studies motivate a broad bracket for the hosting fraction of Population~III in JWST-bright galaxies in the ultra-high redshift universe, f$_{\rm PopIII|bright}\sim$0.01 - 0.7, with the lower bound reflecting semi-analytic expectations and the upper bound optimistic scenarios with persistent pristine pockets.

In the UHZ PRISM sample, GELDAs dominate at high redshift: among the $z\geq9$ targets, 16 out of 19 are classified as GELDAs, corresponding to $f_{\rm GELDA}\simeq0.84$, whereas only $2/7\simeq0.29$ objects at $z<9$ fall in this category. Onto this sample, we get 
N$_{\rm PopIII\text{-}host}\simeq f_{\rm PopIII|bright}\,N_{\rm gal}$ which yields an expected number of $N_{\rm PopIII\text{-}host}\sim0.2$–13 Population~III–hosting systems for $N_{\rm gal}\simeq19$ galaxies at $9<z<15$. Until Population~III galaxies are unambiguously identified and confirmed, it remains unclear whether pessimistic or optimistic theoretical scenarios are closer to reality. Nevertheless, the upper value is very likely too optimistic as we did not clearly identify any safe Population~III system. Population~III hosts would constitute only a subset of the observed GELDAs, supporting the view that the GELDA phenomenon is not uniquely tied to Population~III star formation.

This framework yields testable predictions: GELDAs in a stardust scenario should exhibit low metallicities, reduced dust-to-metal ratios, hard ionising spectra, and dust masses consistent with SN-dominated production and reverse-shock processing \citep{Nozawa2007,Micelotta2016,Kirchschlager2019}. JWST spectroscopy, including MIRI searches for PAHs, together with deeper ALMA and NOEMA observations, will be key to constraining abundances, ionisation conditions, and dust growth mechanisms. 

\begin{figure}
  \centering
  \includegraphics[width=\linewidth]{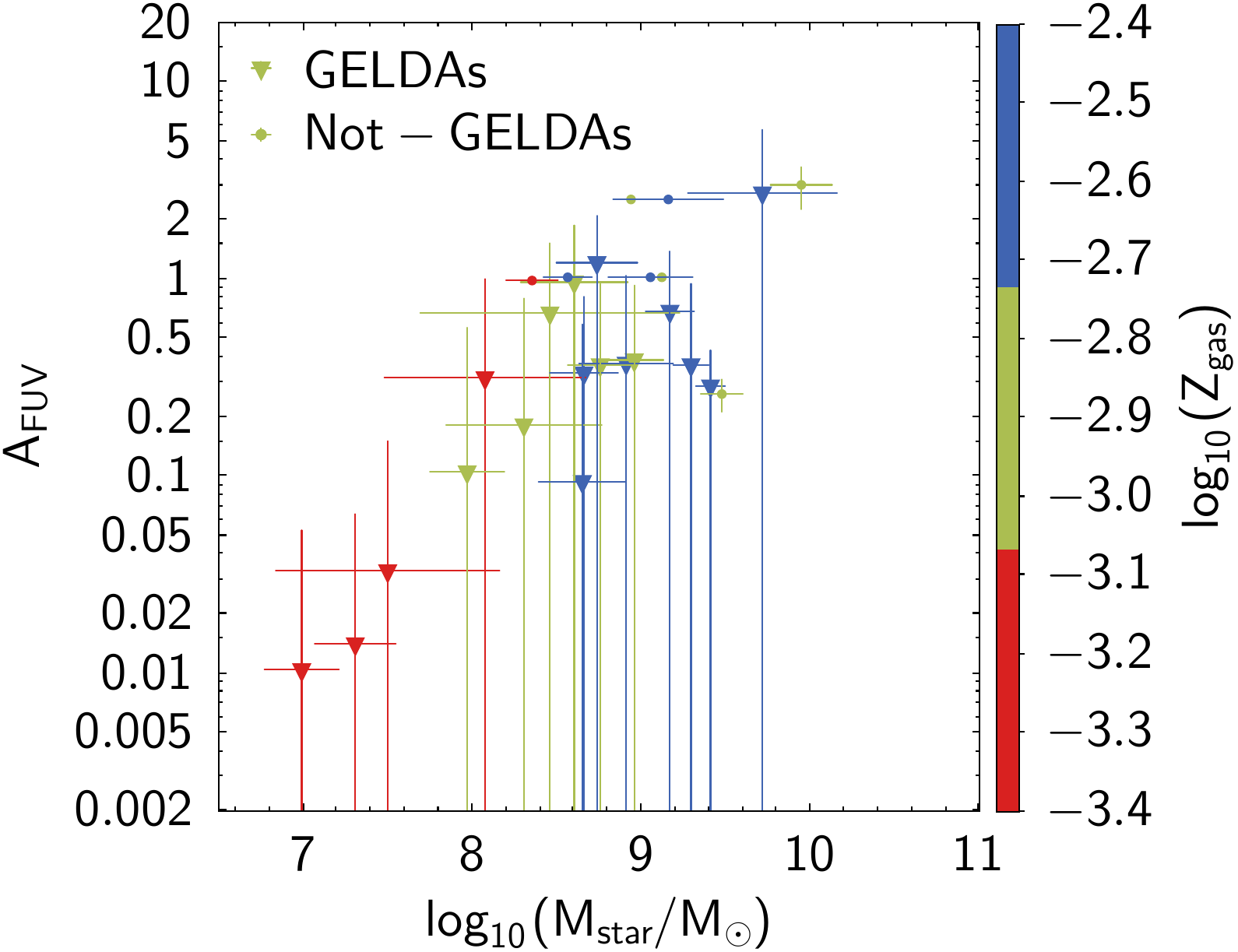}
\caption{
$A_{\rm FUV}$ versus $\log_{10}(M_{\rm star}/M_\odot)$ for the sample presented in Table~\ref{tab:uhz}, color-coded by gas metallicity. GELDAs are shown as large triangles, while the other objects are shown as small circles. We observe a trend toward lower $A_{\rm FUV}$ and lower $M_{\rm star}$ at lower metallicities. Note, however, that a strict correlation might not be expected, as the scatter at $A_{\rm FUV} > 0.1$ may have a physical origin related to the onset of metal accretion onto dust grains. The three lowest-metallicity galaxies, located in the bottom-left region, are UNCOVER37126 at $z = 10.25$, GS-z13-1-LA at $z = 13.01$, and GZ-z13-0 at $z = 13.2$. The figure highlights that several low-mass, metal-poor systems are consistent with very low attenuation. The lowest-metallicity objects in this region represent candidates for possible Population~III descendants.
}
  \label{fig:Afuv_Mstar_UHZ}
\end{figure}

Figure~\ref{fig:Afuv_Mstar_UHZ} shows the ultra-high-redshift sample in the $A_{\rm FUV}$ - $M_\star$ plane, with the points colour-coded by gas metallicity.

\section{Conclusions}
\label{sec:conclusions}

We have investigated the physical origin of the excess of UV-bright galaxies observed by JWST at $z \gtrsim 9$, focusing on the role of dust attenuation in extremely young and metal-poor systems. We developed a physically motivated framework that combines SN-dominated dust production, metallicity- and dust-to-metal–dependent opacity scalings, and porous star-dust geometries to model the emergent far-UV attenuation in early galaxies.

Assuming that reduced dust attenuation is a major contributor to the excess of UV-bright galaxies, we show that our framework provides an alternative to attenuation-free scenarios (\citealt{Ferrara2023, Ferrara2025}) based on large-scale dust removal or UV escape through nearly dust-free channels. Our approach preserves the bulk of the gas and dust reservoirs and instead reduces the effective FUV attenuation through low intrinsic dust opacities combined with porous, mass-conserving star–dust geometries.

The direct requirement emerging from the model comparison is a low effective FUV opacity. Within the baseline physical scenario adopted throughout this paper, this low-opacity regime is naturally explained by SN-produced dust processed by reverse shocks before efficient grain growth in the ISM becomes established. This interpretation provides a coherent physical explanation of GELDAs without requiring extreme feedback or widespread dust evacuation.

Our main results can be summarized as follows:

\begin{enumerate}
    \item The observed $A_{\rm FUV} - M_{\rm star}$ relation is  more compatible with an effective intrinsic FUV dust opacities of $\kappa_{\rm FUV}^{(\rm dust)}\sim10^3-10^4\ {\rm cm^2,g^{-1}}$, substantially below standard ISM-like values. Combined with low metallicities, reduced dust-to-metal ratios, and porous star–dust geometries, these opacities naturally produce A$_{\rm FUV}\lesssim0.3$ even in gas-rich GELDAs. Within our baseline physical scenario, this low-opacity regime is explained by dust produced predominantly in SN ejecta and processed by reverse shocks before efficient ISM grain growth becomes established. The opacity is the direct modelling constraint, while the identification with SNe-processed dust is the physically motivated interpretation tested in this work.

    \item When combined with covering factors $<$1 radiative-transfer geometries (continuous or clumpy), this stardust regime reproduces the observed $A_{\rm FUV}$-$M_{star}$ relation of high-redshift galaxies and naturally explains why GELDAs dominate the galaxy population at $z \gtrsim 9$, while becoming rarer at lower redshift.

    \item The hybrid dust model, in which galaxies transition from SNe-dominated dust at low metallicity to ISM grain-growth–dominated dust above a critical metallicity Z$_{\rm crit}\sim0.1\,Z_\odot$, provides a continuous and physically plausible description of dust evolution across the stellar-mass range probed at high redshift.
    
    \item Applying these attenuation prescriptions to UVLFs from both semi-empirical and semi-analytic galaxy models preferentially suppresses only the very brightest sources. This brings theoretical predictions into agreement with the JWST UVLF measurements which supports the idea that the dust attenuation in these galaxies is extremely low.
    
    \item In this early stardust phase, a minority of GELDAs may trace systems observed shortly after the onset of star formation, whose dust reservoirs are dominated by massive-star SNe and could include contributions from Population~III progenitors. Such a Population~III-descendant interpretation is expected to be transient and not representative of the entire GELDA population, since rapid enrichment quickly drives galaxies toward Population~II regimes and efficient ISM dust growth. Nevertheless, the dust content and attenuation properties of these objects may provide an indirect probe of the ``ashes'' of the first stellar generations and of the earliest phases of chemical enrichment, with testable signatures in low metallicities, reduced dust-to-metal ratios, hard ionising spectra, and
FIR/(sub-)mm dust masses.
\end{enumerate}

In conclusion, we find that a combination of SNe-produced stardust, low dust-to-metal ratios, and porous star - dust geometries along with some redshift evolution by a factor $\lesssim 10$ of the SFE provides a physically grounded and economical explanation for the excess of UV-bright galaxies at z $\gtrsim$ 9. This scenario links the observed GELDA population to the earliest stages of galaxy evolution and offers a coherent framework to interpret the emerging JWST view of the cosmic dawn.

\begin{acknowledgements}
DB and VB thank the Programme National Cosmology and Galaxies (PNCG) and the Centre National d'Etudes Spatiales (CNES) for their financial support.
TTT has been supported by the Japan Society for the Promotion of Science (JSPS) Grants-in-Aid for Scientific Research (24H00247). 
TTT has also been supported in part by the the Collaboration Funding of the Institute of Statistical Mathematics ``Machine-Learning-Based Cosmogony: From Structure Formation to Galaxy Evolution''.
\end{acknowledgements}

\bibliographystyle{aa}

\begin{thebibliography}{}
\bibitem[Adams et al.(2023)]{Adams2023} Adams, N.~J., Conselice, C.~J., Ferreira, L., et al.\ 2023, \mnras, 518, 3, 4755.
\bibitem[Adams et al.(2024)]{Adams2024} Adams, N.~J., Conselice, C.~J., Austin, D., et al.\ 2024, \apj, 965, 2, 169.
\bibitem[Arrabal Haro et al.(2023)]{Arrabal-Haro2023} Arrabal Haro, P., Dickinson, M., Finkelstein, S.~L., et al.\ 2023, \nat, 622, 7984, 707.
\bibitem[Asano et al.(2013)]{Asano2013} Asano, R.~S., Takeuchi, T.~T., Hirashita, H., \& Inoue, A.~K.\ 2013, EPS, 65, 213
\bibitem[Asano et al.(2014)]{Asano2014} Asano R.~S., Takeuchi T.~T., Hirashita H., Nozawa T., 2014, MNRAS, 440, 134.
\bibitem[Atek et al.(2023)]{Atek2023} Atek, H., Shuntov, M., Furtak, L.~J., et al.\ 2023, \mnras, 519, 1, 1201.
\bibitem[Boquien et al.(2019)]{Boquien2019} Boquien, M., Burgarella, D., Roehlly, Y., et al.\ 2019, \aap, 622, A103.
\bibitem[Bouwens et al.(2023a)]{Bouwens2023a} Bouwens, R., Illingworth, G., Oesch, P., et al.\ 2023, \mnras, 523, 1, 1009.
\bibitem[Bouwens et al.(2023b)]{Bouwens2023b} Bouwens, R.~J., Illingworth, G.~D., Oesch, P.~A., et al.\ 2023, \apj, 954, 128.
\bibitem[Boylan-Kolchin(2023)]{BoylanKolchin2023} Boylan-Kolchin, M.\ 2023, Nature Astronomy, 7, 731.
\bibitem[Bromm \& Larson(2004)]{Bromm2004} Bromm, V. \& Larson, R.~B.\ 2004, \araa, 42, 79
\bibitem[Bromm(2013)]{Bromm2013} Bromm, V.\ 2013, Reports on Progress in Physics, 76, 112901.
\bibitem[Bromm \& Yoshida(2011)]{BrommYoshida2011} Bromm, V., \& Yoshida, N.\ 2011, \araa, 49, 373. 
\bibitem[Buat et al.(2012)]{Buat2012} Buat, V., Noll, S., Burgarella, D., et al.\ 2012, \aap, 545, A141. 
\bibitem[Burgarella et al.(2025)]{Burgarella2025} Burgarella, D., Buat, V., Theul{\'e}, P., et al.\ 2025, \aap, 699, A336. 
\bibitem[Calzetti et al.(2000)]{Calzetti2000} Calzetti, D., Armus, L., Bohlin, R.~C., et al.\ 2000, \apj, 533, 682. 
\bibitem[Castellano et al.(2023)]{Castellano2023} Castellano, M., Fontana, A., Treu, T., et al.\ 2023, \apjl, 948, 2, L14. 
\bibitem[Castellano et al.(2024)]{Castellano2024} Castellano, M., Napolitano, L., Fontana, A., et al.\ 2024, \apj, 972, 2, 143. 
\bibitem[Charlot \& Fall(2000)]{Charlot2000} Charlot, S. \& Fall, S.~M.\ 2000, \apj, 539, 718. 
\bibitem[Chemerynska et al.(2024)]{Chemerynska2024} Chemerynska, I., Atek, H., Dayal, P., et al.\ 2024, \apjl, 976, L15
\bibitem[Cherchneff \& Dwek(2009)]{Cherchneff2009} Cherchneff, I., \& Dwek, E.\ 2009, \apj, 703, 642
\bibitem[Chevallard et al.(2013)]{Chevallard2013} Chevallard, J., Charlot, S., Wandelt, B., \& Wild, V.\ 2013, \mnras, 432, 2061. 
\bibitem[Choban et al.(2024)]{Choban2024} Choban, C.~R., Kere{\v{s}}, D., Sandstrom, K.~M., et al.\ 2024, \mnras, 529, 2356. 
\bibitem[Cueto et al.(2024)]{Cueto2024} Cueto, E.~R., Hutter, A., Dayal, P., et al.\ 2024, \aap, 686, A138. 
\bibitem[Cullen et al.(2023)]{Cullen2023} Cullen, F., McLure, R.~J., Donnan, C.~T., et al.\ 2023, \mnras, 520, 14. 
\bibitem[Cullen et al.(2025)]{Cullen2025} Cullen, F., Carnall, A.~C., Scholte, D., et al.\ 2025, \mnras, 540, 3, 2176. 
\bibitem[Curti et al.(2023)]{Curti2023} Curti, M., D'Eugenio, F., Carniani, S., et al.\ 2023, \mnras, 518, 1, 425. 
\bibitem[Dayal et al.(2022)]{Dayal2022} Dayal, P., Ferrara, A., Sommovigo, L., et al.\ 2022, \mnras, 512, 989. 
\bibitem[De Looze et al.(2017)]{DeLooze2017} De Looze, I., Barlow, M.~J., Swinyard, B.~M., et al.\ 2017, \mnras, 465, 3309
\bibitem[De Vis et al.(2019)]{DeVis2019} De Vis, P., Jones, A., Viaene, S., et al.\ 2019, \aap, 623, A5
\bibitem[Di Cesare et al.(2023)]{DiCesare2023} Di Cesare, C., Graziani, L., Schneider, R., et al.\ 2023, \mnras, 519, 4632. 
\bibitem[Donnan et al.(2023)]{Donnan2023} Donnan, C.~T., McLeod, D.~J., McLure, R.~J., et al.\ 2023, \mnras, 518, 6011. 
\bibitem[Donnan et al.(2024)]{Donnan2024} Donnan, C.~T., McLure, R.~J., Dunlop, J.~S., et al.\ 2024, \mnras, 533, 3, 3222. 
\bibitem[Draine(2003)]{Draine2003} Draine, B.~T.\ 2003, \araa, 41, 241
\bibitem[Dwek \& Cherchneff(2011)]{DwekCherchneff2011} Dwek, E., \& Cherchneff, I.\ 2011, \apj, 727, 63. 
\bibitem[Feldmann(2015)]{Feldmann2015} Feldmann, R.\ 2015, \mnras, 449, 3274
\bibitem[Ferrara et al.(1990)]{Ferrara1990} Ferrara, A., Aiello, S., Ferrini, F., \& Barsella, B.\ 1990, \aap, 240, 259
\bibitem[Ferrara(2022)]{Ferrara2022} Ferrara, A.\ 2022, \mnras, 512, 1598. doi:10.1093/mnras/stac620
\bibitem[Ferrara et al.(2023)]{Ferrara2023} Ferrara, A., Pallottini, A., \& Dayal, P.\ 2023, \mnras, 522, 3, 3986. 
\bibitem[Ferrara \& Carniani(2025)]{Ferrara2025} Ferrara, A., \& Carniani, S.\ 2025, \aap, 694, A215
\bibitem[Finkelstein et al.(2022)]{Finkelstein2022} Finkelstein, S.~L., Bagley, M.~B., Arrabal Haro, P., et al.\ 2022, \apjl, 940, L55. 
\bibitem[Finkelstein et al.(2023)]{Finkelstein2023} Finkelstein, S.~L., Bagley, M.~B., Arrabal Haro, P., et al.\ 2023, \apjl, 946, L13. 
\bibitem[Finkelstein et al.(2025)]{Finkelstein2025} Finkelstein, S.~L., Bagley, M.~B., Arrabal Haro, P., et al.\ 2025, \apjl, 983, 1, L4. 
\bibitem[Fujimoto et al.(2024)]{Fujimoto2024} Fujimoto, S., Ouchi, M., Nakajima, K., et al.\ 2024, \apj, 964, 2, 146. 
\bibitem[Gall et al.(2011)]{Gall2011} Gall, C., Andersen, A.~C., \& Hjorth, J.\ 2011, \aapr, 19, 43
\bibitem[Gall \& Hjorth(2018)]{Gall2018} Gall, C. \& Hjorth, J.\ 2018, \apj, 868, 1, 62. 
\bibitem[Gallerani et al.(2010)]{Gallerani2010} Gallerani, S., Maiolino, R., Juarez, Y., et al.\ 2010, \aap, 523, A85
\bibitem[Galliano(2022)]{Galliano2022} Galliano, F.\ 2022, Habilitation Thesis, 1. doi:10.48550/arXiv.2202.01868
\bibitem[Gordon et al.(2003)]{Gordon2003} Gordon, K.~D., Clayton, G.~C., Misselt, K.~A., et al.\ 2003, \apj, 594, 279. 
\bibitem[Graziani et al.(2020)]{Graziani2020} Graziani, L., Schneider, R., Ginolfi, M., et al.\ 2020, \mnras, 494, 1071. 
\bibitem[Greif et al.(2011)]{Greif2011} Greif, T.~H., Springel, V., White, S.~D.~M., et al.\ 2011, \apj, 737, 75. 
\bibitem[Harikane et al.(2022)]{Harikane2022} Harikane, Y., Inoue, A.~K., Mawatari, K., et al.\ 2022, \apj, 929, 1, 1. 
\bibitem[Harikane et al.(2023)]{Harikane2023} Harikane, Y., Ouchi, M., Oguri, M., et al.\ 2023, \apjs, 265, 5. 
\bibitem[Harikane et al.(2024)]{Harikane2024} Harikane, Y., Nakajima, K., Ouchi, M., et al.\ 2024, \apj, 960, 1, 56. 
\bibitem[Heintz et al.(2023)]{Heintz2023} Heintz, K.~E., Gim{\'e}nez-Arteaga, C., Fujimoto, S., et al.\ 2023, \apjl, 944, 2, L30. 
\bibitem[Hirano et al.(2015)]{Hirano2015} Hirano, S., et al.\ 2015, \mnras, 448, 568
\bibitem[Hirashita et al.(2008)]{Hirashita2008} Hirashita, H., Nozawa, T., Takeuchi, T.~T., \& Kozasa, T.\ 2008, \mnras, 384, 1725. 
\bibitem[Inayoshi et al.(2022)]{Inayoshi2022} Inayoshi, K., Harikane, Y., Inoue, A.~K., et al.\ 2022, \apjl, 938, 2, L10. 
\bibitem[Inoue(2005)]{Inoue2005} Inoue, A.~K.\ 2005, \mnras, 359, 171. 
\bibitem[Inoue(2011)]{Inoue2011} Inoue, A.~K.\ 2011, Earth, Planets and Space, 63, 1027. 
\bibitem[Inoue et al.(2020)]{Inoue2020} Inoue, A.~K., Hashimoto, T., Chihara, H., et al.\ 2020, \mnras, 495, 1577. 
\bibitem[Iyer et al.(2024)]{Iyer2024} Iyer, K.~G., Speagle, J.~S., Caplar, N., et al.\ 2024, \apj, 961, 1, 53. 
\bibitem[Jaacks et al.(2019)]{Jaacks2019} Jaacks, J., Finkelstein, S.~L., \& Bromm, V.\ 2019, \mnras, 488, 2, 2202. 
\bibitem[Kataoka et al.(2014)]{Kataoka2014} Kataoka, A., Okuzumi, S., Tanaka, H., et al.\ 2014, \aap, 568, A42. 
\bibitem[Katz et al.(2025)]{Katz2025} Katz, O.~Z., Redigolo, D., \& Volansky, T.\ 2025, \jcap, 2025, 10, 047. 
\bibitem[Kawamata et al.(2018)]{Kawamata2018} Kawamata, R., Ishigaki, M., Shimasaku, K., et al.\ 2018, \apj, 855, 4
\bibitem[Kirchschlager et al.(2019)]{Kirchschlager2019} Kirchschlager, F., Schmidt, F.~D., Barlow, M.~J., et al.\ 2019, \mnras, 489, 4, 4465. 
\bibitem[Klessen \& Glover(2023)]{Klessen2023} Klessen, R.~S. \& Glover, S.~C.~O.\ 2023, \araa, 61, 65. 
\bibitem[Kocevski et al.(2023)]{Kocevski2023} Kocevski, D.~D., Papovich, C., Trump, J.~R., et al.\ 2023, \apjl, 946, L14. 
\bibitem[Langeroodi et al.(2024)]{Langeroodi2024} Langeroodi, D., Hjorth, J., Ferrara, A., et al.\ 2024, arXiv:2410.14671. 
\bibitem[Leung et al.(2023)]{Leung2023} Leung, G.~C.~K., Bagley, M.~B., Finkelstein, S.~L., et al.\ 2023, \apjl, 954, 2, L46. 
\bibitem[Lewis et al.(2023)]{Lewis2023} Lewis, J.~S.~W., Ocvirk, P., Dubois, Y., et al.\ 2023, \mnras, 519, 5987. 
\bibitem[Li et al.(2019)]{Li2019} Li, Q., Narayanan, D., \& Dav{\'e}, R.\ 2019, \mnras, 490, 1425. 
\bibitem[Li et al.(2023)]{Li2023} Li, M., Cai, Z., Bian, F., et al.\ 2023, \apjl, 955, L18
\bibitem[Maiolino et al.(2004)]{Maiolino2004} Maiolino, R., Schneider, R., Oliva, E., et al.\ 2004, \nat, 431, 533
\bibitem[Maiolino \& Mannucci(2019)]{Maiolino2019} Maiolino, R., \& Mannucci, F.\ 2019, \aapr, 27, 3. 
\bibitem[Maiolino et al.(2024)]{Maiolino2024} Maiolino, R., Scholtz, J., Witstok, J., et al.\ 2024, \nat, 627, 8002, 59. 
\bibitem[Mancini et al.(2015)]{Mancini2015} Mancini, M., Schneider, R., Graziani, L., et al.\ 2015, \mnras, 451, L70. 
\bibitem[Mason et al.(2023)]{Mason2023} Mason, C.~A., Trenti, M., \& Treu, T.\ 2023, \mnras, 521, 1, 497. 
\bibitem[Mauerhofer \& Dayal(2023)]{Mauerhofer2023} Mauerhofer, V. \& Dayal, P.\ 2023, \mnras, 526, 2196. 
\bibitem[Mauerhofer et al.(2025)]{Mauerhofer2025} Mauerhofer, V., Dayal, P., Haehnelt, M.~G., et al.\ 2025, \aap, 696, A157. 
\bibitem[McKinney et al.(2025)]{McKinney2025} McKinney, J., Cooper, O.~R., Casey, C.~M., et al.\ 2025, \apjl, 985, L21. 
\bibitem[Micelotta et al.(2016)]{Micelotta2016} Micelotta, E.~R., Dwek, E., \& Slavin, J.~D.\ 2016, \aap, 590, A65. 
\bibitem[Nakajima et al.(2023)]{Nakajima2023} Nakajima, K., Ouchi, M., Isobe, Y., et al.\ 2023, \apjs, 269, 2, 33. 
\bibitem[Naidu et al.(2022)]{Naidu2022} Naidu, R.~P., Oesch, P.~A., van Dokkum, P., et al.\ 2022, \apjl, 940, L14. 
\bibitem[Natta \& Panagia(1984)]{Natta1984} Natta, A. \& Panagia, N.\ 1984, \apj, 287, 228. 
\bibitem[Nenkova et al.(2008)]{Nenkova2008} Nenkova, M., Sirocky, M.~M., Ivezi{\'c}, {\v{Z}}., et al.\ 2008, \apj, 685, 147
\bibitem[Nikopoulos \& Dayal(2024)]{Nikopoulos2024} Nikopoulos, G.~P., \& Dayal, P.\ 2024, arXiv:2409.10613. 
\bibitem[Nozawa et al.(2003)]{Nozawa2003} Nozawa, T., Kozasa, T., Umeda, H., et al.\ 2003, \apj, 598, 785. 
\bibitem[Nozawa et al.(2007)]{Nozawa2007} Nozawa, T., Kozasa, T., \& Habe, A.\ 2007, \apj, 666, 955. 
\bibitem[Parente et al.(2022)]{Parente2022} Parente, M., Ragone-Figueroa, C., Granato, G.~L., et al.\ 2022, \mnras, 515, 2053. 
\bibitem[P{\'e}rez-Gonz{\'a}lez et al.(2023)]{Perez-Gonzalez2023} P{\'e}rez-Gonz{\'a}lez, P.~G., Costantin, L., Langeroodi, D., et al.\ 2023, \apjl, 951, 1, L1. 
\bibitem[Planck Collaboration(2018)]{Planck2018} Planck Collaboration\ 2018, \aap, 641, A6. 
\bibitem[Popping et al.(2017)]{Popping2017} Popping, G., Somerville, R.~S., \& Galametz, M.\ 2017, \mnras, 471, 3152
\bibitem[R{\'e}my-Ruyer et al.(2014)]{RemyRuyer2014} R{\'e}my-Ruyer, A., Madden, S.~C., Galliano, F., et al.\ 2014, \aap, 563, A31
\bibitem[Rydberg et al.(2013)]{Rydberg2013} Rydberg, C.-E., et al.\ 2013, \mnras, 429, 3658
\bibitem[Saldana-Lopez et al.(2022)]{SaldanaLopez2022} Saldana-Lopez, A., Schaerer, D., Chisholm, J., et al.\ 2022, \aap, 663, A59
\bibitem[Salim \& Narayanan(2020)]{Salim2020} Salim, S., \& Narayanan, D.\ 2020, \araa, 58, 529. 
\bibitem[Sanders et al.(2021)]{Sanders2021} Sanders, R.~L., Shapley, A.~E., Jones, T., et al.\ 2021, \apj, 914, 19
\bibitem[Sanders et al.(2023)]{Sanders2023} Sanders, R.~L., Shapley, A.~E., Topping, M.~W., et al.\ 2023, \apj, 955, 1, 54. 
\bibitem[Sarmento et al.(2018)]{Sarmento2018} Sarmento, R., Scannapieco, E., \& Cohen, S.\ 2018, \apj, 854, 1, 75. 
\bibitem[Sarmento \& Scannapieco(2025)]{Sarmento2025} Sarmento, R. \& Scannapieco, E.\ 2025, \apj, 988, 2, 221. 
\bibitem[Schaerer(2003)]{Schaerer2003} Schaerer, D.\ 2003, \aap, 397, 527
\bibitem[Schneider \& Maiolino(2024)]{Schneider2024} Schneider, R. \& Maiolino, R.\ 2024, \aapr, 32, 2. 
\bibitem[Schneider et al.(2004)]{Schneider2004} Schneider, R., Ferrara, A., \& Salvaterra, R.\ 2004, \mnras, 351, 4, 1379. 
\bibitem[Scoville et al.(2017)]{Scoville2017} Scoville, N., Sheth, K., Aussel, H., et al.\ 2017, \apj, 837, 150
\bibitem[Seon \& Draine(2016)]{Seon2016} Seon, K.-I., \& Draine, B.~T.\ 2016, \apj, 833, 201. 
\bibitem[Shen et al.(2023)]{Shen2023} Shen, X., Vogelsberger, M., Boylan-Kolchin, M., et al.\ 2023, \mnras, 525, 3, 3254. 
\bibitem[Shibuya et al.(2015)]{Shibuya2015} Shibuya, T., Ouchi, M., \& Harikane, Y.\ 2015, \apjs, 219, 15
\bibitem[Somerville et al.(2025)]{Somerville2025} Somerville, R.~S., Yung, L.~Y.~A., Lancaster, L., et al.\ 2025, \mnras, 544, 4, 3774. 
\bibitem[Susa et al.(2014)]{Susa2014} Susa, H., et al.\ 2014, \apj, 792, 32
\bibitem[Tacchella et al.(2020)]{Tacchella2020} Tacchella, S., Finkelstein, S.~L., Bagley, M.~B., et al.\ 2020, \mnras, 497, 698. 
\bibitem[Tacconi et al.(2020)]{Tacconi2020} Tacconi, L.~J., Genzel, R., \& Saintonge, A.\ 2020, \araa, 58, 157
\bibitem[Tanaka \& Hasegawa(2021)]{Tanaka2021} Tanaka, T. \& Hasegawa, K.\ 2021, \mnras, 502, 1, 463. 
\bibitem[Todini \& Ferrara(2001)]{Todini2001} Todini, P., \& Ferrara, A.\ 2001, \mnras, 325, 726
\bibitem[Torrey et al.(2019)]{Torrey2019} Torrey, P., Vogelsberger, M., Marinacci, F., et al.\ 2019, \mnras, 484, 5587. 
\bibitem[Triani et al.(2020)]{Triani2020} Triani, D.~P., Sinha, M., Croton, D.~J., et al.\ 2020, \mnras, 493, 2490. 
\bibitem[Trinca et al.(2024)]{Trinca2024} Trinca, A., Schneider, R., Valiante, R., et al.\ 2024, \mnras, 529, 3563
\bibitem[Trussler et al.(2023)]{Trussler2023} Trussler, J., et al.\ 2023, \mnras, 525, 5328
\bibitem[Valiante et al.(2009)]{Valiante2009} Valiante, R., Schneider, R., Bianchi, S., et al.\ 2009, \mnras, 397, 1661. 
\bibitem[Valiante et al.(2016)]{Valiante2016} Valiante, R., Schneider, R., Volonteri, M., et al.\ 2016, \mnras, 457, 3, 3356. 
\bibitem[Vanzella et al.(2023)]{Vanzella2023} Vanzella, E., et al.\ 2023, \aap, 678, A173
\bibitem[Varosi \& Dwek(1999)]{Varosi1999} V{\'a}rosi, F., \& Dwek, E.\ 1999, \apj, 523, 265
\bibitem[Weingartner \& Draine(2001)]{Weingartner2001} Weingartner, J.~C., \& Draine, B.~T.\ 2001, \apj, 548, 296
\bibitem[Wise et al.(2012)]{Wise2012} Wise, J.~H., Turk, M.~J., Norman, M.~L., \& Abel, T.\ 2012, \apj, 745, 50. 
\bibitem[Witstok et al.(2023)]{Witstok2023} Witstok, J., Jones, G.~C., Maiolino, R., et al.\ 2023, \mnras, 523, 3119
\bibitem[Witt \& Gordon(2000)]{Witt2000} Witt, A.~N., \& Gordon, K.~D.\ 2000, \apj, 528, 799. 
\bibitem[Xu et al.(2016)]{Xu2016} Xu, H., Wise, J.~H., Norman, M.~L., et al.\ 2016, \apj, 833, 1, 84. 
\bibitem[Yajima et al.(2023)]{Yajima2023} Yajima, H., Abe, M., Fukushima, H., et al.\ 2023, \mnras, 525, 4, 4832. 
\bibitem[Yung et al.(2024)]{Yung2024} Yung, L.~Y.~A., Somerville, R.~S., Finkelstein, S.~L., et al.\ 2024, \mnras, 527, 3, 5929. 
\bibitem[Yung et al.(2025)]{Yung2025} Yung, L.~Y.~A., Somerville, R.~S., \& Iyer, K.~G.\ 2025, \mnras, 543, 4, 3802. 
\bibitem[Zavala et al.(2023)]{Zavala2023} Zavala, J.~A., Casey, C.~M., Manning, S.~M., et al.\ 2023, \apjl, 943, L9. 

\end{thebibliography}

\begin{appendix}
\section{Appendices description}
\label{appendix:Description}

These appendices provide supporting material for the main analysis. 
Appendix~\ref{appendix:Eqs5a5b} derives Eqs. 5a and 5b, fron \cite{Natta1984}.
Appendix~\ref{appendix:poisson} presents the Poissonian clump formalism. 
Appendix~\ref{appendix:geometry} presents the impact of geometry, covering fraction, and dust opacity in shaping the A$_{\rm FUV}$ - M$_{\rm star}$ relation.
Appendix~\ref{appendix:ext2att} discusses the dust physics underlying the adopted extinction laws and their transformation into effective far-UV attenuation curves in porous media.
Appendix~\ref{appendix:otherkappas} explores models with far-UV dust opacities higher than the fiducial one.
Appendix~\ref{appendix:cigale_params} summarizes the CIGALE configurations and parameter grids used to model the ultra–high-redshift galaxy sample. Finally, appendix~\ref{appendix:scalings} summarizes the scaling relations used in this paper. 
\section{Derivation of leaky-geometry eqs. 5a, 5b from \cite{Natta1984}}
\label{appendix:Eqs5a5b}

Natta \& Panagia (1984) define the transmission of an extended source through an inhomogeneous foreground dust layer as the surface-brightness-weighted mean of the local screen transmission, which, in our notation, can be written as
\[
T_\lambda =
\frac{
\int_{\rm source} e^{-\tau_\lambda(\Omega)} S(\Omega)\,d\Omega
}{
\int_{\rm source} S(\Omega)\,d\Omega
}.
\]
For a source of uniform surface brightness this reduces to an
area average of \(e^{-\tau_\lambda(\Omega)}\). Their one-clump model,
in which a fraction \(f\) of the source is covered by dust of optical
depth \(\tau_c\) and the remaining fraction \(1-f\) is unobscured,
gives
\[
T_\lambda = (1-f) + f e^{-\tau_c}.
\]
Identifying \(f=f_{\rm cov}\), and imposing conservation of the total dust mass so that the optical depth in the covered region is
\(\tau_c=\tau_\lambda/f_{\rm cov}\), gives
\[
T_{\rm L,screen}(\tau_\lambda,f_{\rm cov})
=
(1-f_{\rm cov})
+
f_{\rm cov}
\exp\left(-\frac{\tau_\lambda}{f_{\rm cov}}\right).
\]
In the present work we replace the extinction optical depth by the effective optical depth appropriate for the directed UV radiation field,
\(\tau_{\lambda,\rm eff}=(1-\omega_\lambda g_\lambda)\tau_\lambda\),
which yields
\[
T_{\rm L,screen}(\tau_\lambda,f_{\rm cov})
=
(1-f_{\rm cov})
+
f_{\rm cov}
\exp\left[
-(1-\omega_\lambda g_\lambda)
\frac{\tau_\lambda}{f_{\rm cov}}
\right].
\]

For the mixed case, Natta \& Panagia (1984) also recall the standard solution for internal extinction in a plane-parallel slab with uniformly mixed emitters and absorbers,
\[
T_\lambda = \frac{1-e^{-\tau_\lambda}}{\tau_\lambda}.
\]
We apply this mixed-slab transmission to the covered fraction of the source, again with mass conservation
\(\tau_c=\tau_\lambda/f_{\rm cov}\), and add the unobscured
fraction. This gives
\[
T_{\rm L,mixed}(\tau_\lambda,f_{\rm cov})
=
(1-f_{\rm cov})
+
f_{\rm cov}
\frac{
1-\exp(-\tau_\lambda/f_{\rm cov})
}{
\tau_\lambda/f_{\rm cov}
}.
\]

Thus Eqs.~(5a) and (5b) are obtained by combining the area-averaged attenuation formalism of Natta \& Panagia (1984) with partial covering and conservation of the total dust column.

\section{Poissonian clumps and covering fraction}
\label{appendix:poisson}

We summarize the Poisson clump formalism following \citet{Varosi1999}, used in Sect.~\ref{sec:geometry} to model stochastic leakage of UV photons.

Consider $N_c$ identical dusty clumps randomly distributed over an area $A$, each with projected area $a_c$. The probability that a given sight line intersects a specific clump is $q=a_c/A$. The probability that it intersects none of the $N_c$ clumps is $(1-q)^{N_c}$, so that the covering fraction is 
\begin{equation}
f_{\rm cov} = 1-(1-q)^{N_c}.
\end{equation}
In the rare-clump limit ($q\ll1$ with $N_c q$ finite), this becomes 
\begin{equation}
f_{\rm cov} \simeq 1-e^{-\bar N},
\qquad
\bar N \equiv N_c q ,
\label{eq:fcov_poisson}
\end{equation}
where $\bar N$ is the mean number of clumps intercepted per sight line.  

\paragraph{Transmission through Poisson clumps.}
If the total extinction optical depth at wavelength $\lambda$ is $\tau_\lambda$ and each clump contributes $\tau_\lambda/\bar N$, the ensemble-averaged transmission is
\begin{equation}
T_{\rm P}(\tau_\lambda,f_{\rm cov})
=
\exp\!\left[-\bar N\left(1-e^{-\tau_\lambda/\bar N}\right)\right],
\qquad
\bar N=-\ln(1-f_{\rm cov}),
\label{eq:TP}
\end{equation}
which conserves the total dust mass and accounts for stochastic shadowing \citep{Varosi1999,Nenkova2008}.  

For $\bar N\!\to\!1$, the Poisson model approaches a leaky-screen geometry, while for $\bar N\!\gg\!1$ (many optically thin clumps per sight line) it converges toward the
leaky-mixed limit. 

The fraction of completely unobscured sight lines corresponds to the Poisson zero--intersection probability,
\begin{equation}
P(0) = e^{-\bar N} = 1 - f_{\rm cov},
\end{equation}

\section{Impact of geometry, covering fraction, and dust opacity}
\label{appendix:geometry}

In this appendix, we explore the respective roles and degeneracies of ISM geometry, covering fraction, and intrinsic dust opacity in shaping the $A_{\rm FUV}$-$M_{\rm star}$ relation. Using a series of model grids, we systematically vary these parameters and compare the resulting predictions to the observed galaxy sample. This approach allows us to quantify their relative impact and to assess whether geometry and covering fraction can compensate for variations in intrinsic dust opacity.

\begin{figure}
\centering
\includegraphics[width=\linewidth]{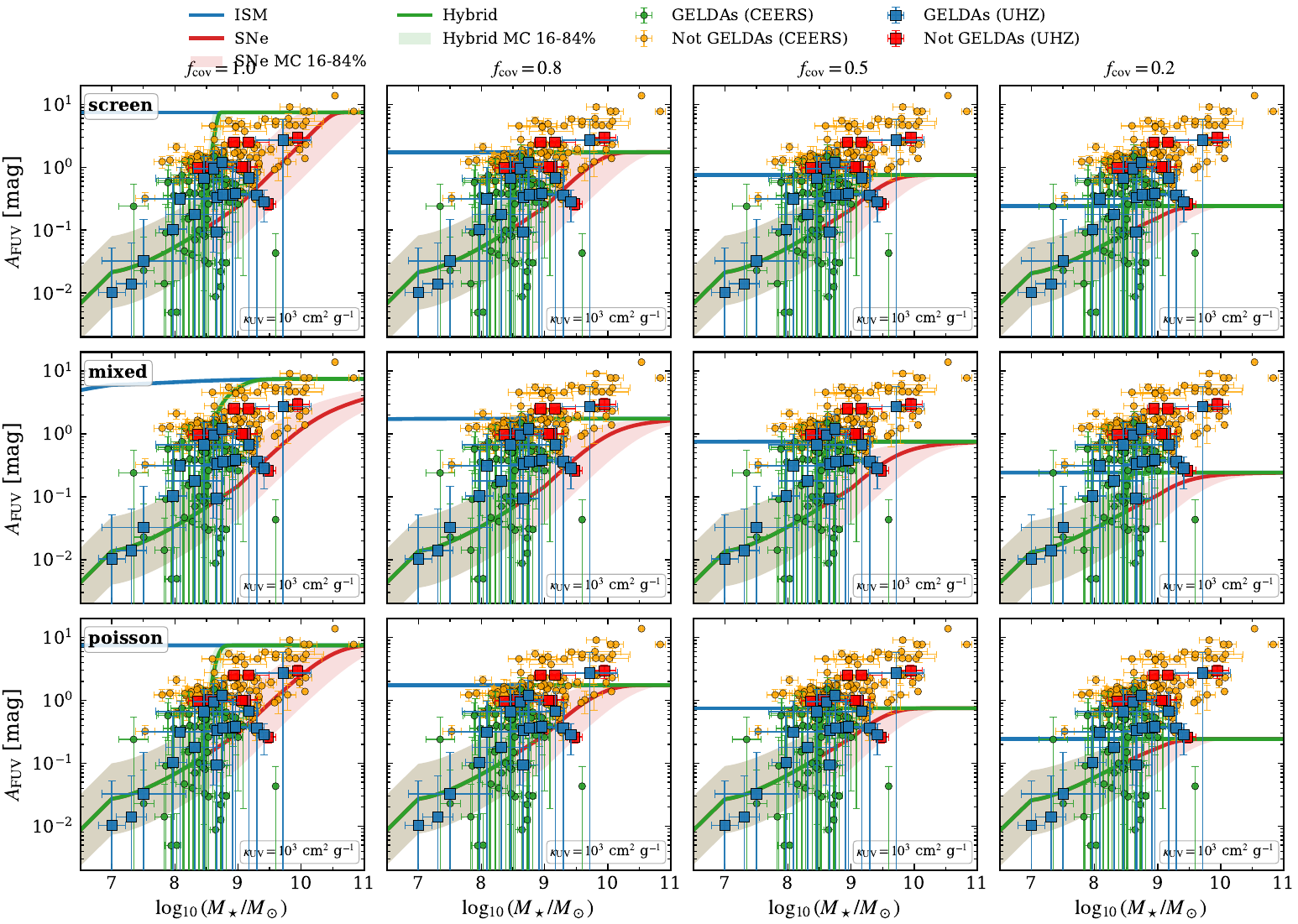}
\caption{
Impact of ISM geometry and covering fraction on the predicted far-UV attenuation as a function of stellar mass, for a fixed intrinsic dust opacity $\kappa_{\rm UV}=10^{3}\ \mathrm{cm^2\,g^{-1}}$. From top to bottom, rows correspond to leaky screen, leaky mixed, and Poisson clump geometries. Columns show decreasing covering fractions ($f_{\rm cov}=1.0$, 0.8, 0.5, and 0.2). In all panels, model predictions for ISM-type, SN-type, and hybrid dust are shown together with their associated Monte Carlo scatter (shaded regions), and compared to the observed galaxy sample. This figure demonstrates that the covering fraction primarily controls the normalization of the $A_{\rm FUV}$--$M_\star$ relation by opening low-opacity sightlines, while the choice of geometry introduces secondary differences in curvature and dispersion. Despite these variations, all geometries converge toward low attenuation at low stellar masses when $\kappa_{\rm UV}$ is low.
}
\label{fig:AppendixC1}
\end{figure}

\begin{figure}
\centering
\includegraphics[width=\linewidth]{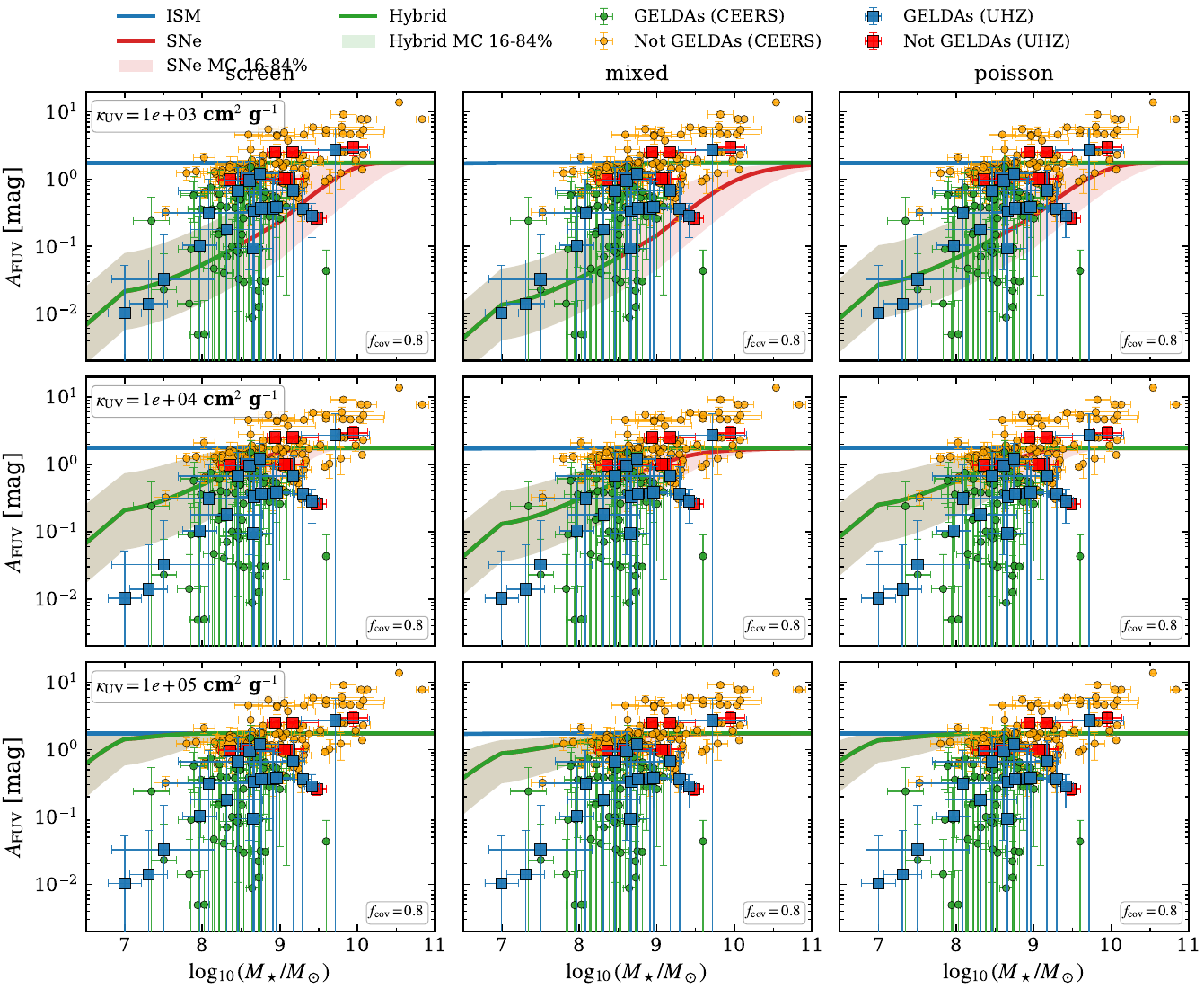}
\caption{
Impact of intrinsic far-UV dust opacity on the predicted attenuation--mass relation, for a fixed covering fraction $f_{\rm cov}=0.8$. 
Rows correspond to increasing opacity ($\kappa_{\rm UV}=10^{3},\,10^{4},\,10^{5}\ \mathrm{cm^2\,g^{-1}}$), while columns show different ISM geometries (leaky screen, mixed, and Poisson clumps). Model predictions are compared to the observed galaxy sample.
This figure illustrates that variations in geometry alone cannot compensate for high intrinsic opacity. Models with $\kappa_{\rm UV}\gtrsim10^{4}\ \mathrm{cm^2\,g^{-1}}$ systematically overpredict the attenuation, even in porous configurations. In contrast, low-opacity models naturally reproduce both the normalization and dispersion of the observed relation across all geometries, demonstrating that $\kappa_{\rm UV}$ is the dominant parameter.
}
\label{fig:AppendixC2}
\end{figure}

\begin{figure}
\centering
\includegraphics[width=\linewidth]{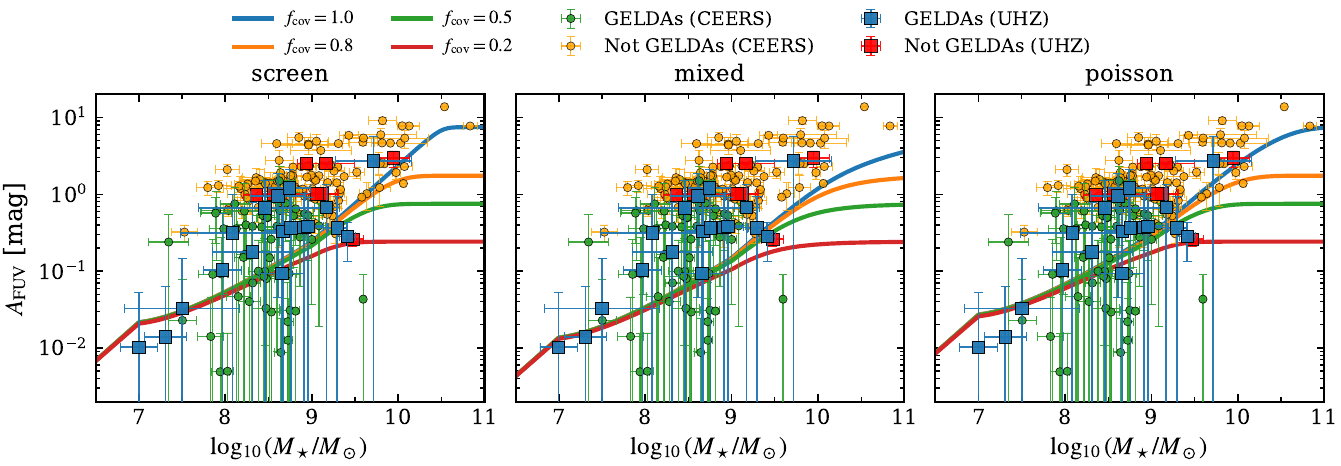}
\caption{
Effect of the covering fraction on the predicted attenuation--mass relation for different ISM geometries. 
Each panel shows a fixed geometry (leaky screen, mixed, and Poisson clumps from left to right), while colored curves correspond to different covering fractions ($f_{\rm cov}=1.0$, 0.8, 0.5, and 0.2). Observed galaxies are overplotted for reference. The covering fraction acts primarily as a normalization parameter, with decreasing $f_{\rm cov}$ lowering the overall attenuation by increasing the fraction of unobscured sightlines. The overall shape of the relation is only weakly affected by geometry, confirming that geometry and $f_{\rm cov}$ introduce secondary variations compared to intrinsic dust opacity.
}
\label{fig:AppendixC3}
\end{figure}

\section{From extinction to attenuation curves}
\label{appendix:ext2att}

Appendix~\ref{appendix:poisson} introduces the geometrical formalism used to model porous dust distributions. In this appendix, we focus on the dust physics that sets the intrinsic extinction curves entering those geometries, and on how these extinction laws are transformed into the effective far-UV attenuation curves adopted in this work.

Dust formed in SNe ejecta is efficiently processed by reverse shocks, which destroy a large fraction of small grains and leave a surviving population dominated by larger sizes \citep[e.g.][]{Nozawa2007,Hirashita2008,Micelotta2016,Kirchschlager2019}. As a result, the post-shock grain-size distribution is significantly flatter than that of the diffuse interstellar medium. 

To provide a compact characterization of the UV dependence of the adopted extinction curves, we define an effective power-law index between $\lambda_1=0.15\,\mu{\rm m}$ and $\lambda_2=0.30\,\mu{\rm m}$,
\begin{equation}
\frac{A_\lambda}{A_{0.30\,\mu{\rm m}}}
 =
\left(
\frac{\lambda}{0.30\,\mu{\rm m}}
\right)^{-b_{\rm UV}} .
\label{eq:buv_definition}
\end{equation}
The corresponding two-point slope is
\begin{equation}
b_{\rm UV}
 =
-\frac{
\ln\!\left[A_{0.15\,\mu{\rm m}}/
           A_{0.30\,\mu{\rm m}}\right]
}{
\ln(0.15/0.30)
}
 =
\frac{
\ln\!\left[A_{0.15\,\mu{\rm m}}/
           A_{0.30\,\mu{\rm m}}\right]
}{
\ln 2
}.
\label{eq:buv_twopoint}
\end{equation}
This index describes the mean change between the FUV reference wavelength and $0.30\,\mu{\rm m}$; it is not intended to reproduce the detailed structure of each extinction curve at intermediate wavelengths.

The extinction curves shown in Fig.~\ref{fig:ExtLaws} have different physical origins. The curves labelled ``mixed $20\,M_\odot$'' and ``mixed $170\,M_\odot$'' are taken from the theoretical calculations of \citet{Hirashita2008}. Those models follow dust formed in supernova ejecta and its subsequent processing by the reverse shock, including the resulting modification of the grain-size distribution. By
contrast, the curve of \citet{Maiolino2004} was inferred empirically from the rest-frame extinction of a high-redshift quasar and is used here as an observationally motivated example of non-standard, SN-like dust.

Applying Eq.~(\ref{eq:buv_twopoint}) gives
$b_{\rm UV}\simeq-0.32$ for the
\citet{Hirashita2008} mixed-$170\,M_\odot$ model,
$b_{\rm UV}\simeq0.77$ for their mixed-$20\,M_\odot$ model,
and $b_{\rm UV}\simeq0.54$ for the
\citet{Maiolino2004} curve. The dashed lines in
Fig.~\ref{fig:ExtLaws} show these two-point power-law representations. The broad range of $b_{\rm UV}$ values reflects the diversity of the adopted curves and, in particular, shows that reverse-shock-processed dust does not imply a unique UV extinction slope. We therefore do not adopt a single representative interval such as $b_{\rm UV}\simeq0.4$ - $0.6$.

The corresponding two-point UV slopes are listed in Table~\ref{tab:ExtLaws} and illustrated in Fig.~\ref{fig:ExtLaws}.

Single-size grain calculations by \citet{Kataoka2014} and \citet{Inoue2020} further show that grains with radii of order $a\sim1\,\mu{\rm m}$ can have FUV mass opacities as low as $\kappa_{\rm FUV}^{(\rm dust)}\sim10^3$ - $10^4\, {\rm cm^2\,g^{-1}}$, whereas substantially smaller grains generally reach values of order $10^5\,{\rm cm^2\,g^{-1}}$. These calculations therefore illustrate that grain populations dominated by sufficiently large grains can produce the low intrinsic FUV opacities explored in this work (Fig.~\ref{fig:kappa_grains}). They do not, however, imply that the SNe-dominated prescription adopted here consists uniquely of grains with $a=1\,\mu{\rm m}$.

For the leaky-screen, leaky-mixed, and Poisson geometries introduced in Sect.~\ref{sec:geometry}, the emergent transmission depends on both the effective optical depth and the covering fraction, such that
\begin{equation}
\tau_\lambda
=
\kappa_\lambda^{(\mathrm{dust})}\,\Sigma_{\rm dust}
=
\kappa_\lambda^{(\mathrm{gas})}\,\Sigma_{\rm gas},
\qquad
\kappa_\lambda^{(\mathrm{gas})}
=
\kappa_\lambda^{(\mathrm{dust})}\,(\mathrm{DTM}\times Z),
\end{equation}
and the attenuation follows from
\[
A_\lambda=-2.5\log_{10}T(\tau_\lambda,f_{\rm cov}).
\]

For moderate effective optical depths, $\tau_{\rm FUV}\sim1$-3, and high covering fractions, $f_{\rm cov}\gtrsim0.9$, the emergent UV flux is dominated by a minority of low-$\tau$ sight lines. This produces Calzetti-like, grey attenuation curves that are largely insensitive to the detailed slope of the intrinsic extinction law \citep{Seon2016}.

\paragraph{Implication for GELDAs.}
When combined with the intrinsically low
$\kappa_{\rm FUV}^{(\mathrm{dust})}\sim10^{3}$ - $10^{4}~\mathrm{cm^2\,g^{-1}}$ adopted for the reverse-shock-processed SN dust baseline, and with the reduced gas opacity
$\kappa_{\rm FUV}^{(\mathrm{gas})}=\kappa_{\rm FUV}^{(\mathrm{dust})}\,(\mathrm{DTM}\,Z)$ at metallicities $Z\lesssim0.1\,Z_\odot$, porous geometries naturally yield $A_{\rm FUV}\lesssim0.3$ - 0.7 even in gas-rich systems. This provides a physically motivated explanation for the extremely low attenuations observed in GELDAs, without invoking large-scale dust removal or powerful outflows. This represents the physical scenario investigated here; the attenuation constraint itself primarily establishes the need for a low effective FUV opacity.

\begin{figure*}
  \centering
  \includegraphics[width=\linewidth]{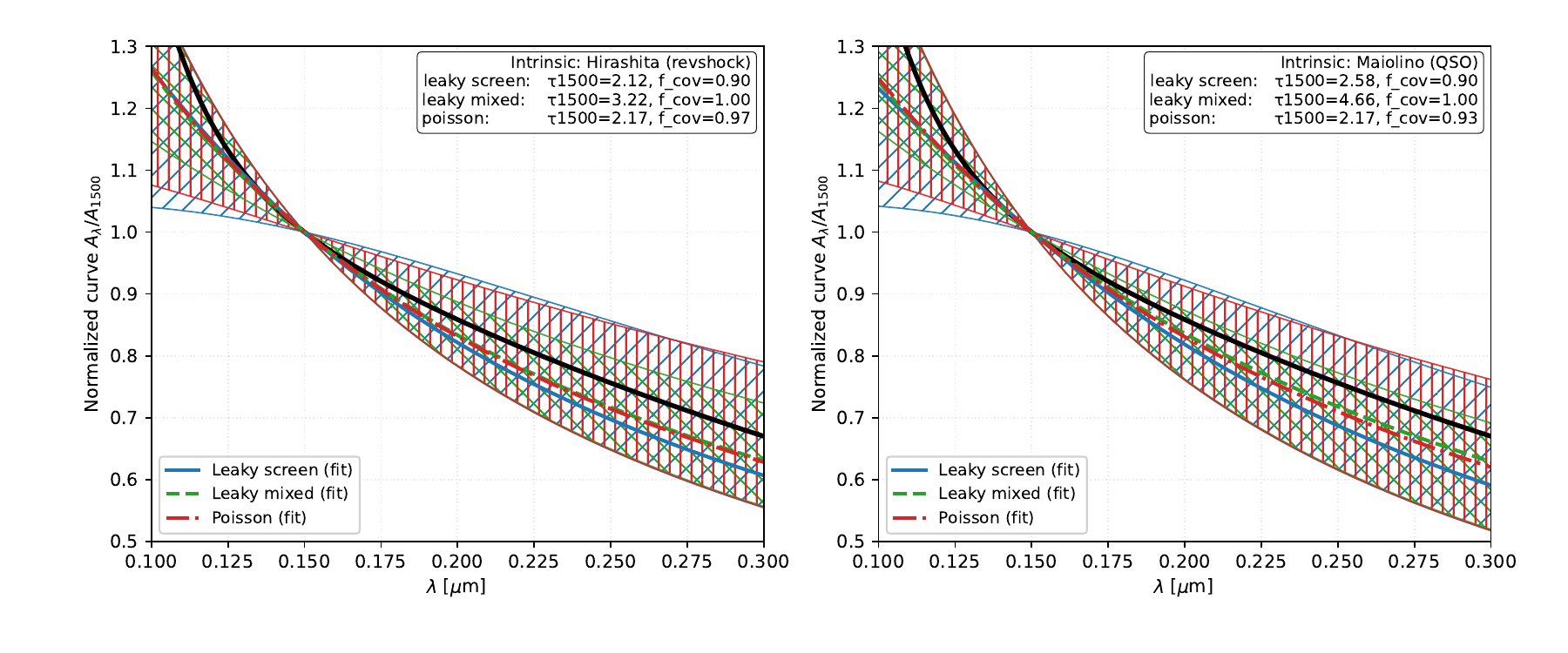}
  \caption{Rest-frame far-UV ($0.10\leq\lambda\leq0.30\,\mu{\rm m}$) attenuation curves produced by porous dust geometries and compared with the Calzetti attenuation law, with all curves normalized at $1500\,\text{\AA}$. The left and right panels adopt different intrinsic dust extinction laws within the porosity kernel: the reverse-shock-processed SNe dust model of \citet{Hirashita2008} (left) and the QSO-type extinction curve of \citet{Maiolino2004} (right). Colored solid, dashed, and dash-dotted lines show the best-fitting leaky-screen, leaky-mixed, and Poisson-clump geometries, respectively. Hatched regions indicate the parameter domains explored for each model: $0.3\leq\tau_{1500}\leq3$ and $0.2\leq f_{\rm cov}\leq1$ for leaky screens; $0.1\leq\tau_{1500}\leq2$ and $0.3\leq f_{\rm cov}\leq1$ for leaky-mixed geometries; and $0.1\leq\tau_{1500}\leq2$ and $0.5\leq f_{\rm cov}\leq1$ for Poisson-clump models. When the Calzetti attenuation law is used as the target curve, the best-fitting solutions tend toward low $\tau_{1500}$ and moderate $f_{\rm cov}$, reflecting the parameter degeneracy in the weak-attenuation regime.} 
  \label{fig:Ext2Att}
\end{figure*}

\begin{table}
\centering
\caption{Effective UV slopes of the adopted extinction curves.}
\begin{tabular}{lcccc}
\hline
Curve &
$b_{\rm UV}^{\rm 2pt}$ &
$\Delta b_{\rm UV}$ &
$A_{\rm FUV}^{\rm 2pt}$ &
$\Delta A_{\rm FUV}$ \\
\hline
Maiolino $-15\%$         & 0.309  & 0.339 & 1.129 & 0.227 \\
Maiolino fiducial        & 0.543  & 0.324 & 1.459 & 0.264 \\
Maiolino $+15\%$         & 0.691  & 0.265 & 1.778 & 0.285 \\
Hirashita $20\,M_\odot$  & 0.747  & 0.134 & 1.696 & 0.010 \\
Hirashita $170\,M_\odot$ & $-0.317$ & 0.155 & 0.804 & 0.044 \\
\hline
\end{tabular}
\tablefoot{Effective UV slopes of the adopted extinction curves. Each curve is normalized to $A(0.30\,\mu{\rm m})=1$. The central two-point slope, $b_{\rm UV}^{\rm 2pt}$, is defined between $\lambda_1=0.15\,\mu{\rm m}$ and $\lambda_2=0.30\,\mu{\rm m}$ through $A(\lambda)/A(0.30\,\mu{\rm m})
=[\lambda/(0.30\,\mu{\rm m})]^{-b_{\rm UV}}$.
Equivalently,
$b_{\rm UV}^{\rm 2pt}
=\ln[A(0.15\,\mu{\rm m})/A(0.30\,\mu{\rm m})]/\ln 2$.
The quantity $A_{\rm FUV}^{\rm 2pt}$ denotes the normalized
extinction at $0.15\,\mu{\rm m}$,
$A_{\rm FUV}^{\rm 2pt}
=A(0.15\,\mu{\rm m})/A(0.30\,\mu{\rm m})$.
The columns $\Delta b_{\rm UV}$ and $\Delta A_{\rm FUV}$ give the absolute differences between the two-point estimates and the values obtained from an unconstrained least-squares power-law fit to all tabulated points over $0.15\leq\lambda\leq0.30\,\mu{\rm m}$. These differences indicate the sensitivity to the fitting procedure and should not be interpreted as formal statistical $1\sigma$ uncertainties. The Maiolino $-15\%$ and $+15\%$ curves represent the lower and upper variants of the empirically inferred extinction curve, whereas only the fiducial Maiolino curve is shown in the corresponding figure for clarity.}
\label{tab:ExtLaws}
\end{table}

\begin{figure*}
\centering
\includegraphics[width=\linewidth]{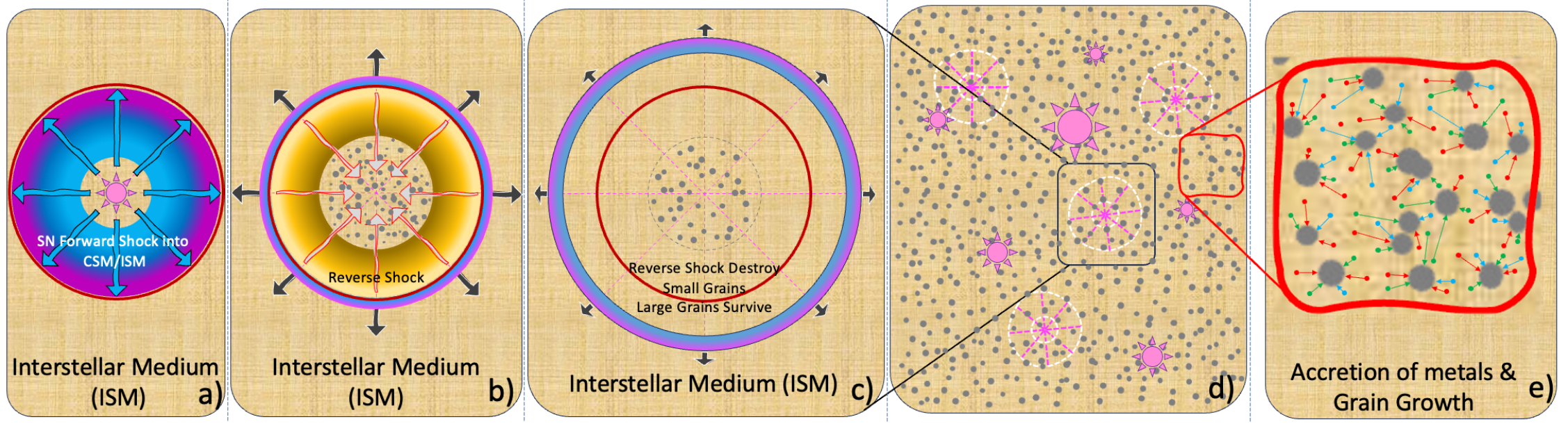}
\caption{
Schematic illustration of the dust cycle: 
(a) following a supernova explosion, a forward shock propagates into the circumstellar and interstellar medium, 
(b) a reverse shock travels back into the ejecta, destroying a substantial fraction of the newly formed dust and potentially modifying the grain-size distribution, 
(c) in the baseline scenario illustrated here, the surviving dust population is biased toward larger grains, although the detailed distribution depends on the ejecta and shock conditions, 
(d) the ISM becomes enriched with the surviving, post–reverse-shock SN-produced grains, 
(e) once the metallicity reaches a critical value, $Z/Z_\odot\sim0.1$, metal accretion onto grain surfaces becomes efficient and the dust mass grows rapidly.
}
\label{fig:flowchart}
\end{figure*}

\begin{figure}
  \centering
    \includegraphics[width=\linewidth]{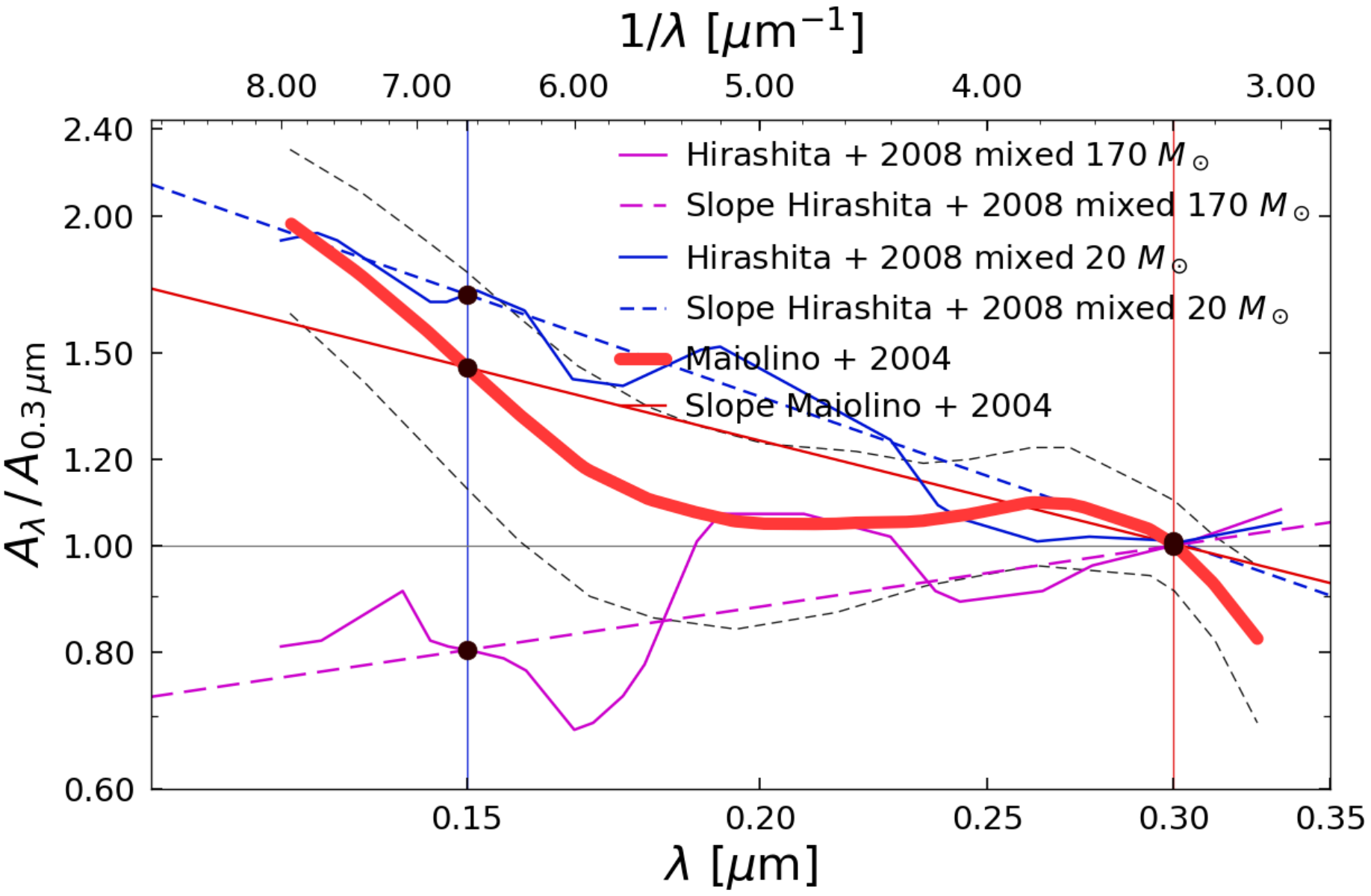}
  \caption{UV extinction curves adopted as representative SNe-like dust prescriptions, normalized to $A_{0.30\,\mu{\rm m}}=1$. The magenta and blue solid curves show the mixed-$170\,M_\odot$ and mixed-$20\,M_\odot$ reverse-shock models of \citet{Hirashita2008}, respectively, while the thick red curve shows the empirically inferred high-redshift quasar extinction curve of \citet{Maiolino2004}. The black points mark the values at $0.15$ and $0.30\,\mu{\rm m}$. Dashed lines show the effective two-point power laws $A_\lambda/A_{0.30\,\mu{\rm m}}\propto \lambda^{-b_{\rm UV}}$, giving $b_{\rm UV}\simeq-0.32$, $0.77$, and $0.54$ for the mixed-$170\,M_\odot$, mixed-$20\,M_\odot$, and Maiolino curves, respectively. These slopes characterize only the mean variation between the two reference wavelengths and do not reproduce the detailed structure of the original curves.}
  \label{fig:ExtLaws}
\end{figure}

The main stages of dust formation, reverse-shock processing, and subsequent ISM evolution considered in this scenario are summarized schematically in Fig.~\ref{fig:flowchart}.

\section{Models with higher far-UV dust opacities}
\label{appendix:otherkappas}

In addition to the fiducial models, which adopt a far-UV dust-opacity normalization of $\kappa_{\rm FUV}^{\rm dust}=10^{3}\ {\rm cm^{2}g^{-1}}$, we explore values of $10^{4}$ and $10^{5}\ {\rm cm^{2}g^{-1}}$. In the dust-mass-based route, increasing the opacity generally increases the predicted attenuation and worsens the agreement with the lowest-attenuation galaxies. The gas-based route is less restrictive because its optical depth also depends on metallicity and the dust-to-metal ratio; values near $10^{4}\ {\rm cm^{2}g^{-1}}$ can therefore remain compatible with part of the observed distribution. Values near $10^{5}\ {\rm cm^{2}g^{-1}}$ generally predict excessive attenuation. The resulting model predictions are shown in Figure~\ref{fig:afuv_mstar}.

For the dust-mass-based prescriptions, increasing the intrinsic opacity from $10^{3}$ to $10^{4}\ {\rm cm^{2}g^{-1}}$ shifts the predicted relations toward higher attenuation and generally worsens the agreement with the lowest-$A_{\rm FUV}$ galaxies. Models with standard ISM-like opacities of order $10^{5}\ {\rm cm^{2}g^{-1}}$ predict excessive attenuation for most of the explored geometries and covering fractions.

The gas-based route behaves differently because the effective opacity per unit gas mass is $\kappa_{\rm FUV}^{\rm gas}=\kappa_{\rm FUV}^{\rm dust}({\rm DTM}\times Z)$. At sufficiently low metallicity and dust-to-metal ratio, the gas-route model with $\kappa_{\rm FUV}^{\rm dust}=10^{4}\ {\rm cm^{2}g^{-1}}$ can therefore reproduce a substantial part of the observed $A_{\rm FUV}$ - $M_{\star}$ distribution. The comparison consequently favours an intrinsic FUV dust-opacity range of approximately $10^{3}$ - $10^{4}\ {\rm cm^{2}g^{-1}}$, rather than uniquely selecting $10^{3}\ {\rm cm^{2}g^{-1}}$. Values near $10^{5}\ {\rm cm^{2}g^{-1}}$ are generally disfavoured within the parameter space explored here.

These tests therefore disfavour standard ISM-like FUV mass absorption coefficients of order $10^5 {\rm cm^2\,g^{-1}}$, while favouring an intrinsic FUV dust-opacity range of approximately $10^3 - 10^4 {\rm cm^2\,g^{-1}}$ within the parameter space explored here.

Figure~\ref{fig:afuv_mstar} is intended as a controlled illustration of how the predicted attenuation changes when the intrinsic FUV opacity and attenuation geometry are varied. Unlike the joint model shown in Figure~\ref{fig:scaling_relations}, where the mass-dependent covering fraction is constrained through the fitting procedure, the covering fractions in Figure~\ref{fig:afuv_mstar} are held fixed at $f_{\rm cov}=0.99$ for the ISM-dominated prescription and $f_{\rm cov}=0.50$ for the SNe-dominated stardust prescription, with the hybrid model interpolating between these values. These fixed endpoint values are adopted only to isolate the effect of changing the opacity and geometry and should not be interpreted as independently fitted or uniquely preferred covering fractions.

To interpret the physical meaning of the FUV dust-opacity normalization explored above, we compare them to theoretical expectations for dust opacity as a function of grain size and composition. 

At FUV wavelengths, small grains ($\lesssim 0.1\,\mu$m) produce high opacities ($\kappa_{\rm UV} \sim 10^{4}$ - $10^{5}\ {\rm cm^{2}\,g^{-1}}$), while larger grains ($\gtrsim 1\,\mu$m) naturally yield lower values ($\kappa_{\rm UV} \sim 10^{2}$ - $10^{3}\ {\rm cm^{2}\,g^{-1}}$). 

\begin{figure}[t]
    \includegraphics[width=\linewidth]{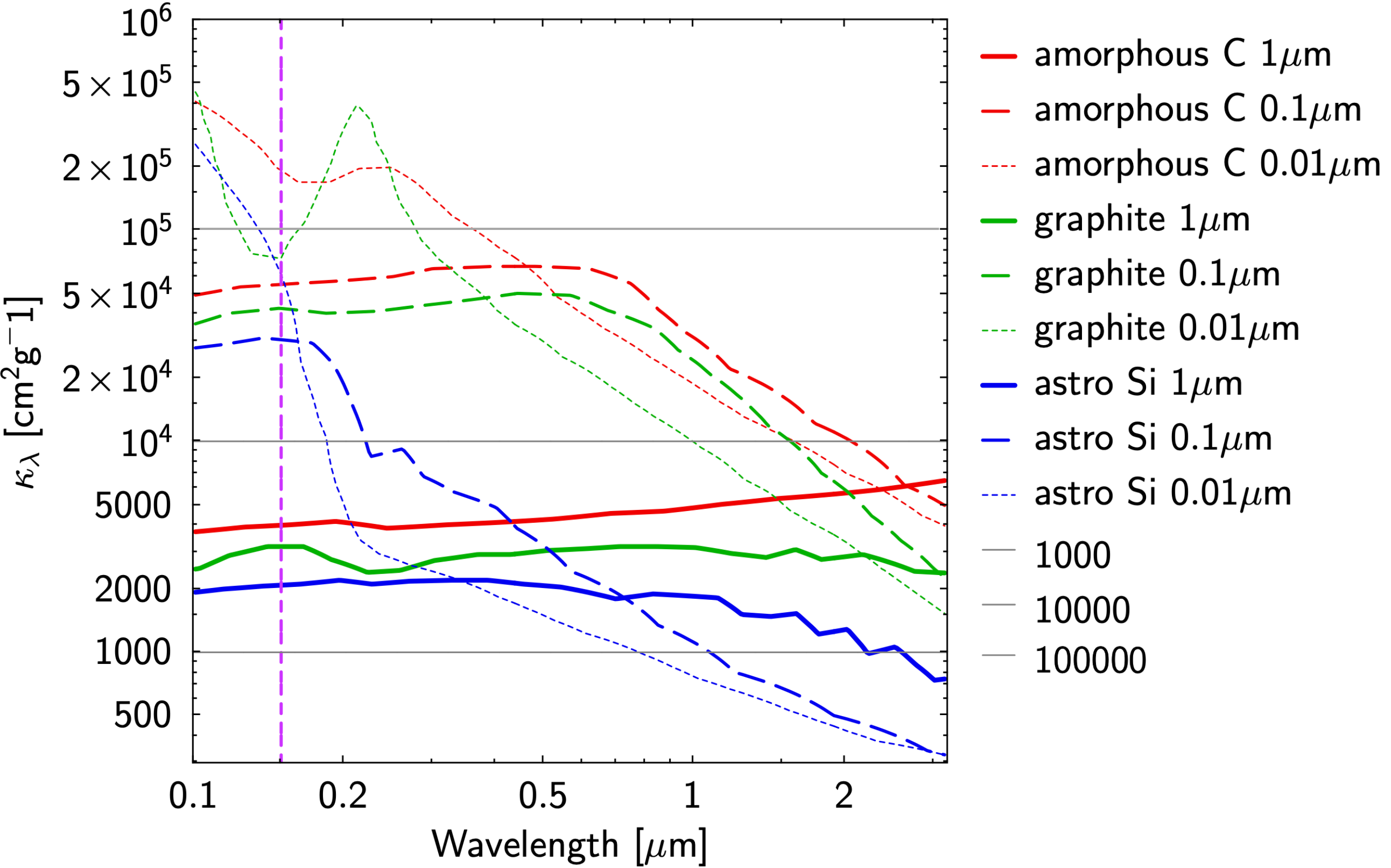}
\caption{
UV-optical dust opacity $\kappa_\lambda$ as a function of wavelength for different grain sizes and compositions, adapted from \citet{Inoue2020}. The vertical violet dashed line marks the FUV wavelength $\lambda=0.15\,\mu$m relevant for this work. Horizontal continuous gray lines indicate reference values of $\kappa_{\rm UV}=10^3$, $10^4$, and $10^5\,{\rm cm^2\,g^{-1}}$.
}
\label{fig:kappa_grains}
\end{figure}

The value required in this work, $\kappa_{\rm UV} \sim 10^{3} - 10^{4}\ {\rm cm^{2}\,g^{-1}}$, therefore points to dust populations dominated by large grains or SN-processed dust, as illustrated in Fig.~\ref{fig:kappa_grains}.

\noindent

This comparison illustrates the physical interpretation adopted for the results presented above. The models disfavour standard ISM-like FUV opacities and require an intrinsically low-opacity dust population within the parameter space explored here. Grain populations biased toward sufficiently large grains can produce opacities in the preferred range. Within the baseline scenario of this work, such a population is associated with SN-produced dust processed by reverse shocks. The grain calculations are illustrative, however, and the opacity constraint alone does not uniquely determine the complete grain-size distribution or dust-production channel.

\section{CIGALE parameters}
\label{appendix:cigale_params}

To model the ultra–high-redshift galaxy sample listed in Tab.~\ref{tab:uhz}, which complements the 173 galaxies analysed by \citet{Burgarella2025}, we performed four CIGALE fits adopting different star formation histories while keeping all other model ingredients fixed. The four SFHs used in this paper are the delayed and periodic ones defined in \citet{Boquien2019} and two new stochastic SFHs, now available in CIGALE (\citealt{Burgarella2025}) and based on \citet{Iyer2024}: a stochastic with a delayed baseline, and a stochastic SFH with an exponential baseline. The parameter grids explored in each configuration are summarised in Tab.~\ref{tab:pcigale_configs}.

\newcommand{\code}[1]{\texttt{\detokenize{#1}}}
\begin{table*}[htp]
\centering
\caption{Summary of the four CIGALE configurations explored.}
\label{tab:pcigale_configs}
\resizebox{\linewidth}{!}{%
\begin{tabular}{l c c c c c}
\hline\hline

\textbf{Parameter} &
\textbf{Symbol} &
\textbf{Run C1} &
\textbf{Run C2} &
\textbf{Run C3} &
\textbf{Run C4}
\tabularnewline
\hline

\multicolumn{6}{c}{\textbf{Star-formation history (SFH)}}
\tabularnewline
\hline

SFH module & --- &
\code{sfhdelayed} &
\code{sfhperiodic} &
\code{sfh_stochasticity_physgaussproc} &
\code{sfh_stochasticity_physgaussproc}
\tabularnewline

SFH type & --- &
Delayed SFH + exponential burst &
Periodic exponential bursts &
Stochastic exponential baseline (GP) &
Stochastic delayed baseline (GP)
\tabularnewline

Baseline code & --- &
--- &
--- &
2 &
0
\tabularnewline

E-folding time (main) & $\tau_{\rm main}$ [Myr] &
500 &
--- &
--- &
-----

\tabularnewline

Age of main population & ${\rm Age}_{\rm main}$ [Myr] &
10, 20, 50, 100, 200, 500 &
--- &
10, 20, 50, 100, 200, 500, 1000, 1500 &
10, 20, 50, 100, 200, 500, 1000, 1500
\tabularnewline

Burst separation & $\delta_{\rm burst}$ [Myr] &
--- &
10, 25, 50, 100, 500 &
--- &
-----

\tabularnewline

Burst e-folding time & $\tau_{\rm burst}$ [Myr] &
100, 500, 2000 &
10, 25, 50, 100, 500, 2000 &
--- &
-----

\tabularnewline

Age of burst / periodic SFH & ${\rm Age}$ [Myr] &
${\rm Age}_{\rm burst}=1,2,5$ &
1, 5, 10, 20, 50, 100, 200, 500, 1000, 2000 &
--- &
-----

\tabularnewline

Burst mass fraction & $f_{\rm burst}$ &
0.0, 0.05, 0.10, 0.20, 0.50 &
--- &
--- &
-----

\tabularnewline

Regular GP amplitude & $\sigma_{\rm reg}$ &
--- &
--- &
0.95 &
0.95
\tabularnewline

Dynamical GP amplitude & $\sigma_{\rm dyn}$ &
--- &
--- &
0.80 &
0.80
\tabularnewline

Exponential baseline timescale & $\tau_{\rm exp}$ [Myr] &
--- &
--- &
$-10,-25,-50,-100,-200,-500,-1000,-5000,-10000$ &
-------------------------------------------------

\tabularnewline

Delayed baseline timescale & $\tau_{\rm delayed}$ [Myr] &
--- &
--- &
500 &
20, 100, 500, 2000, 5000, 10000
\tabularnewline

Inflow timescale & $\tau_{\rm inflow}$ [Myr] &
--- &
--- &
16 &
16
\tabularnewline
\hline

\multicolumn{6}{c}{\textbf{Stellar population synthesis (SSP)}}
\tabularnewline
\hline

SSP model & --- &
\multicolumn{4}{c}{BC03}
\tabularnewline

Initial mass function & IMF &
\multicolumn{4}{c}{Chabrier (CIGALE BC03 \code{imf=1})}
\tabularnewline

Stellar metallicity & $Z_{\star}$ &
\multicolumn{4}{c}{0.0001, 0.0004, 0.004, 0.008, 0.02}
\tabularnewline

Separation age & --- [Myr] &
\multicolumn{4}{c}{10}
\tabularnewline
\hline

\multicolumn{6}{c}{\textbf{Nebular emission}}
\tabularnewline
\hline

Ionisation parameter & $\log U$ &
\multicolumn{4}{c}{$-4.0,-3.8,-3.6,-3.4,-3.2,-3.0,-2.8,-2.6,-2.4,-2.2,-2.0,-1.8,-1.6,-1.4,-1.2,-1.0$}
\tabularnewline

Gas metallicity & $Z_{\rm gas}$ &
\multicolumn{4}{c}{0.0001, 0.0004, 0.001, 0.002, 0.0025, 0.003, 0.004, 0.005}
\tabularnewline

Electron density & $n_{\rm H}$ [cm$^{-3}$] &
100, 1000 &
10, 1000 &
100, 1000 &
100, 1000
\tabularnewline

Escape fraction & $f_{\rm esc}$ &
\multicolumn{4}{c}{0.0}
\tabularnewline

Dust absorption in H,\textsc{ii} regions & $f_{\rm dust}$ &
\multicolumn{4}{c}{0.0}
\tabularnewline

Line width & --- [km s$^{-1}$] &
300 &
100 &
100 &
100
\tabularnewline

Nebular emission enabled & --- &
\multicolumn{4}{c}{\code{True}}
\tabularnewline
\hline

\multicolumn{6}{c}{\textbf{Dust attenuation}}
\tabularnewline
\hline

Attenuation module & --- &
\multicolumn{4}{c}{\code{dustatt_modified_starburst}}
\tabularnewline

Color excess (lines) & $E(B{-}V)_{\rm lines}$ &
\multicolumn{4}{c}{$10^{-6},10^{-5},10^{-4},0.001,0.005,0.010,0.10,0.25,0.50,1.0$}
\tabularnewline

Stellar-to-nebular color-excess ratio & $f_{\rm EBV}$ &
\multicolumn{4}{c}{1.0}
\tabularnewline

UV-bump central wavelength & $\lambda_{\rm bump}$ [nm] &
\multicolumn{4}{c}{217.5}
\tabularnewline

UV-bump width & $\Delta\lambda_{\rm bump}$ [nm] &
\multicolumn{4}{c}{35.0}
\tabularnewline

UV-bump amplitude & --- &
\multicolumn{4}{c}{0.0}
\tabularnewline

Power-law slope & $\delta$ &
\multicolumn{4}{c}{0.0}
\tabularnewline

Emission-line extinction law & --- &
\multicolumn{4}{c}{\code{Ext_law_emission_lines=1}}
\tabularnewline

$R_V$ & $R_V$ &
\multicolumn{4}{c}{3.1}
\tabularnewline
\hline

\multicolumn{6}{c}{\textbf{Dust emission (DL2014)}}
\tabularnewline
\hline

PAH mass fraction & $q_{\rm PAH}$ &
\multicolumn{4}{c}{0.47}
\tabularnewline

Minimum radiation field & $U_{\rm min}$ &
\multicolumn{4}{c}{17}
\tabularnewline

Power-law slope & $\alpha$ &
\multicolumn{4}{c}{2.4}
\tabularnewline

PDR fraction & $\gamma$ &
\multicolumn{4}{c}{0.54}
\tabularnewline
\hline

\multicolumn{6}{c}{\textbf{Other modules and analysis settings}}
\tabularnewline
\hline

Rest-frame parameters & --- &
\multicolumn{4}{c}{\code{restframe_parameters}}
\tabularnewline

AGN emission & --- &
\multicolumn{4}{c}{None}
\tabularnewline

IGM attenuation and redshifting & --- &
\multicolumn{4}{c}{\code{redshifting} (Meiksin 2006 IGM; redshift read from the input catalogue)}
\tabularnewline

Spectroscopic fitting & --- &
\multicolumn{4}{c}{Enabled (\code{use_spectro=True})}
\tabularnewline

Spectral-resolution file & --- &
\multicolumn{4}{c}{\code{jwst_nirspec_prism_disp.fits}}
\tabularnewline

Analysis method & --- &
\multicolumn{4}{c}{\code{pdf_analysis}}
\tabularnewline

\hline\hline
\end{tabular}%
}
\tablefoot{Runs C1 - C4 differ primarily in their SFH parameterisation, although the electron-density grid and adopted nebular line width also differ slightly between some runs. All values listed here are active values in the corresponding CIGALE configuration files. Values appearing after comment characters in those files are not included. The remaining stellar-population, attenuation, dust-emission, and analysis modules are otherwise common to the four configurations. Note that in addition to the fiducial $\delta =$ 0 configuration listed in this table, we performed a dedicated attenuation-curve sensitivity test with $\delta =$ -0.6,-0.3,0.0,0.3, and 0.6. The results are discussed in Sect.~\ref{subsec:AfuvMstar}.}
\end{table*}

\section{Summary of scaling relations and model parameters}
\label{appendix:scalings}

Appendix G summarizes the scaling relations and model parameters adopted in this work. Table~\ref{tab:scalings} distinguishes quantities fixed by construction from those constrained by the joint fit and provides the relevant equations and references. The corresponding observational data and preferred model relations are shown in Figure~\ref{fig:scaling_relations} in the main text.

The scatter around the hybrid dust-mass relation is estimated empirically from the observed residuals. We define

\begin{equation}
\Delta y_i
=
\log_{10}
\left(
\frac{M_{{\rm dust},i}}{M_\odot}
\right)
-
y_{\rm hyb}(x_i),
\label{eq:mdust_residual}
\end{equation}

and estimate the characteristic vertical scatter as
\begin{equation}
\sigma_{\rm dust}
=
\frac{
P_{84}(\Delta y_i)-P_{16}(\Delta y_i)
}{2}.
\label{eq:mdust_scatter}
\end{equation}

We generate 1000 Monte Carlo realisations of the hybrid dust-mass relation. In each realisation, coherent Gaussian offsets are independently applied to the SNe-only and ISM grain-growth branches,
\begin{equation}
y_{{\rm SNe},j}(x)
=
y_{\rm SNe}(x)+\delta_{{\rm SNe},j},
\qquad
y_{{\rm grow},j}(x)
=
y_{\rm grow}(x)+\delta_{{\rm grow},j},
\label{eq:mdust_mc_branches}
\end{equation}
where
\begin{equation}
\delta_{{\rm SNe},j},
\delta_{{\rm grow},j}
\sim
\mathcal{N}(0,\sigma_{\rm dust}).
\label{eq:mdust_mc_offsets}
\end{equation}
At each stellar mass, the shaded envelope denotes the 16th - 84th percentile range of these realisations. The uncertainty bands in the attenuation, gas-fraction, and metallicity panels are obtained by propagating the corresponding adopted scatter terms through their respective model relations.

We emphasize that the dust and radiative-transfer quantities explored in this framework should be interpreted as effective model descriptors rather than as a complete set of mutually independent grain properties. In a physically self-consistent dust model, the grain-size distribution and composition jointly determine the absorption and scattering cross-sections, the intrinsic extinction-curve shape, the single-scattering albedo, the scattering asymmetry parameter, and the opacity per unit dust mass. A change in grain size or composition would therefore generally modify several of these quantities simultaneously.

\begin{table*}
\centering
\scriptsize
\setlength{\tabcolsep}{3pt}
\renewcommand{\arraystretch}{1.15}
\caption{Summary of the scaling relations, fitted parameters, fixed limiting prescriptions, and uncertainty terms used in the full hybrid model.}
\label{tab:scalings}
\resizebox{\textwidth}{!}{%
\begin{tabular}{
p{0.19\textwidth}
p{0.38\textwidth}
p{0.28\textwidth}
p{0.15\textwidth}}
\hline
Quantity
& Adopted prescription
& Status and preferred value / range
& Reference \\
\hline

\multicolumn{4}{l}{\textit{Dust-mass and transition prescriptions}} \\[1mm]

Model variables
& $x\equiv\log_{10}(M_\star/M_\odot)$ and
  $y\equiv\log_{10}(M_{\rm dust}/M_\odot)$
& Definitions
& This work \\

$M_{\rm dust}$--$M_\star$ relation,
SNe-dominated branch
& $y_{\rm SNe}(x)=x-3.422$
& Fixed low-dust limiting prescription
& This work; Witstok et al. (2023) \\

$M_{\rm dust}$--$M_\star$ relation,
ISM grain-growth branch
& $y_{\rm grow}(x)=x-1.296$
& Fixed high-dust limiting prescription
& This work; Witstok et al. (2023) \\

Common transition function
& $\displaystyle
u(x)=\operatorname{clip}\!\left[
\frac{x-x_{\rm lo}}{x_{\rm hi}-x_{\rm lo}},0,1
\right]$,
$\quad
t(x)=u^3(10-15u+6u^2)$
& Quintic smootherstep used for dust mass, opacity,
covering fraction, radius, and metallicity
& This work \\

Transition boundaries
& Common transition from the SNe-dominated to the
ISM grain-growth regime
& $\displaystyle
x_{\rm lo}=7.776^{+0.105}_{-0.149}$;
$\displaystyle x_{\rm hi}=9.000$;
$\Delta x=1.224^{+0.149}_{-0.107}$ dex.
The upper boundary reaches the adopted search limit.
& This work \\

Hybrid dust-mass relation
& $\displaystyle
y_{\rm hyb}(x)=
[1-t(x)]\,y_{\rm SNe}(x)+t(x)\,y_{\rm grow}(x)$
& Derived from the two fixed limiting branches and
the fitted transition
& This work \\

Empirical dust-mass scatter
& $\displaystyle
\sigma_{\rm dust}=
\frac{P_{84}(\Delta y)-P_{16}(\Delta y)}{2}$
& $\sigma_{\rm dust}=1.287$ dex; used for the
Monte Carlo uncertainty envelope
& This work \\

\hline
\multicolumn{4}{l}{\textit{Opacity, geometry, and attenuation}} \\[1mm]

Intrinsic FUV-opacity endpoints
& $\displaystyle
\log_{10}\kappa_{\rm FUV}^{(\rm dust)}(x)
=[1-t(x)]\log_{10}\kappa_{\rm SNe}
+t(x)\log_{10}\kappa_{\rm ISM}$
& Fixed:
$\kappa_{\rm SNe}=10^3$ and
$\kappa_{\rm ISM}=10^5\ {\rm cm^2\,g^{-1}}$
& This work; Inoue et al. (2020) \\

Geometry
& Leaky screen, leaky mixed, and Poisson-clump
geometries are optimized separately
& Selected geometry: mixed.
Objective scores:
mixed $=2.4242$,
screen $=2.4359$,
Poisson $=2.4529$
& This work \\

Covering fraction
& $\displaystyle
f_{\rm cov}(x)=
[1-t(x)]f_{\rm cov,SNe}
+t(x)f_{\rm cov,ISM}$
& $\displaystyle
f_{\rm cov,SNe}=0.050^{+0.050}_{-0.000}$;
$f_{\rm cov,ISM}=0.99$ fixed.
The low-mass value reaches its lower search boundary.
& This work \\

Active-dust fraction
& $\displaystyle
\tau_\lambda=
f_{\rm dust,active}\,
\kappa_\lambda^{(\rm dust)}\Sigma_{\rm dust}$
& $\displaystyle
f_{\rm dust,active}
=0.300^{+0.018}_{-0.000}$;
best fit at the lower search boundary
& This work \\

Effective attenuation radius
& $\displaystyle
R_{\rm e}(x)=
r_{\rm lo}+t(x)(r_{\rm hi}-r_{\rm lo})$
& Fixed endpoints:
$r_{\rm lo}=70$ pc and $r_{\rm hi}=300$ pc,
interpolated over the fitted
$x_{\rm lo}$--$x_{\rm hi}$ interval
& This work \\

Dust surface density
& $\displaystyle
\Sigma_{\rm dust}=
\frac{M_{\rm dust}}{\pi R_{\rm e}^{\,2}}$
& Derived from the adopted dust-mass and
effective-radius relations
& This work \\

Effective transport opacity
& $\displaystyle
\kappa_{\rm eff}=(1-\omega g)\kappa_{\rm ext}$
& $\omega_{\rm FUV}=0.35$ and $g_{\rm FUV}=0.60$
in the fiducial calculation
& Natta \& Panagia (1984);
Inoue (2005) \\

Leaky-screen transmission
& $\displaystyle
T_{\rm screen}=(1-f_{\rm cov})
+f_{\rm cov}
\exp\!\left[-(1-\omega g)
\frac{\tau}{f_{\rm cov}}\right]$
& Applied to the attenuation- and gas-route
optical depths
& This work \\

Leaky-mixed transmission
& $\displaystyle
T_{\rm mixed}=(1-f_{\rm cov})
+f_{\rm cov}
\frac{1-\exp(-\tau/f_{\rm cov})}
{\tau/f_{\rm cov}}$
& Preferred geometry for the full hybrid solution
& This work \\

Poisson-clump transmission
& $\displaystyle
\bar N=-\ln(1-f_{\rm cov}),\qquad
T_{\rm P}=
\exp\!\left[-\bar N
\left(1-e^{-\tau/\bar N}\right)\right]$
& Alternative geometry explored in the model comparison
& Natta \& Panagia (1984);
V\'arosi \& Dwek (1999);
Nenkova et al. (2008) \\

Far-UV attenuation
& $\displaystyle
A_{\rm FUV}=-2.5\log_{10}T_{\rm FUV}$
& Derived for the selected geometry
& This work \\

\hline
\multicolumn{4}{l}{\textit{Gas-fraction and metallicity relations}} \\[1mm]

Gas-fraction transition midpoint
& $\displaystyle
x_b=\frac{x_{\rm lo}+x_{\rm hi}}{2}$
& $x_b=8.388$
& This work \\

Gas-fraction relation
& $\displaystyle
S(x)=w\ln\!\left[
1+\exp\!\left(\frac{x-x_b}{w}\right)\right]$,
\newline
$\displaystyle
f_{\rm gas}(x)=f_b+s_{\rm low}(x-x_b)
+(s_{\rm high}-s_{\rm low})S(x)$
& $\displaystyle
f_b=0.949^{+0.014}_{-0.020}$,
\newline
$\displaystyle
s_{\rm low}=-0.0313^{+0.0221}_{-0.0438}\ {\rm dex}^{-1}$,
\newline
$\displaystyle
s_{\rm high}=-0.0708^{+0.0505}_{-0.0464}\ {\rm dex}^{-1}$,
\newline
$\displaystyle
w=0.250^{+0.100}_{-0.000}$ dex
& This work \\

Total gas mass
& $\displaystyle
M_{\rm gas}=
\frac{f_{\rm gas}}{1-f_{\rm gas}}\,M_\star$
& Derived from the fitted mass-only gas-fraction relation
& This work \\

Gas surface density
& $\displaystyle
\Sigma_{\rm gas}=
\frac{M_{\rm gas}}{\pi R_{\rm e}^{\,2}}$
& Derived from the gas-fraction and radius relations
& This work \\

Gas-fraction comparison points
& $\displaystyle
M_{\rm mol}={\rm SFR}\,t_{\rm dep}$,
$\quad
M_{\rm gas}=
(1+M_{\rm atom}/M_{\rm mol})M_{\rm mol}$
& $t_{\rm dep}=0.48$ Gyr;
fiducial $M_{\rm atom}/M_{\rm mol}=2.0$,
with range $0.7$--$4.5$.
Heintz et al. (2023) points are displayed for
comparison but are not included in the fit.
& This work; Burgarella et al. (2025) \\

Mass--metallicity relation
& Defining $z\equiv Z/Z_\odot$,
\newline
$\displaystyle
\log_{10}z(x)=
[1-t(x)]
\left[\log_{10}z_{9,\rm low}
+s_{\rm MZR,low}(x-9)\right]
+t(x)
\left[\log_{10}z_{9,\rm high}
+s_{\rm MZR,high}(x-9)\right]$
& $\displaystyle
z_{9,\rm low}
=0.0173^{+0.0335}_{-0.0073}$,
\newline
$\displaystyle
z_{9,\rm high}
=0.0606^{+0.0222}_{-0.0080}$,
\newline
$\displaystyle
s_{\rm MZR,low}
=0.0698^{+0.2602}_{-0.0679}$,
\newline
$\displaystyle
s_{\rm MZR,high}
=0.2421^{+0.1847}_{-0.1018}$
& This work; Burgarella et al. (2025) \\

Metallicity at the transition midpoint
& MZR evaluated at
$\displaystyle x_b=(x_{\rm lo}+x_{\rm hi})/2$
& $\displaystyle
Z_{\rm tr}/Z_\odot
=0.0260^{+0.0108}_{-0.0046}$
& This work \\

Oxygen-abundance conversion
& $\displaystyle
12+\log_{10}({\rm O/H})
=8.69+\log_{10}(Z/Z_\odot)$
\newline
or equivalently
$\displaystyle
\log_{10}Z=
[12+\log_{10}({\rm O/H})]-10.410$
& $Z_\odot=0.0142$
& Burgarella et al. (2025);
Asplund et al. (2009) \\

Dust-to-metal ratio
& Empirical broken DTM--metallicity relation
& Anchors at
$12+\log({\rm O/H})=7.4$ and $8.0$;
slopes $\alpha_{\rm low}=1.2$ and
$\alpha_{\rm high}=-0.5$
& Burgarella et al. (2025) \\

Gas opacity and optical depth
& $\displaystyle
\kappa_\lambda^{(\rm gas)}
=\kappa_\lambda^{(\rm dust)}
({\rm DTM}\times Z)$,
\qquad
$\displaystyle
\tau_\lambda^{(\rm gas)}
=\kappa_\lambda^{(\rm gas)}\Sigma_{\rm gas}$
& Derived from the fitted MZR and gas-fraction
relations and the adopted DTM relation
& This work \\

\hline
\multicolumn{4}{l}{\textit{Fit construction and uncertainty estimates}} \\[1mm]

Fit datasets
& Fot on combined \citealt{Burgarella2025} and this paper 
& $N_{A_{\rm FUV}}=143$,
$N_{M_{\rm dust}}=143$,
$N_{\rm MZR}=22$ at $z>9$,
$N_{f_{\rm gas}}=143$, and
$N_{A_{\rm max}}=17$
& This work \\

Objective weights
& Weighted mean of the component objective terms
& $W_{A_{\rm FUV}}=1$,
$W_{M_{\rm dust}}=0.2$,
$W_{\rm MZR}=1$,
$W_{f_{\rm gas}}=1$, and
$W_{A_{\rm max}}=20$
& This work \\

\citealt{Burgarella2025} and this paper fitted offsets
& This paper is the zero-point reference sample
& $\displaystyle
\Delta\log_{10}A_{\rm FUV}
=-0.236^{+0.066}_{-0.056}$ dex,
\newline
$\displaystyle
\Delta\log_{10}M_{\rm dust}
=-0.451^{+0.088}_{-0.114}$ dex,
\newline
$\displaystyle
\Delta{\rm MZR}
=0.499^{+0.001}_{-0.130}$ dex
& This work \\

Likelihood scatter floors
& Fixed sample-dependent intrinsic scatter terms
& \citealt{Burgarella2025}/this paper :
$\sigma_{\log A_{\rm FUV}}=0.35/0.20$ dex,
$\sigma_{\log M_{\rm dust}}=0.50/0.30$ dex,
and $\sigma_{\rm MZR}=0.25/0.15$ dex
& This work \\

Monte Carlo population scatter
& 1000 coherent Monte Carlo realisations
& $\sigma_{\log M_{\rm dust}}=1.287$ dex,
$\sigma_{\log R}=0.20$ dex,
$\sigma_{\rm MZR}=0.20$ dex,
$\sigma_{\operatorname{logit}f_{\rm gas}}=0.35$,
and
$\sigma_{\operatorname{logit}f_{\rm cov}}=2.0$;
opacity endpoints are kept fixed
& This work \\

Parameter uncertainties
& Bootstrap refitting of the continuous parameters
& 16th--84th percentiles from
80/80 successful resamples, conditional on the
selected mixed geometry
& This work \\

\hline
\end{tabular}
}
\tablefoot{The quoted parameter uncertainties are bootstrap 16th--84th-percentile intervals conditional on the selected mixed geometry. The SNe-dominated and ISM grain-growth dust-mass relations, the opacity endpoints, the high-mass covering fraction, and the 70--300 pc radius endpoints are fixed by construction. The transition boundaries, low-mass covering fraction, active-dust fraction, gas-fraction relation, mass--metallicity relation, and \citealt{Burgarella2025} and this paper zero-point offsets are determined by the combined fit.}
\end{table*}

\end{appendix}

\end{document}